\documentclass[twocolumn,showpacs,preprintnumbers,amsmath,amssymb,floatfix]{revtex4}

\usepackage{graphicx}
\usepackage{dcolumn}  % Align table columns on decimal point
\usepackage{bm}  % bold math

\begin{document}
\setlength\arraycolsep{2pt}

\title{Scaling and Universality in the Counterion-Condensation Transition \\
        at Charged Cylinders}

\author{Ali Naji}
\email{naji@ph.tum.de}
\author{Roland R. Netz}
\email{netz@ph.tum.de}

\affiliation{Physics Department, Technical University of Munich, 
		    %James Franck St., 
                  D-85748 Garching, Germany.  }

\date{August 2005}

\begin{abstract}
Counterions at charged rod-like polymers  exhibit  a condensation transition 
at a critical temperature  (or equivalently, at a critical linear 
charge density for polymers), which dramatically influences 
various static and dynamic properties of charged polymers. We address the 
critical and universal aspects of  this transition for counterions at a single charged cylinder 
in both two and three spatial dimensions using numerical and analytical methods. 
By introducing a novel Monte-Carlo sampling method in logarithmic radial scale, 
we are able to numerically simulate the critical limit of infinite system size (corresponding to 
infinite-dilution limit) within tractable equilibration times. 
The critical exponents are determined for the 
inverse moments of the counterionic density profile (which play the role of the order parameters and 
represent the inverse localization length of counterions) both within mean-field theory and  
within Monte-Carlo simulations. 
In three dimensions, we demonstrate that correlation effects (neglected within mean-field theory)
lead to an excessive accumulation of counterions near the charged 
cylinder below the critical temperature (condensation phase), while surprisingly, the critical region
exhibits universal critical exponents in accord with the mean-field theory.  
Also in contrast with  the typical trend in bulk critical phenomena,  where fluctuations
are strongly enhanced in lower dimensions, we demonstrate, using both numerical and analytical
approaches, that  the mean-field theory becomes 
exact for the 2D counterion-cylinder system at all temperatures (Manning parameters), 
when number of counterions tends to infinity. 
For finite particle number, however, the 2D problem displays a series of peculiar singular points 
(with diverging heat capacity), which reflect successive de-localization events 
of individual counterions from the central cylinder.
In both 2D and 3D, the heat capacity shows a universal jump at the critical point, and the energy develops
a pronounced peak. The asymptotic behavior of the 
energy peak location is used to locate the critical point, which is also found to be universal 
and in accordance with the mean-field prediction.
\end{abstract}
 

\maketitle

%%%%%%%%%%%%%%%%%%%%%%%%%%%%%%%%%%%%%%%%%%%%%%%%%%%%%%%%%%%%%%%%

%                               SECTION: INTRODUCTION

%%%%%%%%%%%%%%%%%%%%%%%%%%%%%%%%%%%%%%%%%%%%%%%%%%%%%%%%%%%%%%%%
\section{Introduction}
\label{sec:intro}

Electrostatics of charged polymers is often dominated
by small oppositely charged ions (counterions), 
which maintain the global electroneutrality of charged solutions. 
Many charged polymers, such as tubulin, actin and DNA are stiff and may be 
represented by straight cylinders (on length scales
smaller than the persistence length). Neglecting many-ion effects,
a single counterion is attracted by an 
electrostatic potential that grows logarithmically with the radial distance from the central cylinder axis. 
But since the counterion confinement entropy also shows a logarithmic size dependence, 
it was suggested early by Onsager \cite{Manning69}
that a counterion delocalization transition occurs at a critical 
cylinder charge or equivalently, at a critical temperature. 
Onsager's argument, which is strictly valid for a single particle, was soon corroborated by mean-field studies \cite{Manning69,Oosawa,Oosawa_Imai,Manning77,Manning_Dyn,Manning_rev78,Manning_rev96a,Manning_rev96b,Macgillivray,Ramanathan,Ramanathan_Wood82a,Ramanathan_Wood82b,Zimm}, which demonstrate that a charged cylinder
can indeed bind or condense a finite fraction of counterions 
below a critical temperature (and even in the limit of 
infinite system size with no confining boundaries), 
while above the critical temperature, all counterions de-condense and diffuse to infinity.

This {\em  counterion-condensation transition} (CCT)  dramatically affects a whole number of 
static and dynamic quantities  as observed in recent experiments on  charged polymers
\cite{Oosawa,Manning_rev78,Manning_rev96a,Ikegami,Zana,Zema81,Klein84,Ander84,Penaf92,Hoagland03}: 
upon condensation, the bare polymer charge is screened 
leading, for instance, to a significant reduction in electrophoretic 
mobility \cite{Klein84,Hoagland03} and 
conductivity of polymers \cite{Penaf92}; it also triggers striking static properties 
such as counterion-induced 
attraction between like-charged polymers, which gives rise to compact phases
of F-actin  \cite{Tang}  
and DNA  \cite{Bloom}.  Since its discovery, the CCT has been at the focus of 
numerical \cite{Winkler98,Deserno00,Liu_Muthu02,Liao03}  and analytical  \cite{Fixman79,Stigter,de-la-Cruz95,Tracy97,Levin97,Levin98,Schiess98,Nyquist99,Ray_Mann99,Qian00,Manning01,Rubinstein01,Rudi01,Henle04,Muthu04,Borukhov04,Shaugh,Naji_CC} studies. 
Under particular dispute  has been the connection between CCT and the celebrated 
Kosterlitz-Thouless transition of logarithmically interacting particles in 
two dimensions  \cite{Levin97,Kholod95_KT,Levin98_KT,Suzuki04}.

The CCT at charged cylinders is regulated by a  dimensionless control 
parameter, $\xi=q\ell_{\mathrm{B}} \tau$, known as the Manning parameter \cite{Manning69}, 
which depends on the linear charge density 
of the cylinder, $-\tau e$, charge valency of counterions, $+q$, and the Bjerrum length
$\ell_{\mathrm{B}}=e^2/(4\pi \varepsilon \varepsilon_0 k_{\mathrm{B}}T)$ accommodating
the ambient temperature $T$ and the medium dielectric constant $\varepsilon$. 
%($\tau$ and $q$ are defined to be positive). 
The Manning parameter plays the role of the {\em inverse 
rescaled temperature} and can be varied  experimentally by changing  the linear charge 
density (using synthetic chains or various $p$H) \cite{Zema81,Ander84,Penaf92,Hoagland03}  or  
by varying the dielectric media (mixing different solvents) \cite{Klein84,Hoagland03}. 
According to mean-field theory \cite{Manning69,Oosawa,Macgillivray,Ramanathan,Ramanathan_Wood82a,Ramanathan_Wood82b,Zimm}, condensation occurs above the critical value   $\xi_c=1$. 
In experiments, the critical Manning parameter appears to be about unity, but  
large deviations have also been reported \cite{Penaf92,Hoagland03,Cleland91}, 
and the precise location of the critical point is still debated \cite{Hoagland03}. 

On the other hand, it is known that the critical 
temperature may in general be influenced by correlations and fluctuations, which 
are not captured within the mean-field theory  \cite{critical}. These effects 
typically cause deviations  from mean-field predictions  in both non-universal
and universal quantities below the upper critical dimension. 
Surprisingly, the mean-field prediction for the CCT threshold, $\xi_c$, 
 has not been questioned in literature and apparently assumed to be exact. 
Likewise, the existence of universal scaling relations and critical (scaling) exponents associated with 
the CCT has not been addressed, neither on  the mean-field level nor
in the presence of correlations.

Our chief goal in this paper is to address the following issues: i) what is the exact threshold of the CCT, $\xi_c$, 
and ii) what are the critical exponents associated with this transition both in three and two spatial
dimensions. We shall also address the type of singularities that emerge in thermodynamic quantities as the CCT 
criticality sets in. To establish a systematic investigation of the correlation effects, 
we employ  Monte-Carlo simulations for counterions at a single charged cylinder using 
a novel sampling method (centrifugal sampling), which is realized by mapping the radial coordinate to a logarithmic scale. 
This enables us to investigate the critical limit of infinite system size (that is when the outer boundaries 
confining counterions tend to infinity) within tractable equilibration times in the simulations. 
The importance of taking very large system sizes becomes evident by noting that 
lateral finite-size effects, which mask the critical unbinding behavior of counterions, depend on the {\em logarithm}
of system size in the cylindrical geometry
\cite{Manning69,Manning77,Macgillivray,Ramanathan,Ramanathan_Wood82a,Ramanathan_Wood82b,Zimm,Deserno00,Levin97,Levin98,Qian00,Rubinstein01,Shaugh,Alfrey51,Fuoss51}, 
causing a quite weak convergence to the critical infinite-size limit.

Our simulations provide the first numerical results for the asymptotic critical behavior of CCT 
and systematically incorporate correlation effects (a brief report of some of our results 
has been presented previously \cite{Naji_CC}). 
The relevance of electrostatic correlations 
is in general identified  by a dimensionless coupling parameter,
$\Xi=2\pi q^3 \ell_{\mathrm{B}}^2 \sigma_{\mathrm{s}}$ with $\sigma_{\mathrm{s}}=\tau/(2\pi R)$
being the surface charge density and $R$ the radius of the cylinder. 
The mean-field theory represents the limit  $\Xi\rightarrow 0$ \cite{Netz,Andre}, while 
in the converse limit of strong  coupling, $\Xi\gg 1$,  correlations become significant and 
typically lead to drastic changes  \cite{Netz,Andre,Low_T,Naji04}.
In order to investigate scaling properties of the CCT in various regimes of the coupling parameter, 
we focus on the inverse moments of the counterionic density profile, which play the 
role of the ``order parameters" for this transition and represent the mean inverse 
localization length of counterions.  Using 
a combined finite-size-scaling analysis with respect to both
lateral size of the system and the number of counterions, we show that the order parameters adopt 
scale-invariant forms in the vicinity of the critical point. The critical exponents 
associated with the reduced temperature and the size parameters are determined both within the simulations and 
also analytically within two limiting theories of mean field and strong coupling. 
As a main result, we find that  the critical exponents of the CCT 
are {\em universal} (that is independent of the coupling parameter varied over several 
decades $0.1<\Xi<10^5$)  and appear to be in close agreement with the
mean-field prediction.  
Surprisingly, we find that  the critical Manning parameter is also 
universal and given by  the mean-field value  $\xi_c=1$.  
The transition threshold, $\xi_c$, is determined with high accuracy from the 
asymptotic behavior of the location of a singular peak that emerges in average internal energy 
 of the system. The excess heat capacity is found to vanish 
at small Manning parameters (de-condensation phase) and exhibits a {\em universal  
 jump}  at the transition point indicating that the 
CCT may be regarded as a second-order phase transition as also suggested  in a previous 
mean-field  study  \cite{Rubinstein01}. 

As will be shown, the validity of mean-field predictions in 3D breaks down as the Manning parameter increases 
beyond the critical value (i.e. in the condensation phase), where inter-particle 
correlations become significant at large couplings. This leads to an enhanced accumulation of counterions 
near the cylinder surface and a crossover to the strong-coupling theory predictions
 \cite{Netz,Andre,Naji_PhysicaA}.  
 
In order to bring out possible role of fluctuations, we also study the CCT in a 2D
counterion-cylinder system (equivalent to a 3D system composed of a central charged cylinder and parallel 
cylindrical ``counterions" with logarithmic Coulomb interactions, as may be 
applicable to an experimental system of oriented cationic and anionic polymers, e.g., DNA with polylysine \cite{Jason}).
For finite number of counterions,  a peculiar series of  singular points  emerge  that reflect 
 delocalization events of individual counterions as the Manning parameter varies. 
For increasing particle number, the singular points tend to merge and eventually 
in the thermodynamic limit, the 2D results tend to universal values determined by the mean-field
theory. Therefore, in contrast to the typical trend in critical phenomena, the
CCT  in 2D is found to be in {\em exact} agreement with the mean-field theory for the whole
range of Manning parameters (or temperatures) 
when  the number of counterions tends to infinity.  
As will be shown, the simulation results in 2D can be reproduced using an  approximate analytical
approach. A more systematic method has been developed recently  \cite{Burak_unpub} 
supporting the present analytical results. 

The organization of the paper is as follows: 
in Sections \ref{sec:CCT_intro}-\ref{sec:sim_data}, we focus
on the counterion-cylinder system in three spatial dimensions. 
Our model is introduced in Section \ref{sec:CCT_intro},
where we shall also outline the general method proposed for the investigation of  the CCT.  
In Section \ref{sec:CCT_PB}, 
we derive the scaling relations for order parameters and determine the 
asymptotic behavior of  thermodynamic quantities
within the mean-field theory (which is valid in all dimensions). In Section \ref{sec:CCT_SC}, 
analytical results are obtained within the strong-coupling theory. The
 numerical analysis of the CCT for various coupling strength will be presented 
in Sections \ref{sec:sim_data} and \ref{sec:2D} in three and two dimensions, respectively.

%%%%%%%%%%%%%%%%%%%%%%%%%%%%%%%%%%%%%%%%%%%%%%%%%%%%%%%%%%%%%%%%%%%%

%                               SECTION: CCT 3D

%%%%%%%%%%%%%%%%%%%%%%%%%%%%%%%%%%%%%%%%%%%%%%%%%%%%%%%%%%%%%%%%%%%%
\section{Counterion-condensation transition (CCT) in three dimensions}
\label{sec:CCT_intro}

%%%%%%%%%%%%%%%%%%%%%%%%%%%%%%%%%%%%
%                  SUBSECTION: MODEL
%%%%%%%%%%%%%%%%%%%%%%%%%%%%%%%%%%%%
\subsection{Cell model for charged rod-like polymers}
\label{subsec:model}

\begin{figure}[t]
\begin{center}
\includegraphics[angle=0,width=4cm]{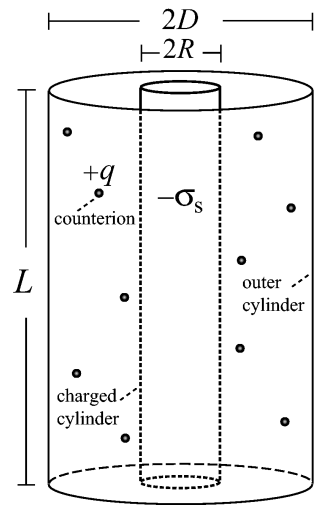}
\caption{The three-dimensional model consists of a charged
cylinder of infinite length, $H$,  and its neutralizing counterions confined in an outer cylindrical box 
(see the text for parameters).}
\label{fig:model}
\end{center}
\end{figure}

We consider a primitive cell model \cite{Alfrey51,Fuoss51,Lifson54}, which consists of 
a single charged cylinder of radius $R$ 
and point-like neutralizing counterions  of charge valency $+q$ that are 
confined  laterally  in an outer (co-axial) cylindrical box of 
radius $D$--see Figure \ref{fig:model}. 
The cylinder has infinite length, $L$, and a uniform (surface) charge distribution,  
$-\sigma({\mathbf x})e$, where $\sigma({\mathbf x})=\sigma_{\mathrm{s}}\delta(r-R)$. 
(Note that $q$ and $\sigma_{\mathrm{s}}$ are given in units of the elementary charge, $e$, and 
are positive by definition.)  
The cylinder is assumed to be rigid and  impenetrable to counterions and the dielectric medium
is represented by a uniform dielectric constant, $\varepsilon$. In the three dimensions, 
electric charges interact via  bare Coulombic interaction
\begin{equation}
	v_{\mathrm{3D}}({\mathbf x})=1/|{\mathbf x}|.  
\end{equation}

The electroneutrality condition holds globally inside the cell and entails the relation 
\begin{equation}
    qN=\tau L, 
\label{eq:ENC_v1}
\end{equation}
where $N$ is the number of counterions per cell and $\tau=2\pi R \sigma_{\mathrm{s}}$ represents 
the linear charge density of the cylinder. 
The system is described by the Hamiltonian
\begin{eqnarray}
   \frac{{\mathcal H}_N}{k_{\mathrm{B}}T} &=& 
        q^2 \ell_{\mathrm{B}} \sum_{\langle ij \rangle} v_{\mathrm{3D}}({\mathbf x}_i-{\mathbf x}_j)     
        \nonumber\\
        &-& q \ell_{\mathrm{B}} \sum_{i=1}^{N}\int\, 
               v_{\mathrm{3D}}({\mathbf x}-{\mathbf x}_i) \,\sigma({\mathbf x})  \, {\mathrm d}{\mathbf x}
        \nonumber\\
        &+&\frac{ \ell_{\mathrm{B}} }{2} \int \,  \sigma({\mathbf x})\,v_{\mathrm{3D}}
                 ({\mathbf x}-{\mathbf x}')\, \sigma({\mathbf x}') \, {\mathrm d}{\mathbf x} \, {\mathrm d}{\mathbf x}', 
\label{eq:H_full}
\end{eqnarray}
which comprises mutual repulsions between counterions located at 
$\{{\mathbf x}_i\}$ (first term), the counterion-cylinder attraction (second term) and
the self-energy of the cylinder (last term). It can be written as 
\begin{equation}
   \frac{{\mathcal H}_N}{k_{\mathrm{B}}T} = 
        q^2 \ell_{\mathrm{B}} \sum_{\langle ij \rangle} v_{\mathrm{3D}}({\mathbf x}_i-{\mathbf x}_j)     
        + 2 \xi \sum_{i=1}^{N} \ln \left(\frac{r_i}{R}\right)+C_0,
\label{eq:Hamilt}
\end{equation}
where $\xi$ is the {\em Manning parameter} of the system \cite{Manning69,Manning77},
\begin{equation}
  \xi=q \ell_{\mathrm{B}} \tau
\label{eq:xi}
\end{equation}
with $\ell_{\mathrm{B}}=e^2/(4\pi \varepsilon\varepsilon_0 k_{\mathrm{B}}T)$ being the Bjerrum length
(in water and at room temperature $\ell_{\mathrm{B}}\simeq 7$\AA),  
and $r_i=(x_i^2+y_i^2)^{1/2}$ being the radial coordinate of the $i$-th counterion from the cylinder axis,
which coincides with $z$-axis. 
The additive term $C_0$ in Eq. (\ref{eq:Hamilt}) is related to the cylinder self-energy, 
which will be important in obtaining a convergent energy expression for the system in the simulations
(Section \ref{subsec:sim_para} and Appendix \ref{app:H_periodic}).

%%%------Fig Model

%%%%%%%%%%%%%%%%%%%%%%%%%%%%%%%%%%%%
%     SUBSECTION: Dimensionless
%%%%%%%%%%%%%%%%%%%%%%%%%%%%%%%%%%%%
\subsection{Dimensionless description}
\label{subsec:d_less}

The parameter space of the system may be spanned by a minimal set of independent 
dimensionless parameters obtained from the ratios between 
characteristic length scales. These length scales are
the rescaled Bjerrum length, $q^2 \ell_{\mathrm{B}}$, the Gouy-Chapman length 
\begin{equation}
  \mu=\frac{1}{2\pi q \ell_{\mathrm{B}} \sigma_{\mathrm{s}}},
\label{eq:mu}
\end{equation}
and the radius of the charged cylinder, $R$,  and that of the 
outer boundary, $D$. The rescaled cylinder radius 
\begin{equation}
   \tilde R=\frac{R}{\mu}=\xi
\label{eq:xi_Rtilde}
\end{equation} 
equals the Manning parameter, $\xi$.
The ratio between the rescaled Bjerrum length and the Gouy-Chapman length, $\mu$, 
gives the so-called {\em electrostatic coupling parameter} \cite{Netz}, 
\begin{equation} 
 \Xi=\frac{q^2 \ell_{\mathrm{B}}}{\mu}=2\pi q^3\ell_{\mathrm{B}}^2\sigma_{\mathrm{s}},
\label{eq:Xi_coupl}
\end{equation}
which can identify the importance of electrostatic correlations 
in a charged system \cite{Netz,Andre,Naji04,Naji_PhysicaA}, and 
the ratio between $D$ and $R$, which enters only through the 
{\em lateral extension parameter}
\begin{equation}
   \Delta\equiv \ln \left(\frac{D}{R}\right) 
\label{eq:Delta}
\end{equation}
characterizing lateral finite-size effects. 
The relevant infinite-system-size limit is obtained for 
$\Delta\rightarrow \infty$ \cite{Zimm,Alfrey51,Fuoss51}.  

We shall use the dimensionless form of the Hamiltonian 
obtained by rescaling the spatial coordinates
as ${\tilde {\mathbf x}}={\mathbf x}/\mu$ \cite{Netz}, that is 
\begin{equation}
   \frac{{\mathcal H}_N}{k_{\mathrm{B}}T} = 
        \Xi \sum_{\langle ij \rangle} v_{\mathrm{3D}}({\tilde {\mathbf x}}_i-{\tilde {\mathbf x}}_j) 
        + 2 \xi \sum_{i=1}^{N} \ln \left(\frac{{\tilde r}_i}{\tilde R}\right)+C_0.
\label{eq:Hamilt_rescaled}
\end{equation}
The electroneutrality condition (\ref{eq:ENC_v1}) in rescaled units reads 
\begin{equation}
   2\pi\xi {\tilde L}=2\pi\Xi N,
\label{eq:ENC}
\end{equation} 
where the left hand side is simply the rescaled area of the cylinder covered
by the electric charge. The thermodynamic limit  is obtained for  
$N\rightarrow \infty$ and $L\rightarrow \infty$, but keeping $N/L=\tau/q$ (or equivalently, $N/\tilde L=\xi/\Xi$) fixed.

%%%%%%%%%%%%%%%%%%%%%%%%%%%%%%%%%%%%
%     SUBSECTION:BINDING-UNB. TRANS. 
%%%%%%%%%%%%%%%%%%%%%%%%%%%%%%%%%%%%
\subsection{CCT as a generic binding-unbinding process}
\label{subsec:generic}

The statistical physical properties of the system may 
be investigated using the canonical partition function,
\begin{equation}
  {\mathcal Z}_N=\frac{\mu^{3N}}{N!}
        \int_{\tilde V} \left[ \prod_{i=1}^{N} {\mathrm{d}}{\tilde z}_i \,{\mathrm{d}}\phi_i\, {\mathrm{d}}{\tilde r}_i\,
                                        {\tilde r}_i\right]
                 \exp\bigg\{- \frac{{\mathcal H}_N}{k_{\mathrm{B}}T}\bigg\}
\label{eq:Z}
\end{equation}
represented in cylindrical coordinates ${\tilde {\mathbf x}}_i=({\tilde r}_i, \phi_i, {\tilde z}_i)$, with 
the spatial integral running over the volume, $\tilde V$, of the space accessible for counterions, i.e.
 ${\tilde R}\leq {\tilde r} \leq {\tilde D}$. 

Naively, one may conjecture that the partition function (\ref{eq:Z})
diverges in a certain range of Manning parameters, 
when the upper boundary of the radial integrals, ${\tilde D}$,  tends to infinity,  as
may be indicated by the logarithmic form of the counterion-cylinder 
interaction, which gives rise to algebraic prefactors of the form
${\tilde r}_i^{1-2\xi}$ in the integrand. 
The possible emergence of a divergency
in a charged cylindrical system was first pointed out
by Onsager and the connection with the counterion-condensation
transition was discussed by Manning \cite{Manning69}. 

Here we demonstrate this peculiar point 
using a  transformation of coordinates, which provides the basis 
for our numerical simulations considered later in Sections \ref{sec:sim} and  \ref{sec:sim_data}. 
The radial coordinate is transformed as
\begin{equation}   
  y=\ln \left(\frac{\tilde r}{\tilde R}\right),
\label{eq:y}
\end{equation}
upon which the partition function in (\ref{eq:Z}) transforms as
\begin{equation}
  {\mathcal Z}_N=\frac{\mu^{3N}{\tilde R}^{2N}}{N!}
        \int_{\tilde V} \left[\prod_{i=1}^{N} {\mathrm{d}}{\tilde z}_i\, {\mathrm{d}}\phi_i\, {\mathrm{d}} y_i\right]
                 \exp\bigg\{-\frac{{\mathcal H}_N^\ast}{k_{\mathrm{B}}T}\bigg\},
\label{eq:Z_y}
\end{equation}
where the volume integral  runs over the region 
$0<y<\Delta=\ln (D/R)$, and 
\begin{equation}
  \frac{{\mathcal H}_N^\ast}{k_{\mathrm{B}}T}=\sum_{i=1}^{N} W(y_i)
                +\Xi \sum _{\langle ij \rangle} v_{\mathrm{3D}}({\tilde {\mathbf x}}_i-{\tilde {\mathbf x}}_j)+C_0
\label{eq:H_virtual}
\end{equation}
is the transformed Hamiltonian of the system with
\begin{equation}
  W(y)=2(\xi-1)y.
\label{eq:W_y}
\end{equation}
As  seen,  the original partition function is
now mapped to the partition function of a system of 
interacting (repelling) particles in a {\em linear}  potential well, $W(y)$. 
This virtual potential includes the contributions associated with the
cylindrical boundary, namely, the bare counterion-cylinder attraction (i.e. $2\xi y$) and 
an entropic (repulsive) term from the measure of the radial integral (i.e.  $2y$), which 
may be regarded as an induced {\em centrifugal} component.

For small Manning parameter, $\xi<1$, 
the potential well, $W(y)$, becomes purely repulsive suggesting that counterions 
{\em unbind} (or ``de-condense")  from the central cylinder 
departing  to infinitely large distances as the outer confining boundary tends to infinity,
$\Delta=\ln (D/R) \rightarrow \infty$.
In contrast  for $\xi>1$, the potential well exerts an attractive
force upon counterions, which 
might lead to {\em partial binding} (or  ``condensation") of counterions even in the absence of 
confining walls. The new representation of ${\mathcal Z}_N$ in Eq. (\ref{eq:Z_y}), therefore,  
reflects the interplay between energetic and entropic factors on a microscopic level. 

Note that the rigorous analytical derivation of the aforementioned properties for counterions
based on the full partition function is still an open problem, and only 
approximate limiting cases have been examined analytically (Section \ref{subsec:beyond_ons}).

%%%%%%%%%%%%%%%%%%%%%%%%%%%%%%%%%%%%
%        SUBSECTION: ONSAGER INSTAB. 
%%%%%%%%%%%%%%%%%%%%%%%%%%%%%%%%%%%%
\subsection{Onsager instability}
\label{subsec:onsager}

As a simple illustrative case, let us consider a ``hypothetical'' system, 
in which mutual counterionic repulsions are 
 switched off. The partition function (\ref{eq:Z}) 
thus factorizes as ${\mathcal Z}_N\sim {\mathcal Z}_1^N$, where 
\begin{equation}
  {\mathcal Z}_1=
        \int_0^\Delta {\mathrm{d}}y\,e^{(2-2\xi)y}
        =\frac{e^{(2-2\xi)\Delta}-1}{2-2\xi}
\label{eq:Z_single}
\end{equation}
is the single-particle partition function. It diverges for  
$\xi<1$, when the lateral extension parameter, $\Delta$, tends to infinity, which implies 
complete de-condensation of counterions,
i.e. the probability, $P(r)\sim \exp(-2\xi \ln r)/{\mathcal Z}_1$, of finding counterions 
at any finite distance, $r$, from the cylinder
tends to zero (equivalent to a vanishing density profile, $\rho(r)=N P(r)$). 
But ${\mathcal Z}_1$ and the counterionic density profile remain finite for $\xi>1$, 
indicating that the
Manning parameter $\xi_c=1$ is the onset of the CCT 
on the {\em one-particle} level, which we term here as the {\em Onsager instability} (in the
spirit of Onsager's original argument  \cite{Manning69}). 
Onsager instability captures the basic features of the CCT. It exhibits the weak logarithmic 
convergence (via $\Delta=\ln D/R$)   to the critical limit 
as the volume per polymer ($\sim D^2$) goes to infinity   \cite{Note_Ons}, and as shown 
in Appendix \ref{app:Onsager},  displays algebraic singularities 
in energy and heat capacity (at $\xi_c=1$) that may be identified by a set of scaling exponents.
Such scaling relations are crucial in our analysis of the CCT in the following sections.

We emphasize here that the results obtained within Onsager instability 
are by no means conclusive as soon as 
inter-counterionic interactions are switched on, which, as will be shown, 
lead to qualitative differences. In particular, it turns out that a diverging partition function is {\em not}
necessarily an indication of the {\em onset} of the CCT
as asserted by the single-particle argument \cite{Manning69}.

%%%%%%%%%%%%%%%%%%%%%%%%%%%%%%%%%%%%
%        SUBSECTION: BEYOND ONSAGER  
%%%%%%%%%%%%%%%%%%%%%%%%%%%%%%%%%%%%
\subsection{Beyond the Onsager instability: many-body effects 
and electrostatic correlations}
\label{subsec:beyond_ons}

Many-body terms  involved in the full partition function (\ref{eq:Z})
render the systematic analysis of the CCT quite difficult. 
The analytical results are available in the asymptotic limits of
i) vanishingly small coupling parameter $\Xi\rightarrow 0$, 
which leads to the  mean-field  or Poisson-Boltzmann (PB) theory, 
and ii) for infinitely large coupling parameter $\Xi\rightarrow \infty$, which leads
to the strong-coupling (SC) theory \cite{Netz}. 
In the mean-field approximation (case i), 
statistical correlations among counterions are systematically neglected. 
In the opposite limit of strong coupling (case ii),
the leading contribution to the partition function takes a very simple form
comprising only the one-particle (counterion-cylinder) 
contributions, which is associated with strong electrostatic correlations (pronounced correlation hole)
between counterions at the surface \cite{Netz,Andre,Naji04,Naji_PhysicaA,Burak04}. 
We study the mean-field predictions for the CCT in Section \ref{sec:CCT_PB}. 
The SC description ($\Xi\rightarrow \infty$) 
resembles the Onsager instability and will be discussed  in Section \ref{sec:CCT_SC} and  
Appendix \ref{app:Onsager}. 
The perturbative improvement of these two limiting theories in a system of {\em finite}
coupling parameter, $\Xi$, is formally possible by computing higher-order
correction terms as previously performed for planar charged walls \cite{Netz,Andre},
but will not be considered here. 

Interestingly, in both limits, 
the onset of the CCT is obtained as $\xi_c=1$, which is due to 
the simplified form of the counterionic correlations. 
An important question is whether the critical value, $\xi_c$, 
varies with the coupling parameter. Such a behavior may be expected since 
the Manning parameter represents the rescaled inverse temperature
of the system (i.e. $\xi=T_\ast/T$ with 
$T_\ast\equiv q\tau e^2/(4\pi\varepsilon\varepsilon_0k_{\mathrm{B}})$), which,   
as known from bulk critical phenomena \cite{critical}, can be shifted from its mean-field value
due to inter-particle correlations for large couplings. 
Also it is interesting to examine whether  the CCT exhibits 
scale-invariant properties near $\xi_c$ and if it can be classified in terms
of a universal class of critical exponents. Such scaling relations are known to represent relevant 
statistical characteristics of systems close to continuous phase transitions \cite{critical}.

To address these issues, one has to define quantities which can serve as 
{\em order parameters} of the CCT. In the following section, we shall introduce such quantities and, 
by considering the mean-field theory, show that the order parameters  indeed exhibit 
scaling behavior near the CCT critical point. 
We return to the influence of electrostatic correlations on the critical Manning parameter and 
 scaling exponents of the CCT in the subsequent  sections.

%%%%%%%%%%%%%%%%%%%%%%%%%%%%%%%%%%%%%%%%%%%%%%%%%%%%%%%%%%%%%%%%%%%%%%

%                               SECTION: PB

%%%%%%%%%%%%%%%%%%%%%%%%%%%%%%%%%%%%%%%%%%%%%%%%%%%%%%%%%%%%%%%%%%%%%%
\section{Mean-field theory for the counterion-condensation transition}
\label{sec:CCT_PB}

%%%%%%%%%%%%%%%%%%%%%%%%%%%%%%%%%%%%
%              SUBSECTION: PB THEORY 
%%%%%%%%%%%%%%%%%%%%%%%%%%%%%%%%%%%%
\subsection{Non-linear Poisson-Boltzmann (PB) equation}
\label{subsec:PB_cyl}

The  mean-field theory can be derived systematically using a saddle-point analysis in the limit
 $\Xi\rightarrow 0$ \cite{Netz}. It  is governed by the well-known Poisson-Boltzmann (PB) equation  \cite{Alfrey51,Fuoss51}, which, 
in rescaled units, reads (Appendix \ref{app:PB_rescal})
\begin{equation}
            \nabla^2_{{\tilde {\mathbf x}}} \psi=
                        2{\tilde \sigma}({\tilde {\mathbf x}})
                       -{\tilde \kappa}^2\,{\tilde \Omega}({\tilde {\mathbf x}})
                            \,e^{-\psi({\tilde {\mathbf x}})}
\label{eq:PB}
\end{equation}
for the dimensionless potential field $\psi({\tilde {\mathbf x}})$. Here
\begin{equation}
 	{\tilde \sigma}({\tilde {\mathbf x}})=\delta({\tilde r}-{\tilde R})
\label{eq:sigma_res}
\end{equation}
is the rescaled charge distribution of the cylinder and 
\begin{equation}
    {\tilde \Omega}({\tilde {\mathbf x}})= {\tilde \Omega}({\tilde r})
          = \left\{
             \begin{array}{ll}
               1
              & {\,\,\,\,\,\,\,\,{\tilde R}\leq {\tilde r} \leq {\tilde D},}\\
            \\
               0
              & {\,\,\,\,\,\,\,\,{\mathrm{otherwise}}}
      \end{array}
        \right. 
\label{eq:Omega}
\end{equation}
specifies the volume accessible to  counterions. 
In the canonical ensemble, one has
\begin{equation}
  \frac{{\tilde \kappa}^2}{2}
         =\frac{2\pi\xi{\tilde L}}  
         {\int {\mathrm{d}}{\tilde {\mathbf x}}\, {\tilde \Omega}({\tilde {\mathbf x}})\,\exp(-\psi)}.
\label{eq:Lambda_kappa}
\end{equation}

Assuming the cylindrical symmetry (for an infinitely long cylinder) and using Eq. (\ref{eq:PB})
and the global electroneutrality condition (\ref{eq:ENC}), one obtains 
\begin{eqnarray}
          \left({\tilde r}\frac{{\mathrm{d}}\psi}{{\mathrm{d}}{\tilde r}}\right)_{{\tilde r}={\tilde R}}=
                                               2\xi 
                                               \,\,\,\,\,{\mathrm{and}}\,\,\,\,\,
          \left({\tilde r}\frac{{\mathrm{d}}\psi}{{\mathrm{d}}{\tilde r}}\right)_{{\tilde r}={\tilde D}}=0, 
\label{eq:PBboundary}
\end{eqnarray}
which are used to solve the PB equation (\ref{eq:PB})
 in the non-trivial region ${\tilde R}\leq {\tilde r} \leq {\tilde D}$  \cite{Alfrey51,Fuoss51}.
Thereby, one obtains 
both the free energy (Section \ref{subsubsec:EC_PB}) and the rescaled radial 
density profile of counterions around the charged cylinder
\begin{equation}
  {\tilde \rho}({\tilde r}) = \frac{{\tilde \kappa}^2}{2}
           {\tilde \Omega}({\tilde r})\, e^{-\psi({\tilde r})}.
\label{eq:density}
\end{equation}
The rescaled density profile, ${\tilde \rho}({\tilde r})$, is related to the actual {\em number} density
of counterions, $\rho(r)$, through ${\tilde \rho}({\tilde r})=\rho(r)/(2\pi\ell_{\mathrm{B}}\sigma_{\mathrm{s}}^2)$
\cite{Netz} (Appendix \ref{app:PB_rescal}).  

As shown by  Alfrey et al. \cite{Alfrey51} and Fuoss et al. \cite{Fuoss51}, the 
PB solution takes different functional forms depending on whether $\xi$ 
lies below or above the Alfrey-Fuoss threshold
\begin{equation}
  \Lambda_{\mathrm{AF}}=\frac{\Delta}{1+\Delta}, 
\label{eq:lambda}
\end{equation}
that is 
\begin{eqnarray}
     \psi_{\mathrm{PB}}({\tilde r})= \left\{
             \begin{array}{ll}
             \ln[
                 \frac{{\tilde \kappa}^2{\tilde r}^2}{2\beta^2}
                    \sinh^2(\beta \ln \frac{{\tilde r}}{{\tilde R}}
                               +\coth^{-1}\frac{\xi-1}{\beta})] 
             & {\xi\leq\Lambda_{\mathrm{AF}},}\\
        \\
             \ln[
                 \frac{{\tilde \kappa}^2{\tilde r}^2}{2\beta^2}
                    \sin^2(\beta \ln \frac{{\tilde r}}{{\tilde R}}
                               +\cot^{-1}\frac{\xi-1}{\beta})] 
             & {\xi\geq\Lambda_{\mathrm{AF}},}
      \end{array}
        \right. 
\label{eq:psi}
\end{eqnarray}
where $\beta$ is given by the transcendental equations
\begin{eqnarray}
    \xi = \left\{
             \begin{array}{ll}
               \frac{1-\beta^2}
                  {1-\beta\coth(-\beta\Delta)}
              & {\,\,\,\,\,\xi\leq\Lambda_{\mathrm{AF}},}\\
            \\
               \frac{1+\beta^2}
                  {1-\beta\cot(-\beta\Delta)}
              & {\,\,\,\,\,\xi\geq\Lambda_{\mathrm{AF}}.}
      \end{array}
        \right. 
\label{eq:beta}
\end{eqnarray}
The PB density profile of counterions, Eq. (\ref{eq:density}), 
is then obtained for ${\tilde R}\leq {\tilde r} \leq {\tilde D}$ as
\begin{eqnarray}
     {\tilde \rho}_{\mathrm{PB}}({\tilde r})= \frac{\beta^2}{{\tilde r}^2}\times
            \left\{
             \begin{array}{ll}
              \sinh^{-2}(\beta \ln \frac{{\tilde r}}{{\tilde R}}
                               +\coth^{-1}\frac{\xi-1}{\beta})
             & {\xi\leq\Lambda_{\mathrm{AF}},} \\
            \\
            \sin^{-2}(\beta \ln \frac{{\tilde r}}{{\tilde R}}
                               +\cot^{-1}\frac{\xi-1}{\beta})
            & {\xi\geq\Lambda_{\mathrm{AF}},}
      \end{array}
        \right. 
\label{eq:rho}
\end{eqnarray}
where we have arbitrarily chosen $\psi_{\mathrm{PB}}({\tilde r}={\tilde R})=0$ to fix the reference 
of the potential. This condition also fixes ${\tilde \kappa}$ in Eq. (\ref{eq:psi}) as well as the 
radial density of counterions {\em at contact} with the cylinder using Eq. (\ref{eq:density}), i.e.
\begin{eqnarray}
    \frac{{\tilde \kappa}^2}{2} = {\tilde \rho}_{\mathrm{PB}}({\tilde R})= 
          \frac{1}{\xi^2}\times\left\{
             \begin{array}{ll}
              (\xi-1)^2-\beta^2
              & {\,\,\,\,\xi\leq\Lambda_{\mathrm{AF}},}\\
            \\
               (\xi-1)^2+\beta^2
              & {\,\,\,\,\xi\geq\Lambda_{\mathrm{AF}}.}
      \end{array}
        \right. 
\label{eq:kappa}
\end{eqnarray}
The density profiles given in Eq. (\ref{eq:rho}) 
are in fact normalized to the total number of counterions, $N$, a condition 
imposed via Eq. (\ref{eq:Lambda_kappa}).  Using Eq. (\ref{eq:density}), the 
normalization condition in rescaled units reads  (Appendix \ref{app:PB_rescal})
\begin{equation}
   \int_{\tilde R}^{\tilde D}{\mathrm{d}}{\tilde r}\,{\tilde r}\,{\tilde \rho}_{\mathrm{PB}}({\tilde r})=\xi.
\label{eq:density_norm}
\end{equation}

%%%%%%%%%%%%%%%%%%%%%%%%%%%%%%%%%%%%
%              SUBSECTION: PB ONSET 
%%%%%%%%%%%%%%%%%%%%%%%%%%%%%%%%%%%%
\subsection{Onset of the CCT within mean-field theory}
\label{subsec:PB_onset}

The threshold of CCT  within the mean-field PB  theory was considered by several workers 
\cite{Macgillivray,Ramanathan,Ramanathan_Wood82a,Ramanathan_Wood82b,Zimm,Alfrey51,Fuoss51}. 
It may  be obtained from the asymptotic behavior of the density profile
($\Delta\rightarrow \infty$) as  reviewed below. 

First note that for $\Delta\gg 1$, the Alfrey-Fuoss threshold 
$\Lambda_{\mathrm{AF}}$, Eq. (\ref{eq:lambda}), tends to unity from below, i.e. 
\begin{equation}
  \Lambda_{\mathrm{AF}} = 1-\frac{1}{\Delta}+{\mathcal O}(\Delta^{-2}).
\label{eq:lambda_limiting}
\end{equation}
Therefore, for Manning parameter $\xi<1$, one may use
the first relation in Eq. (\ref{eq:beta}) to obtain the limiting behavior of 
the integration constant $\beta$ as (Appendix \ref{app:PB_beta})
\begin{equation}
   \beta = (1-\xi)+{\mathcal O}\bigg(e^{-2\Delta(1-\xi)}\bigg), 
\label{eq:beta_exp_low}
\end{equation}
when $\Delta\rightarrow \infty$. 
Using this into Eq. (\ref{eq:kappa}), one finds that 
the density of counterions at contact, ${\tilde \rho}_{\mathrm{PB}}({\tilde R})$,
asymptotically vanishes. Hence,
the density profile (\ref{eq:density}) at any finite
distance from the cylinder  tends to zero 
for $\xi\leq1$, i.e.
\begin{equation}
  {\tilde \rho}_{\mathrm{PB}}({\tilde r}) \rightarrow 0,
\label{eq:density_asyp_low}
\end{equation} 
representing the de-condensation regime in the limit  $\Delta\rightarrow \infty$.
For $\xi\geq1$, on the other hand, one has  
$\beta\rightarrow 0$ for increasing $\Delta$
(Appendix \ref{app:PB_beta}), and thus using 
Eq. (\ref{eq:kappa}), 
\begin{equation}
   {\tilde \rho}_{\mathrm{PB}}({\tilde R})\rightarrow \frac{(\xi-1)^2}{\xi^2}.
\label{eq:density_R_high}
\end{equation}
Using the second relation in Eq. (\ref{eq:rho}) and expanding for small
$\beta$, the radial density profile follows as  
\cite{Ramanathan,Netz_Joanny}
\begin{equation}
  {\tilde \rho}_{\mathrm{PB}}({\tilde r})\rightarrow \frac{(\xi-1)^2}{\xi^2}
                        \left[\frac{\tilde r}{\tilde R}\right]^{-2}
                        \left[1+(\xi-1)\ln \frac{\tilde r}{\tilde R}\right]^{-2}
\label{eq:density_asyp_high}
\end{equation}
in the limit $\Delta\rightarrow \infty$ (see also Appendix \ref{app:PB_Delta_infty}), which is finite
and indicates condensation of counterions.
This proves that the mean-field critical point is given by 
\begin{equation}
  \xi_c^{\mathrm{PB}}=1, 
\end{equation}
corresponding to the mean-field  critical temperature
\begin{equation}
T_c^{\mathrm{PB}}=\frac{q \tau e^2 }{4\pi \varepsilon \varepsilon_0 k_{\mathrm{B}} }.
\end{equation}

%%%%%%%%%%%%%%%%%%%%%%%%%%%%%%%%%%%%
%       SUBSECTION: PB SCALING EXPS.
%%%%%%%%%%%%%%%%%%%%%%%%%%%%%%%%%%%%
\subsection{Critical scaling-invariance: Mean-field exponents}
\label{subsec:PB_exp}

It is readily seen from Eqs. (\ref{eq:density_R_high}) and (\ref{eq:density_asyp_high}) that 
the asymptotic density of counterions admits a scale-invariant or 
homogeneous form with respect to the {\em reduced Manning parameter},
\begin{equation}
  \zeta = 1-\frac{\xi_c^{\mathrm{PB}}}{\xi},
\end{equation}
close to the critical value $\xi_c^{\mathrm{PB}}=1$. 
Note that the reduced Manning parameter equals the {\em reduced temperature} of the system, 
$t=1-(T/T^{\mathrm{PB}}_c)$, when other quantities such as 
the dielectric constant, $\varepsilon$, and the linear charge density of the cylinder, 
$\tau$, are kept fixed. (Experimentally, however, the Manning parameter may be varied by
changing  $\varepsilon$ \cite{Klein84,Hoagland03}
or $\tau$ \cite{Zema81,Ander84,Penaf92,Hoagland03} at constant temperature,
in which case, $\zeta$ can be related to
the reduced dielectric constant or the reduced linear charge density.)

In a finite confining volume (finite $\Delta$), such scaling forms 
with respect to $\zeta$ do not hold since the true
CCT is suppressed. Yet as a general trend \cite{critical}, we  
expect that for {\em sufficiently large} $\Delta$, 
the reminiscence of such scaling relations appears in the form of 
 finite-size-scaling relations near the transition point. These relations 
would involve both $\zeta$ and the lateral extension 
parameter, $\Delta$, (as the only relevant parameters in the mean-field limit) 
in a scale-invariant fashion as will be shown below.

%------------------------------------
%            SUBSUBSECTION: PB ORDER PARAMETER 
%------------------------------------
\subsubsection{The CCT order parameters}
\label{subsubsec:order_p}

As possible candidates for the CCT ``order parameter", 
we use the inverse moments of the counterionic density profile
\begin{equation}
   S_n(\xi, {\tilde D})\equiv \bigg\langle \frac{1}{{\tilde r}^n}\bigg\rangle
   = \frac{\int_{\tilde R}^{\tilde D} \,{\tilde r}\,{\mathrm{d}}{\tilde r}\,\,{\tilde r}^{-n}\,{\tilde \rho}({\tilde r})}
          {\int_{\tilde R}^{\tilde D}\, {\tilde r}\,{\mathrm{d}}{\tilde r}\,{\tilde \rho}({\tilde r})}
\label{eq:S_n_def}
\end{equation}
where $n>0$ \cite{Note_Sn}. Note that these quantities reflect {\em mean inverse localization length} of counterions. 
In the condensation phase (where counterions adopt a finite density profile), one has $S_n>0$, 
reflecting a finite localization length. But at the critical point and 
in the de-condensation phase (with vanishing counterionic density profile),  one has $S_n=0$
in the limit of infinite system size $\Delta\rightarrow \infty$, which 
indicates  a diverging counterion localization length.  

In order to derive the mean-field finite-size-scaling relations for $S_n$ near $\xi_c^{\mathrm{PB}}=1$, 
we focus on the PB solution in the regime of Manning parameters 
$\xi\geq\Lambda_{\mathrm{AF}}$, since
for any finite $\Delta$, we have $\Lambda_{\mathrm{AF}}\leq\xi_c^{\mathrm{PB}}=1$
from Eq. (\ref{eq:lambda_limiting}).
Inserting the first relation in Eq. (\ref{eq:rho}) into
Eq. (\ref{eq:S_n_def}), we obtain
\begin{equation}
   S_n^{\mathrm{PB}}=\frac{\beta^2}{\xi}
             \int_{\tilde R}^{\tilde D} {\mathrm{d}}{\tilde r}\,\, {\tilde r}^{-n-1} 
                \sin^{-2}(\beta \ln \frac{{\tilde r}}{{\tilde R}}
                               +\cot^{-1}\frac{\xi-1}{\beta}).
\end{equation}
Changing the integration variable as $y=\ln ({\tilde r}/{\tilde R})$, we get  
\begin{equation}
   S_n^{\mathrm{PB}}=\frac{\beta^2}{\xi^{n+1}}
             \int_0^\Delta {\mathrm{d}}y \, e^{-ny} 
                \sin^{-2}(\beta y +\cot^{-1}\frac{\xi-1}{\beta}).
\label{eq:S_n_PB}
\end{equation} 
For $\Delta\gg1$, 
the above relation may be approximated by a simple analytic expression 
as (Appendix  \ref{app:PB_S_n})
\begin{equation}
   S_n^{\mathrm{PB}}(\zeta, \Delta)\simeq 
                           \frac{1}{n}\bigg[\zeta^2+\beta^2(\zeta, \Delta)\bigg]
\label{eq:S_n_PB_approx}
\end{equation}
for $\xi$ being sufficiently close to the critical value $\xi_c^{\mathrm{PB}}=1$. 

Using the above result, we may distinguish two limiting cases, where different 
scaling relations are obtained, namely, i) when $\Delta\rightarrow \infty$ 
but $\zeta = 1-\xi_c^{\mathrm{PB}}/\xi$  is {\em finite} and close to the critical value $\zeta^{\mathrm{PB}}=0$,
and ii) when $\Delta$ is finite and large, but the system tends towards the critical point, 
$\zeta\rightarrow \zeta_c^{\mathrm{PB}}=0$.

In the first case, as stated before, we have $\beta\rightarrow 0$ 
for the above-threshold regime, $\zeta\geq 0$; thus using Eq. (\ref{eq:S_n_PB_approx}), we obtain 
\begin{equation}
  S_n^{\mathrm{PB}}(\zeta, \Delta\rightarrow \infty)\simeq \frac{\zeta^2}{n}.
\label{eq:Sn_PB_zeta}
\end{equation}
On the other hand, $S_n^{\mathrm{PB}}$ 
vanishes for $\zeta\leq 0$ (Appendix \ref{app:PB_S_n}). Hence, the following
scaling relation is obtained in the infinite-system-size limit  $\Delta\rightarrow \infty$, 
\begin{equation}
     S_n^{\mathrm{PB}}(\zeta, \infty)\simeq \left\{
             \begin{array}{ll}
             \zeta^{\chi_{\mathrm{PB}}}/n   & {\,\,\,\, 0\leq \zeta\ll 1,}\\
            \\
             0 & {\,\,\,\,\zeta \leq 0,}
      \end{array}
        \right. 
\label{eq:scal_rel_chi_PB}
\end{equation}
which introduces the mean-field critical  exponent
associated with the reduced Manning parameter, $\zeta$ 
(or the reduced temperature, $t$)  as
\begin{equation}
 \chi_{\mathrm{PB}}=2. 
\label{eq:chi_PB}
\end{equation}
The mean-field counterion-condensation transition is therefore characterized by a
diverging (localization) length scale $1/S_1^{\mathrm{PB}}\sim\zeta^{-2}$, 
as the critical point is approached from above. 
The scaling relation (\ref{eq:scal_rel_chi_PB}) may also be derived 
in a direct way by considering a strictly infinite system ($\Delta=\infty$) 
as shown in Appendix \ref{app:PB_Delta_infty}. 

In the limiting case (ii) with $\zeta\rightarrow \zeta_c^{\mathrm{PB}}=0$, we 
have from Eq. (\ref{eq:beta}) that $\beta\simeq \pi/(2\Delta)$
when $\Delta$ is finite but large, $\Delta\gg 1$ (Appendix \ref{app:PB_beta}). 
Therefore, using  Eq. (\ref{eq:S_n_PB_approx}) we obtain
\begin{equation}
  S_n^{\mathrm{PB}}(0, \Delta)\simeq \frac{\pi^2}{4n\Delta^2},
\label{eq:Sn_PB_del}
\end{equation}
which introduces a new scaling relation
\begin{equation}
  S_n^{\mathrm{PB}}(0, \Delta)\sim \Delta^{-\gamma_{\mathrm{PB}}}
\end{equation}
with the mean-field critical exponent
\begin{equation}
 \gamma_{\mathrm{PB}}=2
\label{eq:gamma_PB}
\end{equation}
associated with the lateral extension parameter, $\Delta$. 
This relation shows that the approach to the 
true CCT limit (when $S_n^{\mathrm{PB}}$ vanishes at the critical point) 
is {\em logarithmically} weak as the box size, 
$D$, increases to infinity, i.e. 
$S_n^{\mathrm{PB}}(\zeta=0)\sim 1/(\ln D/R)^2$.

The scaling relations (\ref{eq:Sn_PB_zeta}) and (\ref{eq:Sn_PB_del})
indicate that $S_n^{\mathrm{PB}}$ takes a general scale-invariant form with respect
to $\zeta$ and $\Delta$ as
\begin{equation}
  S_n^{\mathrm{PB}}\simeq
  \Delta^{-\gamma_{\mathrm{PB}}}{\mathcal D}_n(\zeta\Delta^{\gamma_{\mathrm{PB}}/\chi_{\mathrm{PB}}})
\label{eq:Sn_FSS_PB}
\end{equation}
for sufficiently large $\Delta$ and in the vicinity of the mean-field critical point.
% (from above).
The scaling function, ${\mathcal D}_n(u)$, has the following 
asymptotic behavior
\begin{equation}
    {\mathcal D}_n(u) \sim \left\{
             \begin{array}{ll}
             {\mathrm{const.}}    & {\,\,\,\,u\rightarrow 0,}\\
            \\
             u^{\chi_{\mathrm{PB}}}/n & {\,\,\,\,u\rightarrow +\infty.}
      \end{array}
        \right. 
\label{eq:An_FSS} 
\end{equation}

In general, the scale-invariant relations such as Eq. (\ref{eq:Sn_FSS_PB}) 
may be obtained within the PB frame-work using the fact that the integration constant $\beta(\zeta, \Delta)$
takes a scale-invariant form as
\begin{equation}
  \beta \simeq \Delta^{-1}{\mathcal B}(\zeta\Delta).
\label{eq:beta_FSS}
\end{equation}
Here ${\mathcal B}(u)$ is a scaling function which 
behaves asymptotically as (Appendix \ref{app:PB})
\begin{equation}
    {\mathcal B}(u) \sim \left\{
             \begin{array}{ll}
             {\mathrm{const.}}    & {\,\,\,\,u\rightarrow 0,}\\
            \\
             \sqrt{u} & {\,\,\,\,u\rightarrow +\infty.}
      \end{array}
        \right. 
\label{eq:Bn_FSS} 
\end{equation}
Combining Eqs. (\ref{eq:S_n_PB_approx}) and (\ref{eq:beta_FSS}), 
the scaling function ${\mathcal D}_n(u)$ 
is obtained in terms of ${\mathcal B}(u)$ as
\begin{equation}
  {\mathcal D}_n(u)\simeq \frac{1}{n}\left[u^2+{\mathcal B}^2(u)\right],
\label{eq:A_B_FSS} 
\end{equation}
which reproduces Eq. (\ref{eq:An_FSS}) when combined with Eq. (\ref{eq:Bn_FSS}). 

The mean-field critical exponents $\chi_{\mathrm{PB}}$ and $\gamma_{\mathrm{PB}}$ appear to 
be independent of the order of the density moments, $n$.
They may be used to characterize the mean-field universality class of the CCT
in all dimensions, since the PB results are independent of the space dimensionality. 

%------------------------------------
%       SUBSUBSECTION:    E & C in PB
%------------------------------------
\subsubsection{Mean-field energy and heat capacity}
\label{subsubsec:EC_PB}

As shown in a previous work \cite{Naji03}, the mean-field
canonical free energy of the counterion-cylinder system may be obtained using
a saddle-point analysis from the field-theoretic representation of the partition function
when $\Xi\rightarrow 0$ \cite{Netz}. 
The rescaled PB free energy defined as
${\tilde {\mathcal F}}^{\mathrm{PB}}\equiv {\mathcal F}^{\mathrm{PB}}_{N}/(Nk_{\mathrm{B}}T)$, 
is given by (up to the trivial kinetic energy part)
\begin{eqnarray}
   {\tilde {\mathcal F}}^{\mathrm{PB}} =
                    &-&\frac{1}{\xi}
                       \int {\tilde r}\,{\mathrm{d}}{\tilde r}\,
                          \bigg[\frac{1}{4}\bigg(\frac{{\mathrm{d}} \psi_{\mathrm{PB}}}{{\mathrm{d}}\tilde r}\bigg)^2
                            +\delta({\tilde r}-{\tilde R})
                                \psi_{\mathrm{PB}}({\tilde r})\bigg]
                       \nonumber\\
                        &-&\ln 
                             \bigg[\frac{1}{\xi}\int_{\tilde R}^{\tilde D}\, {\tilde r}\,{\mathrm{d}}{\tilde r}\,
                                   e^{-\psi_{\mathrm{PB}}({\tilde r})}\bigg]-\ln\bigg(\frac{2V_{\mathrm{cyl}}}{N\xi}\bigg),\,\,\,
\label{eq:FE_PB}
\end{eqnarray}
where $V_{\mathrm{cyl}}=\pi R^2L$ is the actual volume of the cylinder.
In the thermodynamic limit $N\rightarrow\infty$, the ratio $V_{\mathrm{cyl}}/N$
is a constant and will be dropped in what follows. 

Inserting the PB potential field, 
Eq. (\ref{eq:psi}), into the free energy expression (\ref{eq:FE_PB}), we
find that for $\xi>\Lambda_{\mathrm{AF}}$ 
\begin{eqnarray}
   {\tilde {\mathcal F}}^{\mathrm{PB}}=
                   &-&\frac{1}{\xi}
                      \bigg[(1-\beta^2)\Delta+
                          \ln \left(\frac{(\xi-1)^2+\beta^2}
                            {1+\beta^2}\right)
                     +\xi\bigg]\nonumber\\
                     &+&\ln [(\xi-1)^2+\beta^2]
                     -\ln (2\xi).
\label{eq:freePB_high}
\end{eqnarray}
While for $\xi<\Lambda_{\mathrm{AF}}$, we have 
\begin{eqnarray}
   {\tilde {\mathcal F}}^{\mathrm{PB}}=
                  &-&\frac{1}{\xi}
                      \bigg[(1+\beta^2)\Delta+
                          \ln \left(\frac{(\xi-1)^2-\beta^2}
                            {1-\beta^2}\right)
                   +\xi\bigg]\nonumber\\
                     &+&\ln [(\xi-1)^2-\beta^2]
                     -\ln (2\xi).
\label{eq:freePB_low}
\end{eqnarray} 
These expressions (up to some additive constants) 
have also been obtained by Lifson et al.  \cite{Lifson54} using a charging process method. 
 
The rescaled (internal) energy, ${\tilde E}^{\mathrm{PB}}\equiv E_N^{\mathrm{PB}}/(Nk_{\mathrm{B}}T)$,
and the rescaled excess heat capacity, ${\tilde C}^{\mathrm{PB}}\equiv C_N^{\mathrm{PB}}/(Nk_{\mathrm{B}})$, 
can be calculated  using the thermodynamic relations 
\begin{eqnarray}
   {\tilde E}^{\mathrm{PB}} &=& \xi\frac{\partial {\tilde {\mathcal F}}^{\mathrm{PB}}}{\partial \xi},
\label{eq:E_PB}
\\
  {\tilde C}^{\mathrm{PB}}  &=& -\xi^2\frac{\partial^2 {\tilde {\mathcal F}}^{\mathrm{PB}}}{\partial \xi^2}, 
\label{eq:C_PB}
\end{eqnarray}
where the derivatives are taken at fixed volume, number of particles, and also for fixed 
charges and dielectric constant.
A closed-form expression may be obtained for energy
using the relation $E=(\varepsilon\varepsilon_0/2)\int {\mathrm{d}}{\mathbf x}\, (\nabla \psi_{\mathrm{elec}})^2$,
where $\psi_{\mathrm{elec}}=k_{\mathrm{B}}T \psi_{\mathrm{PB}}/q e$ is the potential field in actual units.
In rescaled units, the result is
\begin{eqnarray}
	\tilde E&=&\frac{1}{4\xi}\int_{\tilde R}^{\tilde D} \,\tilde r\, 
			{\mathrm{d}}\tilde r\,\bigg(\frac{{\mathrm{d}}\psi}{{\mathrm{d}}\tilde r}\bigg)^2=
	\label{eq:E_rescal_def}
	 \\
	 &=& \frac{1}{\xi}\times\left\{
             \begin{array}{ll}
              (1+\beta^2)\Delta+
                          \ln \bigg[\frac{(\xi-1)^2-\beta^2}
                            {1-\beta^2}\bigg]
                   +\xi\,
                    & {\,\,\,\,\xi\leq \Lambda_{\mathrm{AF}},}\\
            \\ 
                (1-\beta^2)\Delta+
                          \ln \bigg[\frac{(\xi-1)^2+\beta^2}
                            {1+\beta^2}\bigg]
                     +\xi\,
              & {\,\,\,\,\xi\geq\Lambda_{\mathrm{AF}}.}
      \end{array}
        \right. 
\label{eq:energyPB_full}
\end{eqnarray}

In general, the above quantities can be calculated numerically using the transcendental equation 
(\ref{eq:beta}). But in the limit of
$\Delta\rightarrow \infty$, one may use the asymptotic results for $\beta$ (Appendix \ref{app:PB})
to derive the asymptotic form of the rescaled PB free energy as \cite{Naji03}
\begin{eqnarray}
    {\tilde {\mathcal F}}^{\mathrm{PB}} =\left\{
             \begin{array}{ll}
               (\xi-2) \Delta
              & {\,\,\,\,\,\,\xi\leq \xi_c^{\mathrm{PB}}=1,}\\
            \\
              -\Delta/\xi
              & {\,\,\,\,\,\,\xi\geq\xi_c^{\mathrm{PB}}=1.}
      \end{array}
        \right. 
\label{eq:freePB_asym}
\end{eqnarray}
The rescaled  PB energy asymptotically behaves as
\begin{eqnarray}
    {\tilde E}^{\mathrm{PB}} =\left\{
             \begin{array}{ll}
               \xi \Delta
              & {\,\,\,\,\,\,\xi\leq \xi_c^{\mathrm{PB}},}\\
            \\
              \Delta/\xi
              & {\,\,\,\,\,\,\xi\geq\xi_c^{\mathrm{PB}},}
      \end{array}
        \right. 
\label{eq:energyPB_asym}
\end{eqnarray}
and the rescaled PB excess heat capacity as
\begin{eqnarray}
    {\tilde C}^{\mathrm{PB}} =\left\{
             \begin{array}{ll}
               0
              & {\,\,\,\,\,\,\xi< \xi_c^{\mathrm{PB}},}\\
            \\
              2\Delta/\xi
              & {\,\,\,\,\,\,\xi>\xi_c^{\mathrm{PB}}.}
      \end{array}
        \right. 
\label{eq:heatPB_asym}
\end{eqnarray} 

The above results show that both energy and excess heat capacity 
develop a singular peak at   the Manning parameter 
$\xi_c^{\mathrm{PB}}=1$ when the critical limit $\Delta\rightarrow \infty$
is approached. The PB results also show that  
the free energy diverges with $\Delta$  {\em both} above and below the mean-field critical point, 
in contrast with the behavior obtained 
within the (one-particle) Onsager instability \cite{Manning69}, which 
suggests a connection between the onset of the counterion condensation and the divergence
of the partition function (Section \ref{sec:CCT_intro} and Appendix \ref{app:Onsager}). 

Another important point is that the PB heat capacity  exhibits a discontinuity at $\xi_c^{\mathrm{PB}}=1$. 
Therefore, the CCT may be considered as a second-order transition as also pointed out 
in a previous mean-field study  \cite{Rubinstein01}.
We shall return to the  singular behavior of energy and heat capacity  later in our numerical studies.

%%%%%%%%%%%%%%%%%%%%%%%%%%%%%%%%%%%%%%%%%%%%%%%%%%%%%%%%%%%%%%%%%%%%%%

%                               SECTION: SC

%%%%%%%%%%%%%%%%%%%%%%%%%%%%%%%%%%%%%%%%%%%%%%%%%%%%%%%%%%%%%%%%%%%%%%
\section{Strong-coupling theory for the CCT}
\label{sec:CCT_SC}

In the limit of large coupling parameter, $\Xi\rightarrow \infty$, 
the partition function of a charged system adopts an expansion 
in powers of $1/\Xi$, the  leading term of  which 
comprises only single-particle contributions, i.e. a single counterion interacting with
fixed charged objects \cite{Netz,Andre}.  This leading-order theory, referred to as the 
asymptotic strong-coupling (SC) theory, 
describes the complementary limit to the mean-field regime, $\Xi\gg1$, where 
inter-particle correlations are expected to become important \cite{Low_T,Naji_PhysicaA}. 
  
The rescaled SC density profile for counterions is obtained as   \cite{Netz}
\begin{equation}
  {\tilde \rho}_{\mathrm{SC}}({\tilde r})=\lambda_0\,{\tilde \Omega}({\tilde r})\,
                e^{-{\tilde u}({\tilde r})}
   \label{eq:SCprofile}
\end{equation}
where ${\tilde u}({\tilde r})=2\xi\ln ({\tilde r}/{\tilde R})$  is
 the single-particle interaction energy and 
$\lambda_0$ is a normalization factor, which is fixed with 
the total number of counterions.  Thus we have
\begin{equation}
   \lambda_0 = \frac{2(\xi-1)}{\xi}\bigg[1-e^{-2(\xi-1)\Delta}\bigg]^{-1}, 
\end{equation}
in the cell model considered here. 
Note that for $\Delta\rightarrow \infty$, $\lambda_0$, and therefore the whole 
density profile,  vanishes for $\xi\leq 1$.
But for $\xi\geq 1$, we get $\lambda_0\rightarrow 2(\xi-1)/\xi$ and hence a
finite limiting density profile as 
\begin{equation}
{\tilde \rho}_{\mathrm{SC}}({\tilde r})						 
 		  \rightarrow  \frac{2(\xi-1)}{\xi}
                \left(\frac{\tilde r}{\tilde R}\right)^{-2\xi}.
               % +{\mathcal O}\bigg(e^{-2(\xi-1)\Delta}\bigg).
\label{eq:rho_SClimit}
\end{equation}
This shows that  the CCT is reproduced within the SC theory as well, and surprisingly, the
critical value is found to be $\xi_c^{\mathrm{SC}}=1$ 
in coincidence with the mean-field prediction. 
Note however that the SC profile for $\xi>1$ 
indicates a {\em larger} contact density for counterions
as compared with the mean-field theory, e.g., for
$\Delta\rightarrow \infty$, one has
\begin{equation}
{\tilde \rho}_{\mathrm{SC}}(\tilde R)						 
 		  =   \frac{2(\xi-1)}{\xi}, 
\label{eq:rho_SCcontact}
\end{equation}
which is larger than the PB value (\ref{eq:density_R_high}) by a factor 
of ${\tilde \rho}_{\mathrm{SC}}(\tilde R)/{\tilde \rho}_{\mathrm{PB}}(\tilde R)=2(1-1/\xi)^{-1}$. 
The SC  density profile also decays faster than the PB profile indicating a more
compact counterionic layer at the cylinder. This reflects strong ionic correlations in the condensation
phase ($\xi>1$) for $\Xi\gg 1$ as will be discussed further in the numerical studies below. 
 
Using Eq. (\ref{eq:SCprofile}), the SC order parameters can be calculated as 
\begin{equation}
   S_n^{\mathrm{SC}}(\xi, \Delta)=\frac{2(\xi-1)}{\xi^n(2\xi-2+n)}\frac{1-e^{-(2\xi-2+n)\Delta}}
   									{1-e^{-2(\xi-1)\Delta}}
 \label{eq:Sn_SC_full}
\end{equation}
for arbitrary $\xi$ and $\Delta$.   For $\Delta\rightarrow \infty$, 
$S_n^{\mathrm{SC}}$ vanishes for  $\xi\leq \xi_c^{\mathrm{SC}}=1$, but tends to
\begin{equation}
    S_n^{\mathrm{SC}}(\zeta, \infty)  = \frac{2(\xi-1)}{\xi^n(2\xi-2+n)}
\label{eq:app_Sn_SC}
\end{equation}
for $\xi\geq \xi_c^{\mathrm{SC}}=1$.
 In the vicinity of the critical point, $S_n^{\mathrm{SC}}$ exhibits  the scaling relation 
\begin{equation}
    S_n^{\mathrm{SC}}(\zeta, \infty) \simeq \frac{2\zeta}{n},
\end{equation}
which gives the SC critical  exponent  associated with the 
reduced Manning parameter, $\zeta=1-\xi_c/\xi$, as $\chi_{\mathrm{SC}}=1$. 
In a finite system and right at the critical point $\xi =  \xi_c^{\mathrm{SC}}$, $S_n^{\mathrm{SC}}$ exhibits  the
finite-size-scaling relation
\begin{equation}
    S_n^{\mathrm{SC}}(0, \Delta) \simeq \frac{1}{n\Delta},
\end{equation}
which gives the SC critical exponent   associated with the lateral extension parameter, 
$\Delta$, as $\gamma_{\mathrm{SC}}=1$. 

These exponents are different from the corresponding mean-field values, Eqs. (\ref{eq:chi_PB}) and
(\ref{eq:gamma_PB}), and in fact coincide with the Onsager instability results. 
As will be shown later, the SC predictions in fact break down {\em near} the 
CCT critical point.

%%%%%%%%%%%%%%%%%%%%%%%%%%%%%%%%%%%%%%%%%%%%%%%%%%%%%%%%%%%%%%%%%%

%                               			MC SIMULATION

%%%%%%%%%%%%%%%%%%%%%%%%%%%%%%%%%%%%%%%%%%%%%%%%%%%%%%%%%%%%%%%%%%
\section{Monte-Carlo study of the CCT in 3D}
\label{sec:sim}

The preceding analysis within the mean-field and the strong-coupling theory
reveals a set of new scaling relations 
associated with the counterion-condensation transition
(CCT) in the limit of infinitely large (lateral) system size. 
In the following sections, we shall use numerical methods 
to examine  the  critical behavior
in various regimes of the coupling parameter, and thereby, to examine 
the validity of the aforementioned analytical results. 
 
%%%%%%%%%%%%%%%%%%%%%%%%%%%%%%%%%%%%
%       SUBSECTION: PB SCALING EXPS.
%%%%%%%%%%%%%%%%%%%%%%%%%%%%%%%%%%%%
\subsection{The centrifugal sampling method}
\label{subsec:sampling}

The major difficulty in studying the CCT numerically goes back to the lack of 
an efficient sampling technique. Poor sampling problem 
arises for  counterions at charged curved surfaces
in the infinite-confinement-volume limit because, contrary to charged plates, 
a finite fraction of counterions always tends to unbind from curved boundaries and diffuse to 
infinity as the system relaxes toward its equilibrium state.
This situation is, of course, not tractable in numerical simulations; hence to achieve proper equilibration
within a reasonable time,
charged cylinders are customarily considered in a confining box (in lateral directions) of 
{\em practically} large size.  As well known \cite{Manning77,Ramanathan,Ramanathan_Wood82a,Ramanathan_Wood82b,Deserno00,Liao03}, 
lateral finite-size effects are quite small 
for sufficiently large Manning parameter ($\xi>\xi_c$). 
But at small Manning parameters ($\xi\sim \xi_c$), these effects become significant  
and suppress the de-condensation of counterions.

The mean-field results already reveal 
a very weak asymptotic convergence to the critical transition controlled by the
logarithmic size of the confining box $\Delta=\ln (D/R)$. 
Hence one needs to consider a confinement volume of 
extremely large radius, $D$, to establish the large-$\Delta$ regime,
where  the  scaling (and possibly universal) properties of the CCT emerge. For this purpose, 
clearly, the simple-sampling methods within Monte-Carlo
or Molecular Dynamics schemes \cite{Winkler98,Deserno00,Liu_Muthu02,Liao03,Bratko82,Zimm84,Rossky85}
are not useful at {\em low} Manning parameter as
they render an infinitely long relaxation time. 

We shall therefore introduce a novel sampling method within the Monte-Carlo 
scheme, which enables one to properly span the relevant parts of the phase space 
for large confining volumes. In three dimensions, we use the 
configurational Hamiltonian (\ref{eq:Hamilt_rescaled}) 
in rescaled coordinates. 
The sampling method, which we refer to as the {\em centrifugal sampling}, 
is obtained by mapping the radial coordinate to a logarithmic scale according to 
Eq. (\ref{eq:y}), i.e. $y=\ln ({\tilde r}/{\tilde R})$, 
which leads to the transformed partition function (\ref{eq:Z_y}). 
As explained before (Section \ref{subsec:generic}), the entropic (centrifugal) factor, $\exp(2y)$,
is absorbed from the measure of the radial integrals into the Hamiltonian, yielding the transformed 
Hamiltonian $ {\mathcal H}_N^\ast$ in Eq.  (\ref{eq:H_virtual}).

We thus simulate the system using Metropolis algorithm  \cite{Metropolis}, 
but making use of the transformed Hamiltonian (\ref{eq:H_virtual}). 
The entropic factors, which cause 
unbinding of counterions, are hence incorporated
into the transition probabilities of the associated Markov chain of states,
that generates equilibrium states with the distribution function 
$\sim\exp(-\beta {\mathcal H}_N^\ast)$. 
The averaged quantities, say ${\bar Q}$, follow by extracting
a set of $T$ values $\{Q_1,\ldots, Q_T\}$ in the course of the simulations as
${\bar Q}=\sum_{t=1}^{T} Q_t/T$, which, for sufficiently large $T$, produces the
desired ensemble average $\langle Q\rangle$, i.e.
\begin{eqnarray}
 {\bar Q} &=& \frac{1}{T}\sum_{t=1}^{T} Q_t 
          \simeq 
          \frac{\mu^{3N}{\tilde R}^{2N}}{N!{\mathcal Z}_N}
            \int_{\tilde V} \left[\prod_{i=1}^{N} {\mathrm{d}}{\tilde z}_i\, 
            			{\mathrm{d}} \phi_i \, {\mathrm{d}} y_i\right]\nonumber\\
                 &\times&    
                 Q({y_i, \phi_i, {\tilde z}_i})\,
                 e^{-\beta {\mathcal H}_N^\ast}=\langle Q\rangle,
\end{eqnarray}
up to relative corrections of the order $1/\sqrt{T}$.

%%%%--------Fig snapshots:
\begin{figure*}
\begin{center}
\includegraphics[angle=0,width=15cm]{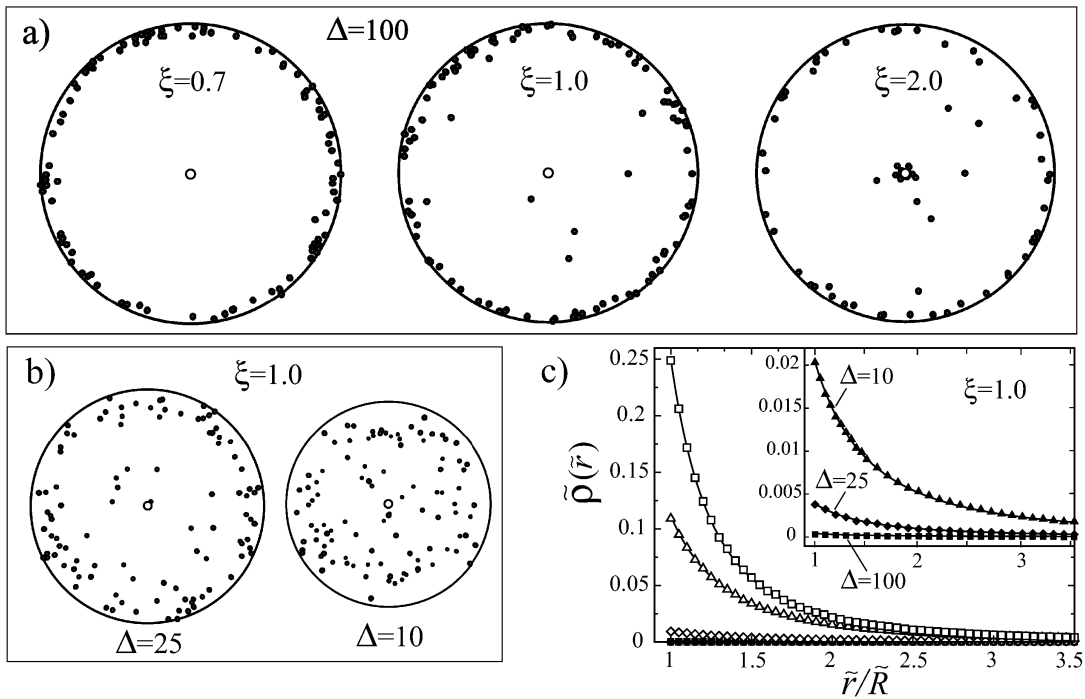}
\caption{Typical snapshots from the simulations on the counterion-cylinder system in 3D
for a) lateral extension parameter
$\Delta=\ln (D/R)=100$ and three different Manning parameters $\xi=0.7, 1.0$ and 2.0 as indicated 
on the graph, and b) for Manning
parameter $\xi=1.0$ and smaller lateral extension parameters $\Delta=10$ and 25. 
The snapshots show top-views
of the simulation box (see Section \ref{subsec:sim_para} 
and Figure \ref{fig:model}) with radial distances shown
in logarithmic scale $y=\ln ({\tilde r}/{\tilde R})$. Point-like counterions are shown by 
black spheres and the charged cylinder by a circle in the middle. 
Figure c) gives the simulated radial density
of counterions in rescaled units, $\tilde \rho(\tilde r)=\rho(r)/(2\pi \ell_{\mathrm{B}} \sigma_{\mathrm{s}}^2)$, 
 as a function of the (linear) distance from the cylinder
axis. Main set shows the data for $\Delta=100$ and
Manning parameters $\xi=2.0$ (open square), 1.5 (open triangle-ups), 1.1 (open diamonds), 
1.0 (filled squares) and 0.7 (filled circles) from top to bottom. 
Inset shows the same for $\xi=1.0$, but for various
lateral extension parameters $\Delta=$10,  25 and 100 (top to bottom).
Solid curves represent  the mean-field PB prediction Eq. (\ref{eq:rho}). 
Number of counterions here is $N=100$ and the coupling parameter $\Xi=0.1$.
Error-bars are smaller than the size of symbols.}
\label{fig:snapshots}
\end{center}
\end{figure*}

%%%%%%%%%%%%%%%%%%%%%%%%%%%%%%%%%%%%%%
%              SIMULATION   PARAMETERS
%%%%%%%%%%%%%%%%%%%%%%%%%%%%%%%%%%%%%%
\subsection{Simulation model and parameters}
\label{subsec:sim_para}

The geometry of the counterion-cylinder system in our simulations 
is similar to what we have sketched in Figure \ref{fig:model}. 
We use typically between $N=25$ to 300 counterions
(most of the results are obtained with $N$=100 and 200 particles) and 
increase the lateral extension parameter, $\Delta=\ln (D/R)$, up to $\Delta=300$.
We also vary the Manning parameter, $\xi$, and consider  
a wide range of values for the electrostatic coupling parameter, $\Xi$, 
from $\Xi=0.1$ (close to the mean-field regime)
up to $\Xi=10^5$ (close to the strong-coupling regime). 

The cylindrical simulation box has a finite height, ${\tilde L}$, 
which is set by the electroneutrality  condition (\ref{eq:ENC}),  i.e. 
${\tilde L}=N\Xi/\xi$. 
In order to mimic the thermodynamic limit and reduce the finite-size effects due 
to the finiteness of the cylinder height,
 we apply periodic boundary conditions in  $z$ direction
(parallel to the cylinder axis) by replicating the main
simulation box infinitely many times in that direction. 
The long-range character of the Coulomb interaction in such a periodic system
leads to summation of infinite series over all periodic images.
These series are not generally convergent, but in an electroneutral system, the divergencies
cancel and the series can be 
converted to fast-converging series. 
We use the summation techniques due to
Lekner \cite{Lekner} and Sperb \cite{Sperb}, which are utilized to the one-dimensionally periodic 
system considered here--see Appendix \ref{app:H_periodic} (similar methods are developed in Ref. \cite{Arnold}). 
Finally in order to obtain reliable values for the error-bars, the 
standard block-averaging scheme is used \cite{Block_Average}. 
The simulations typically run for $\sim 1.1\times 10^6$ Monte-Carlo steps per particle
with $\sim 10^5$ steps used for the equilibration purpose.

%%%%%%%%%%%%%%%%%%%%%%%%%%%%%%%%%%%%%%%%%%%%%%%%%%%%%%%%%%%%%%%%%%%%

%                             SIMULATION RESULTS in 3D

%%%%%%%%%%%%%%%%%%%%%%%%%%%%%%%%%%%%%%%%%%%%%%%%%%%%%%%%%%%%%%%%%%%%
\section{Simulation results in 3D} 
\label{sec:sim_data}

\subsection{Overall behavior in the infinite-system-size limit}
\label{subsec:sim_overall}

\subsubsection{Distribution of counterions}  
\label{subsubsec:ci_dis}

Let us start with the distribution of counterions as generated by the centrifugal sampling 
method. 
In Figure \ref{fig:snapshots} typical simulation snapshots are shown 
together with the  counterionic density profile
for small coupling parameter  $\Xi=0.1$. Counterion distribution is shown for 
large ($\Delta=100$), intermediate ($\Delta=25$) and small ($\Delta=10$) 
lateral extension parameter. The counterion-condensation transition is clearly
reproduced for large $\Delta$ (Figure \ref{fig:snapshots}a): 
counterions are ``de-condensed" 
and gather at the outer boundary at small Manning parameter (shown for $\xi=0.7$), 
while they partially ``condense" and accumulate near the cylinder 
surface for large Manning parameter (shown for $\xi=2$). The Manning parameter
$\xi=1$, as seen, represents an intermediate situation. This trend 
is demonstrated on a quantitative level
by the radial density profile of counterions $\tilde \rho (\tilde r)$ 
(Figure \ref{fig:snapshots}c, main set), which tends to zero by decreasing $\xi$ 
down to about unity. Note that relatively large 
fluctuations occur at low $\xi$ making $\tilde \rho (\tilde r)$ an inconvenient
quantity to precisely locate the critical value $\xi_c$, which will be considered later.
 The data moreover follow the mean-field PB density prediction,
Eq. (\ref{eq:rho}), shown by solid curves, 
as expected since the chosen coupling parameter is small.

The transition regime at intermediate $\xi$ exhibits strong finite-size effects.
 As may be seen from the snapshots in Figure \ref{fig:snapshots}b, 
 the de-condensation process at $\xi=1$ is strongly suppressed
 for small logarithmic sizes $\Delta=\ln (D/R)=10$ and 25. The corresponding density profiles 
(inset of Figure \ref{fig:snapshots}c) indicate a sizable accumulation of counterions
near the cylinder surface, which is washed away only by taking a sufficiently 
large $\Delta$. Such finite-size effects at low $\xi$ are also observed in previous numerical studies, which 
have devised simulations in linear scale and thus considered only small confinement 
volumes per polymer (typically $\Delta<10$) \cite{Deserno00,Liao03,Bratko82,Zimm84,Rossky85}. 
In some studies \cite{Rossky84}, 
these effects have been interpreted as an evidence for counterion condensation 
at {\em small} $\xi$, leading to the incorrect conclusion that no condensation transition exists.

%----------------------------------------------
\subsubsection{Condensed fraction of counterions}
\label{subsubsec:CondRatio}

Our results for large $\xi$  exhibit a counterionic density 
profile that extends continuously from the cylinder surface to larger distances. 
This indicates that making a  distinction  between 
two layers of  condensed and de-condensed counterions, in the sense of 
two-fluid models frequently used in literature \cite{Manning69,Oosawa,Oosawa_Imai,Manning77,Manning_Dyn,Manning_rev78,Manning_rev96a,Manning_rev96b,de-la-Cruz95,Levin97,Levin98,Schiess98,Nyquist99,Ray_Mann99,Henle04,Muthu04}, requires a criterion. 

The two-fluid description predicts a fraction of 
 \begin{eqnarray}
    \alpha_{\mathrm{M}}  = \left\{
             \begin{array}{ll}
               0
              & {\,\,\,\,\xi\leq 1}\\
              	\\
              1-1/\xi
              & {\,\,\,\,\xi\geq 1}
      \end{array}
        \right. 
\label{eq:alpha_PB}
\end{eqnarray}
of counterions to reside in the condensed layer (which is considered as a layer with small thickness at the polymer surface), 
when the infinite-dilution limit is reached.
Previous studies \cite{Macgillivray,Ramanathan,Ramanathan_Wood82a,Zimm,Deserno00,Shaugh} 
show that the Manning condensed fraction, $\alpha_{\mathrm{M}}$, may also be  
identified systematically within the Poisson-Boltzmann theory by employing
an {\em inflection-point criterion} \cite{Deserno00,Shaugh}. 
This can be demonstrated using the PB cumulative density  
(the number of counterions inside a cylindrical region of radius $r$), $n_{\mathrm{PB}}(r)$, obtained as 
\begin{eqnarray}
 && \frac{n_{\mathrm{PB}}(r)}{N} = \frac{2\pi L}{N} \int_{R}^{r} r' \, {\mathrm{d}}r'\, \rho_{\mathrm{PB}} (r')=
  		\nonumber\\
	       &&= \frac{1}{\xi}\times \left\{
             \begin{array}{ll}
               (\xi-1)-\beta\coth \bigg[\beta y+\coth^{-1}\frac{\xi-1}{\beta}\bigg]
              & {\xi\leq \Lambda_{\mathrm{AF}},}\\
              (\xi-1)-\beta\cot \bigg[\beta y+\cot^{-1}\frac{\xi-1}{\beta}\bigg]
              & {\xi\geq \Lambda_{\mathrm{AF}},}
      \end{array}
        \right. 
 \label{eq:PB_cumu}
\end{eqnarray}
using Eq. (\ref{eq:rho}). 
For $\xi\geq \Lambda_{\mathrm{AF}}$, $n_{\mathrm{PB}}(r)$ 
exhibits an inflection point at a radial distance $r_\ast$  when 
plotted as a function of $y=\ln (r/R)$ \cite{Deserno00}. 
 One can show that for $\Delta\rightarrow \infty$, 
only the fraction of counterions, that lie within the cylindrical region $r\leq r_\ast$, remains associated with
the cylinder and tends to the Manning condensed fraction, i.e.
\begin{equation}
    \frac{n_{\mathrm{PB}}(r_\ast)}{N}\rightarrow \alpha_{\mathrm{M}}.
\end{equation}
In other words, only this fraction of counterions contribute to the asymptotic density profile and 
the rest ($1/\xi$ of all) is pushed to infinity (Appendix \ref{app:PB_Delta_infty}).
%This result can also be understood by considering the PB solution for a
%strictly unbounded system with $\Delta=\infty$ \cite{Netz_Joanny} (Appendix \ref{app:PB_Delta_infty}).

%%----------- Fig Cumulative density
\begin{figure}[t]
\begin{center}
\includegraphics[angle=0,width=8cm]{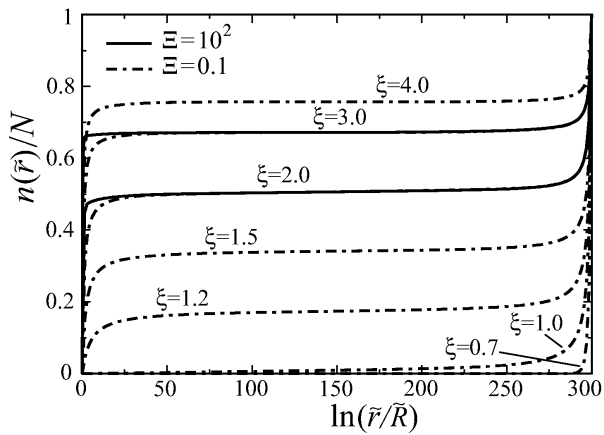}
\caption{
Cumulative density, $n({\tilde r})$, per total number $N$, of counterions
as a function of the logarithmic distance from the
charged cylinder, $\ln ({\tilde r}/{\tilde R})$. Dot-dashed curves are simulation results for
$\Xi=0.1$, $N=70$, and $\Delta=300$ and for various Manning parameters as shown on the graph.  
These curves also closely represent the PB prediction (\ref{eq:PB_cumu}), which are not explicitly shown. 
Solid curves show the simulation 
data for large coupling parameter $\Xi=10^2$ and for $\xi=3.0$ and $\xi=2.0$.}
\label{fig:cumu}
\end{center}
\end{figure}

The simulations results for the cumulative density 
as a function of the logarithmic radial distance $y=\ln (\tilde r/\tilde R)$ 
 are  shown in Figure  \ref{fig:cumu} 
for various Manning parameters (solid and dot-dashed curves). Here we have chosen a
very large lateral extension parameter $\Delta=\ln (D/R)=300$, which exhibits the concept of
condensed fraction more clearly. 
The data  show an inflection point, which is located approximately at 
$y_\ast=\ln (r_\ast/R)\simeq \Delta/2$ for large $\xi$
(for small $\xi\rightarrow 1$, the location of the inflection point, 
$r_\ast$, tends to $R$--see Appendix \ref{app:PB_cumu}).
The rapid increase of  $n(\tilde r)$ at
small ($r\sim R$) and at large  distances ($r\sim D$) reflects the two counterion-populated regions 
at the inner and outer boundaries, which are separated by an extended plateau (compare with Figure \ref{fig:snapshots}). 
For small $\Delta$, the inflection point has a non-vanishing slope 
and the two regions are not quite separated (data not shown) \cite{Deserno00,Henle04}
%It is only for $\Delta\gg 1$ that  
%the so-called condensed and de-condensed region may be distinguished 
(see Ref. \cite{Ray_Mann99}  for a similar trend in an extended two-fluid model).

%%---------Fig Condensed fraction
\begin{figure}[t]
\begin{center}
\includegraphics[angle=0,width=8cm]{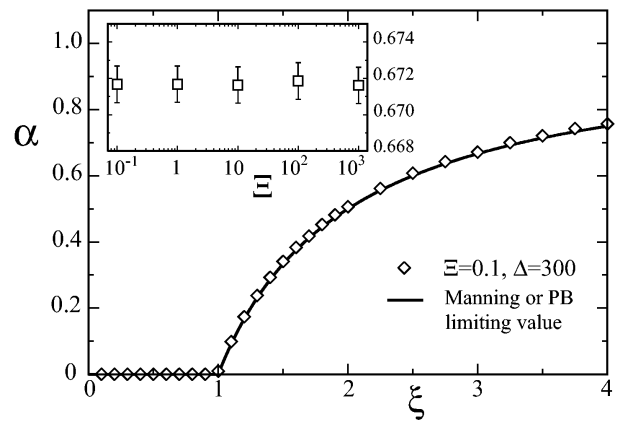}
\caption{
Main set: Simulated condensed fraction of counterions, $\alpha$, as defined via 
   the inflection-point criterion,
   as a function of Manning parameter, $\xi$, for $\Xi=0.1$ (diamonds). 
   Solid curve displays Manning (or the PB) limiting value, $\alpha_{\mathrm{M}}$, 
   for $\Delta=\ln (D/R)\rightarrow\infty$ (Eq. (\ref{eq:alpha_PB})). 
   Inset: Condensed fraction as a function of the coupling parameter, $\Xi$, for
   $\xi=3.0$.  These data are obtained for
   $\Delta=300$ and $N=70$.}
\label{fig:condr}
\end{center}
\end{figure}

Using the inflection-point criterion,  the condensed fraction, $\alpha$, may be estimated as 
$\alpha=n(r_\ast)/N$  \cite{Deserno00}, which roughly corresponds to the plateau level in Figure \ref{fig:cumu}. 
Simulation results are shown in Figure \ref{fig:condr} for large $\Delta=300$. 
Let us first consider the case of a small coupling parameter $\Xi=0.1$, where  
 the simulated cumulative density, $n(\tilde r)$ (dot-dashed curves in Figure \ref{fig:cumu}), closely
follows the PB prediction (\ref{eq:PB_cumu}) (PB curves are
not explicitly shown). The calculated condensed fraction (diamonds in Figure \ref{fig:condr}) agrees 
already quite well (within $<1$\%) with the Manning or PB limiting value $\alpha_{\mathrm{M}}$ (solid curve in Figure  \ref{fig:condr}).
%Taking a smaller value of $\Delta$ (data not shown) leads to somewhat larger estimate for the condensed 
%fraction \cite{Deserno00}. 

An important question is whether the form of the cumulative density profile, $n(\tilde r)$, 
and the condensed fraction
are influenced by electrostatic correlations for increasing coupling parameter $\Xi$. 
In Figure \ref{fig:cumu}, we show  $n(\tilde r)$ from the simulations for  $\Xi=10^2$ and for
two values of Manning parameter $\xi=2.0 $ and 3.0 (solid curves). This coupling strength
generally falls into  the strong-coupling regime for charged systems, where electrostatic correlations are 
expected to matter \cite{Naji_PhysicaA} (note that DNA with trivalent counterions represents 
$\Xi\sim 10^2$, but  with a larger $\xi\sim 12$). 
As seen, $n(\tilde r)$ shows a more rapid increase at small distances from the cylinder
(condensed region) indicating a stronger accumulation of counterions near the surface. 
This trend is also observed in previous simulations \cite{Deserno00,Zimm84,Bratko82,Rossky85} 
and in experiments with multivalent counterions \cite{Zakha99},  
and will be analyzed in more detail in the following sections. 

However, in contrast to previous 
conclusions (obtained based on small values of $\Delta$) \cite{Deserno00,Henle04}, 
the aforementioned behavior for large $\Xi$ does not imply a larger condensed fraction as
defined within the inflection-point criterion.  Since as seen in Figure \ref{fig:cumu}, 
the {\em large-distance} behavior of the density profile is not influenced
by electrostatic correlations, and so remains the condensed 
fraction (plateau level) unaffected for increasing coupling strength (inset of Figure  \ref{fig:condr}). This 
result can be appreciated only when the asymptotic behavior for $\Delta\gg 1$ is considered.

%--------Fig Order parameter S1
\begin{figure}[t]
\begin{center}
\includegraphics[angle=0,width=8cm]{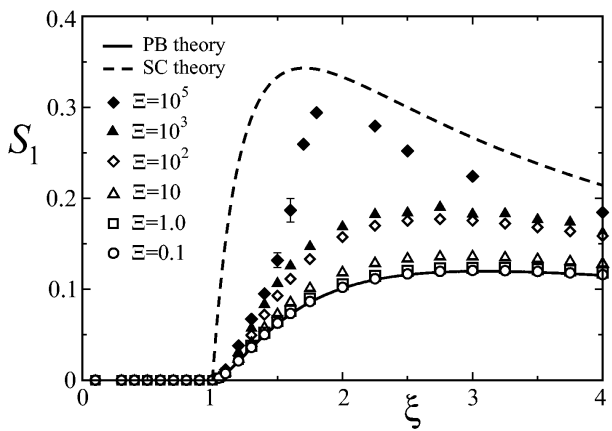}
\caption{Simulation data for 
the order parameter $S_1=\langle 1/{\tilde r}\rangle$ as a function of Manning parameter, $\xi$,
for various coupling parameters $\Xi=0.1$ up to $10^5$ as indicated on the graph. 
The mean-field PB and the strong-coupling predictions 
 are calculated from Eqs. (\ref{eq:S_n_PB}) and (\ref{eq:Sn_SC_full}) 
 (solid and dashed curves respectively). 
The lateral extension parameter is $\Delta=300$ and the number of counterions is 
$N=200$ for $\Xi=0.1$, $N=50$ for $\Xi=10^5$, and $N=70$ for other coupling parameters.}
\label{fig:s1}
\end{center}
\end{figure}

%------------------------------------
%       SUBSUBSECTION:    S_n
%------------------------------------
\subsubsection{The order parameters $S_n$}
\label{subsubsec:S_n}

The $n$-th-order inverse moment of the counterionic density profile may be calculated numerically using
\begin{equation}
       S_n=\frac{1}{N}\sum_{i=1}^{N}{\overline{{\tilde r}_i^{-n}}} 
\end{equation}
for $n>0$, where ${\tilde r}_i$ is the radial distance of the $i$-th counterion from the cylinder axis and 
the bar sign denotes the Monte-Carlo time average after 
proper equilibration of the system. 
The overall behavior is shown in Figure \ref{fig:s1} for $S_1$ as a function of Manning parameter, $\xi$.
Recall that a vanishing order parameter, $S_1$,  indicates
the complete de-condensation of counterions, 
while a finite $S_1$ reflects a finite degree of counterion binding to the charged cylinder
(corresponding to a finite localization length $\sim 1/S_1$).

As seen from the figure,  de-condensation can occur  
in all relevant regimes of the coupling parameter $\Xi$. For large Manning parameter, 
electrostatic coupling effects become important and shift the order parameter to larger values
exhibiting a crossover from the mean-field prediction (solid curve), which is verified  for small $\Xi<1$,
to the strong-coupling prediction (dashed curve) at very large values of $\Xi$ \cite{Netz,Andre,Naji04}. 
The mean-field result follows from Eq. (\ref{eq:S_n_PB}) and 
the strong-coupling  prediction is obtained using Eq. (\ref{eq:Sn_SC_full}). 
As seen, in the transition regime $\xi\sim 1$,  the order parameter data remain close to the mean-field
curve and deviate from the SC prediction. A close examination of correlation effects as well as finite-size effects 
in this region is quite important in determining the scaling behavior and will be considered later.
Here we concentrate on the correlation-induced crossover  behavior in the condensation phase.

%------------------------------------
%       SUBSUBSECTION:    CORRELATIONS
%------------------------------------
\subsubsection{Electrostatic correlations at surface and for large $\xi$}
\label{subsubsec:corr}

%%%---------Fig Density for var Xi
\begin{figure}[t]
\begin{center}
\includegraphics[angle=0,width=8cm]{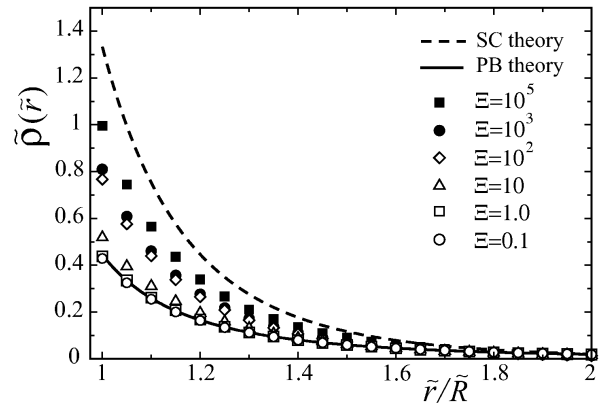}
\caption{Radial density of counterions in rescaled units, 
$\tilde \rho(\tilde r)=\rho(r)/(2\pi \ell_{\mathrm{B}} \sigma_{\mathrm{s}}^2)$, as a function of
the (linear) distance from the cylinder axis for  Manning parameter
$\xi=3.0$ and various coupling parameters ($\Xi=0.1$ up to $10^5$) as shown on the graph. 
Here $\Delta=300$ and the number of counterions is
$N=50$ for $\Xi=10^5$, and $N=70$ for other values of $\Xi$.
The mean-field (solid curve) and the strong-coupling (dashed curve) predictions are obtained from
Eqs. (\ref{eq:rho}) and (\ref{eq:SCprofile}), which,  for $\Delta=300$, roughly coincide with the asymptotic 
expressions (\ref{eq:density_asyp_high}) and (\ref{eq:rho_SClimit}). }
\label{fig:densX}
\end{center}
\end{figure}

In Figure \ref{fig:densX}, we plot
the simulated radial density profile of counterions, $\tilde \rho({\tilde r})$, for $\xi=3.0$ and consider
several different coupling parameters. In agreement with the preceding results, the
counterionic density  in the immediate vicinity of the charged cylinder 
increases for increasing  $\Xi$ exhibiting  large deviations from the mean-field prediction
(see Ref. \cite{Lau_Safran} for a similar trend at charged plates). 
For a given surface charge density $\sigma_{\mathrm{s}}$, 
the observed trend is predicted, e.g., for increasing counterion valency, $q$,
since the coupling parameter scales as $\Xi\sim q^3$ 
(Eq. (\ref{eq:Xi_coupl})). 
The crossover from the mean-field PB prediction (solid curve) to the 
strong-coupling prediction (dashed curve) appears to be quite weak, in agreement
with the situation observed for counterions at
planar charged walls \cite{Andre}.  These  limiting profiles are calculated from 
Eqs. (\ref{eq:rho}) and (\ref{eq:SCprofile}) respectively, and 
both exhibit an {\em algebraic}
decay with the radial distance, $\tilde r$. But the SC profile shows a faster decay
and thus a more compact counterion layer near the surface 
at large coupling strength (compare Eqs. (\ref{eq:density_asyp_high}) and (\ref{eq:rho_SClimit})).

An interesting point is that the simulated density at contact
with the cylinder shows a more rapid increase 
 when the coupling parameter increases from $\Xi=10$ to
$\Xi=100$ as compared with other ranges of $\Xi$ (Figure \ref{fig:densX}). 
This is in fact accompanied by the
formation of correlation holes around counterions 
near the surface as we show now. 
 
\begin{figure}[t]
\begin{center}
\includegraphics[angle=0,width=8cm]{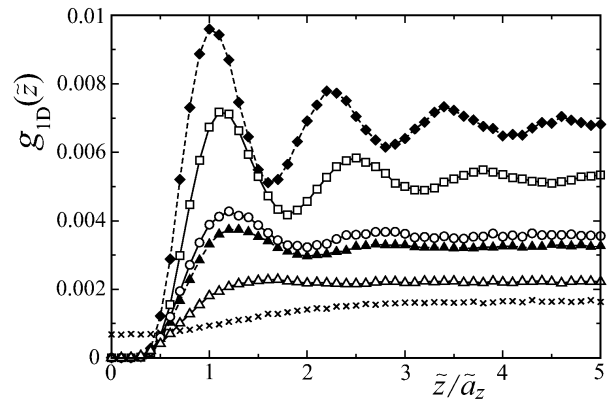}
\caption{The one-dimensional pair distribution function of counterions at contact with 
 cylinder as defined in Eq. (\ref{eq:g_1D}). Symbols show simulation data for
$\Xi=10^5$ and $\xi=4.0$ (filled diamonds), $\Xi=10^3$ and $\xi=2.0, 3.0$ and 4.0 
(open symbols from bottom to top), and for coupling parameters 
$\Xi=10$ and $\Xi=100$ with Manning parameter $\xi=3.0$ 
(cross symbols and filled triangle-ups respectively).}
\label{fig:pdf}
\end{center}
\end{figure}

In order to examine counterion-counterion correlations at the surface, 
we consider the one-dimensional pair distribution of counterions, $g_{\mathrm{1D}}({\tilde z})$,
which measures the probability of finding two counterions lined-up along 
$z$-axis (i.e. along the cylinder axis with equal 
azimuthal angles $\phi$) at a distance  ${\tilde z}$ from each other. 
In Figure \ref{fig:pdf}, we plot the unnormalized pair distribution function defined via
\begin{equation}
   g_{\mathrm{1D}}({\tilde z}) 
            \equiv \frac{1}{N} 
                 \sum_{i\neq j}' \bigg\langle \delta({\tilde z}_i-{\tilde z}_j-{\tilde z})\,
                                 \delta(\phi_i-\phi_j)\bigg \rangle,
\label{eq:g_1D}
\end{equation}
where the prime mark indicates that the sum runs only over counterions at   the
surface (defined in the simulations as counterions residing in a 
shell of thickness about the Gouy-Chapman length, $\mu$, around the cylinder).
At small coupling parameter 
($\Xi=10$, cross symbols), the pair distribution function only shows a very weak depletion 
zone at small distances along the cylinder axis. 
For larger values of $\Xi$, one observes a pronounced correlation hole at 
small distances around counterions, where the 
distribution function vanishes over a finite range.
This correlation hole develops in the range of coupling parameters $10<\Xi<100$, 
which marks the crossover regime between the mean-field and the strong-coupling regime
(compare cross symbols and filled triangle-ups) \cite{Andre}. 
The correlation hole appears only for sufficiently large Manning parameter $\xi$
(large enough number of condensed counterions) and is distinguishable 
in our simulations for $\xi>1.2$.   

%%---------Fig pdf

%%------Fig Energy and Heat capacity
\begin{figure*}[t]
\begin{center}
\includegraphics[angle=0,width=17cm]{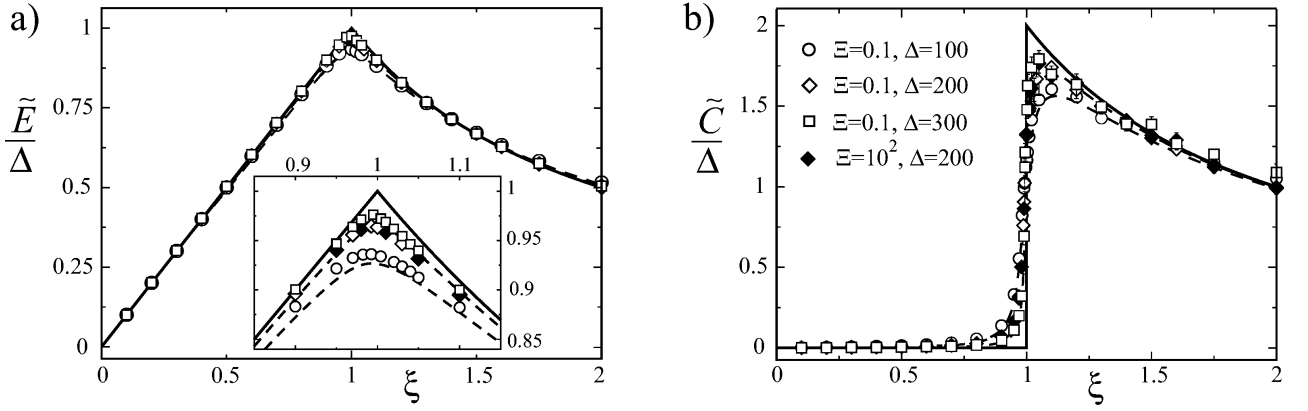}
\caption{a) The rescaled internal energy, ${\tilde E}=E_N/(Nk_{\mathrm{B}}T)$ and b) the 
rescaled excess heat capacity,
${\tilde C}=C_N/(Nk_{\mathrm{B}})$, of the counterion-cylinder system as a function of 
Manning parameter, $\xi$.
Open symbols show the data for small coupling parameter $\Xi=0.1$ and for increasing
lateral extension parameters $\Delta=100, 200$ and 300 as indicated on the graph. 
Filled symbols are the data for large coupling
$\Xi=10^2$ and $\Delta=200$. Here number of counterions is $N=100$. 
Solid curves show the asymptotic PB prediction for $\Delta\rightarrow \infty$, 
Eqs. (\ref{eq:energyPB_asym}) and (\ref{eq:heatPB_asym}).
Dashed curves are the full PB result for $\Delta=100$ and 300 (from bottom to top), which are 
calculated numerically using Eqs. (\ref{eq:energyPB_full}) and (\ref{eq:E_C_relation}). 
The inset shows a closer view of the energy peak. }
\label{fig:EC}
\end{center}
\end{figure*}

The small-separation correlation hole is followed by an oscillatory behavior 
for elevated $\xi$ indicative of a short-ranged liquid-like order 
among counterions line-up along the cylinder axis (distinguishable from the data
for $\xi>2.0$ in the large-coupling regime $\Xi>100$).
The location of the first peak of $g_{\mathrm{1D}}$ gives a rough measure of the 
typical distance between lined-up counterions, $a_z$, at the cylinder surface. 
This distance is set by the local 
electroneutrality condition  $a_z\tau=q$. 
In rescaled units, we obtain  
\begin{equation}
  {\tilde a_z}\equiv \frac{a_z}{\mu}=\frac{\Xi}{\xi}, 
\end{equation}
from Eqs. (\ref{eq:xi}), (\ref{eq:mu}) and (\ref{eq:Xi_coupl}), 
which is used to rescale the horizontal axis of the graph in Figure \ref{fig:pdf}. 

Note that the correlation hole size increases with the coupling parameter and thus
counterions at the surface become highly isolated, reflecting dominate single-particle contributions 
for $\Xi\gg 1$ \cite{Netz,Andre}. In fact, as discussed 
elsewhere \cite{Netz,Andre,Naji04,Naji_PhysicaA,Burak04}, the single-particle form of the SC theory 
(obtained formally for $\Xi\rightarrow\infty$) is a direct consequence of  large correlation hole size 
around counterions at the surface. 
Clearly, for the counterion-cylinder system, this can be the case only for sufficiently 
large Manning parameter, where a sizable fraction of counterions can gather near the surface. 
Consequently in this regime,  the data tend to the strong-coupling predictions for elevated $\Xi$ 
(Figures \ref{fig:s1} and \ref{fig:densX}) as also verified in the simulations of charged plates, 
where all counterions are bound to the surface \cite{Andre}, 
and two charged cylinders with large $\xi$ \cite{Naji04}. 
This also explains why the SC theory, though being able to reproduce the CCT on a qualitative level, 
fails to capture the quantitative features {\em near the critical point} (except for the value of 
the critical Manning parameter), where counterion are mostly de-condensed.

%%%%%%%%%%%%%%%%%%%%%%%%%%%%%%%%%%%%%%
%       SUBSECTION:    THRESHOLD \xi
%%%%%%%%%%%%%%%%%%%%%%%%%%%%%%%%%%%%%%
\subsection{Critical Manning parameter $\xi_c$}
\label{subsec:xi_c}

We now turn our attention to the behavior of counterions near the critical point
and begin with determining the precise location of the critical Manning parameter, $\xi_c$. 

To this end, we shall employ  a procedure similar to the method of locating the transition temperature in 
bulk critical phenomena  \cite{critical,Landau91}. Namely,  one expects that the transition point is reflected 
by a singular behavior in thermodynamic quantities such as energy or heat capacity
as already indicated by the mean-field results obtained in Section \ref{subsubsec:EC_PB}. 
The mean (internal) energy, $E$, and the 
excess heat capacity, $C$,  may be obtained directly from the simulations and 
in rescaled units as
\begin{eqnarray}
&& {\tilde E}=\frac{E_N}{Nk_{\mathrm{B}}T}=\bigg\langle 
						\frac{{\mathcal H}_N}{Nk_{\mathrm{B}}T}\bigg\rangle,
\label{eq:E_def}\\
&& {\tilde C}=\frac{C_N}{Nk_{\mathrm{B}}}=N\bigg\langle 
					\left(\frac{{\delta \mathcal H}_N}{Nk_{\mathrm{B}}T}\right)^2\bigg\rangle,
\label{eq:C_def}
\end{eqnarray}
where the configurational Hamiltonian ${\mathcal H}_N$ is defined through Eq. (\ref{eq:Hamilt_rescaled})
and $\delta {\mathcal H}_N={\mathcal H}_N-\langle {\mathcal H}_N\rangle$.

Simulation results for the rescaled energy, $\tilde E$, and the rescaled excess heat capacity, $\tilde C$, in Figure \ref{fig:EC}
(symbols)  show a non-monotonic behavior as a function of $\xi$.  The energy develops a pronounced peak
 and the heat capacity exhibits a jump at intermediate 
Manning parameters, which become  singular for $\Delta$ increasing to infinity. 
The general behavior of energy and 
heat capacity can be understood using simple arguments as follows. 

For sufficiently small $\xi$, counterions are all unbound 
and the electrostatic potential in space is roughly given by the bare potential of the charged 
cylinder, i.e. $\psi({\tilde r})\simeq 2\xi\ln ({\tilde r}/\tilde R)$. 
This yields the rescaled internal energy, $\tilde E$,  
(via integrating over the square electric field, Eq. (\ref{eq:E_rescal_def})) as 
\begin{equation}
 	\tilde E=\frac{1}{4\xi}\int_{\tilde R}^{\tilde D} \,\tilde r\, {\mathrm{d}}\tilde r\, 
				\bigg(\frac{ {\mathrm{d}}\psi}{ {\mathrm{d}}\tilde r}\bigg)^2
		     \simeq \xi\Delta
\label{eq:E_asymp1}
\end{equation}
for $\Delta=\ln(D/R)\gg 1$. Intuitively, this result may be obtained also by assuming that 
counterions experience the potential of the cylinder at the outer boundary; thus one simply
has $\tilde E\simeq \psi(\tilde D)/2 \simeq  \xi \Delta$, 
 which explains the linear increase of the left tail of the 
energy curve with  both $\xi$ and $\Delta$ (Figure \ref{fig:EC}a).  
Now using the following thermodynamic relation
\begin{equation}
  \xi\frac{\partial \tilde E}{\partial \xi}={\tilde E}-{\tilde C},
\label{eq:E_C_relation}
\end{equation}  
the excess heat capacity is obtained to vanish in the 
de-condensation regime, i.e. $ \tilde C\simeq 0$ (Figure \ref{fig:EC}b).
Hence, the heat capacity reduces to that of an ideal gas of particles.

For large $\xi$, the electrostatic potential of the cylinder is screened
due to counterionic binding. If we estimate the screened electrostatic potential of the cylinder
as $\psi({\tilde r})\simeq 2\ln ({\tilde r}/{\tilde R})$, which can be verified systematically within the 
PB theory  \cite{Fuoss51,Netz_Joanny},  we obtain the energy and the heat capacity as 
\begin{equation}
 \tilde E\simeq \Delta/\xi\,\,\,\,{\mathrm{and}}\,\,\,\,
 \tilde C\simeq 2\Delta/\xi.
\label{eq:EC_asymp2}
\end{equation} 
These  results may also be obtained  by noting that only the fraction $1/\xi$ of de-condensed 
counterions (Section \ref{subsubsec:CondRatio}) contributes to the energy on the leading order; thus  
$\tilde E\simeq \psi(\tilde D)/(2\xi) \simeq  \Delta/\xi$.
The above asymptotic estimates  in fact coincide with the asymptotic ($\Delta\rightarrow \infty$)  
PB results (\ref{eq:energyPB_asym}) and
(\ref{eq:heatPB_asym}), which are shown by solid curves in Figure \ref{fig:EC}.

%%--------Fig Energy peak
\begin{figure}[t]
\begin{center}
\includegraphics[angle=0,width=8cm]{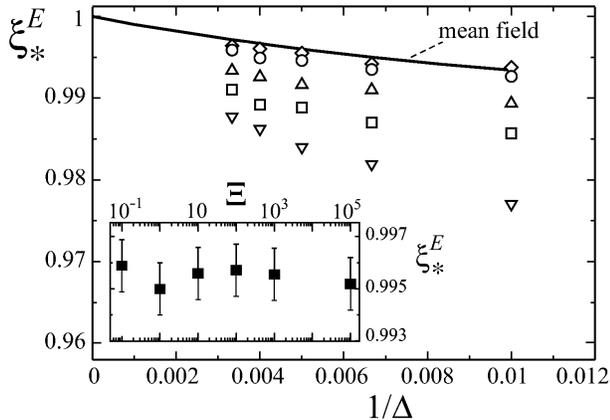}
\caption{Main set: Location of the peak of energy, $\xi_\ast^E$, as a function of the 
inverse lateral extension parameter, $1/\Delta=1/\ln (D/R)$. Open symbols are simulation results
for small coupling parameter $\Xi=0.1$ and for various number of counterions (from bottom):
$N=25$ (triangle-downs), 50 (squares), 100 (triangle-ups), 200 (circles) and 300 (diamonds). 
Solid curve shows the mean-field PB prediction for the peak location as obtained
 by numerical evaluation of the full PB energy, Eq. (\ref{eq:energyPB_full}). 
Inset: Location of the energy peak as
a function of the coupling parameter, $\Xi$, for $N=200$ and $\Delta=300$.}
\label{fig:Emax}
\end{center}
\end{figure}

The preceding considerations demonstrate that the non-monotonic behavior of the energy 
and excess heat capacity results directly from the screening effect due to the condensation of
counterions for increasing $\xi$. Hence, the singular peaks emerging in both quantities reflect 
the onset of the counterion-condensation transition, $\xi_c$, which occurs 
in the thermodynamic infinite-system-size limit $N\rightarrow \infty$ and $\Delta\rightarrow \infty$.  
Within the PB theory (solid and dashed  curves in Figure \ref{fig:EC}), 
the location of the peak of energy, $\xi_\ast^{E, {\mathrm{PB}}}(N, \Delta)$, 
 tend to the mean-field critical value $\xi_c^{\mathrm{PB}}=1$ 
 from {\em below} as $\Delta$ increases obeying the finite-size-scaling relation
\begin{equation}
 \xi_c^{\mathrm{PB}}-\xi_\ast^{E, {\mathrm{PB}}}(\Delta)\simeq \frac{1}{\Delta},
\label{eq:Emax_asymp}
\end{equation}
which is obtained using the full PB energy  (\ref{eq:energyPB_full}). 
On the other hand, the location of the peak of the PB heat capacity  approaches
$\xi_c^{\mathrm{PB}}$ from {\em above}. 
%Both trends agree with the simulation data in Figure \ref{fig:EC}.

We locate the critical point from the asymptotic behavior of the energy peak, $\xi_\ast^E$, for
increasing $N$ and $\Delta$.  (The heat capacity peak is found to be located further 
away from the critical point than the energy peak, resembling the well-known behavior of the 
heat capacity peak in finite Magnetic systems \cite{Landau91}, which  makes it inconvenient  for our purpose). 
In Figure \ref{fig:Emax}, we show the simulation results for $\xi_\ast^E$  (symbols)
as a function of $\Delta^{-1}$ for $\Xi=0.1$ and for various number of particles (main set). 
These data are obtained using the thermodynamic relation (\ref{eq:E_C_relation}), which allows us to calculate the first
derivative of energy, $\partial \tilde E/\partial \xi$,  
%(including its error-bars)
directly from the energy and the heat capacity data without referring to numerical differentiation
methods which typically generate large errors near the peak.
As seen, for increasing
$N$, the data converge to and closely follow the mean-field prediction 
(solid curve) within the estimated error-bars; for $N>100$, $\xi_\ast^E$ lies 
within about 1\% of the PB critical Manning parameter $\xi_c^{\mathrm{\mathrm{PB}}}=1$. Since in the simulations we have 
used $\Delta\leq 300$, the behavior of  $\xi_\ast^E$ for very small $\Delta^{-1}\rightarrow 0$
is not obtained, nevertheless, the excellent convergence of the data for $\Xi=0.1$ to the PB prediction 
gives an accurate estimate for  the critical Manning parameter as
\begin{equation}
  \xi_c=1.00\pm 0.002. 
\end{equation}

Our results for larger values of the coupling parameter, $\Xi$, in the inset of Figure \ref{fig:Emax}
show that  the location of the energy peak does not vary with the coupling 
parameter. 
Therefore, we find that the critical Manning parameter is
{\em universal} and given by the mean-field value $\xi_c=1.0$. 
Recall that the same threshold is obtained within the Onsager instability and the strong-coupling
analysis (Sections \ref{subsec:beyond_ons} and \ref{sec:CCT_SC}).

Another important result is that 
the CCT is not associated with a diverging singularity, in contrast 
to the Onsager instability prediction  \cite{Manning69}. But, the
energy at any finite value of $\xi$, and also the heat capacity for $\xi>1$, tend to infinity (as $\sim \Delta$) when
the lateral extension parameter, $\Delta$, increases to infinity, which, as illustrated before, reflects the logarithmic 
divergency of the effective electrostatic potential in a charged cylindrical system.
The CCT, however, exhibits a {\em universal} discontinuous jump for the excess heat capacity 
at $\xi_c$, and thus indicates a second-order phase transition (Figure \ref{fig:EC}).

%%%%%%%%%%%%%%%%%%%%%%%%%%%%%%%%%%%%%%
%       SUBSECTION:    FS EFFECTS and GFSS
%%%%%%%%%%%%%%%%%%%%%%%%%%%%%%%%%%%%%%
\subsection{Scale-invariance near the critical point}
\label{subsec:scaling}

Now that the precise location of the critical Manning parameter is determined, 
a finite-size analysis, similar to what we presented within the 
mean-field theory,
may  be used to determine the near-threshold properties of the CCT
order parameters from the simulation data. 

Note that in the simulations, finite size effects arise both from the finiteness
of the system size (via the lateral extension parameter, $\Delta$), and also from the finiteness 
of the number of counterions, $N$; the latter being related to the finiteness of the height
of the main simulation box $L=Nq/\tau$ (Section \ref{sec:sim}), which has a sizable 
influence on the transition, although the implemented periodic boundary condition already
reduces its effects. In what follows, we present the numerical evidence for
scaling relations with respect to both $N$ and $\Delta$. The 
asymptotic behavior for increasing $N$ and $\Delta$ to infinity provides us
with the scaling behavior with respect to the reduced Manning parameter, $\zeta$ 
(or the reduced temperature, $t$), which characterizes the CCT universality class in 3D.   

%--------------------------
\subsubsection{Finite-size effects near $\xi_c$}
\label{subsubsec:FS_effect}

\begin{figure}[t]
\begin{center}
\includegraphics[angle=0,width=8cm]{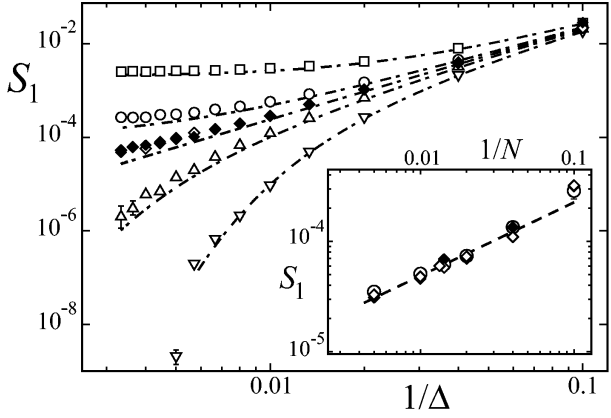}
\caption{Main set: Order parameter $S_1=\langle 1/\tilde r\rangle$ 
as a function of the inverse lateral extension parameter $1/\Delta$.  Open symbols
are data for  $\Xi=0.1$ and Manning parameters (from top): 
$\xi=1.05$ (squares), 1.01 (circles), 1.0 (open diamonds), 0.99 (triangle-ups) and 0.97 (triangle-downs). 
Filled diamonds are the data for 
large coupling parameter $\Xi=10^2$ and  $\xi=1.0$. Number of
counterions $N=100$ is fixed. Dot-dashed  curves are
the PB result, Eq. (\ref{eq:S_n_PB}), for the corresponding $\xi$.  Inset:  $S_1$
as a function of the inverse number of counterions $1/N$ for $\xi=1.0$ and for the
coupling parameters $\Xi=0.1$ (open diamonds), $10^2$ (filled diamonds) and $10^3$ (open circles). 
Dashed line shows the power-law exponent 2/3. Here $\Delta=300$.}
\label{fig:varDN}
\end{center}
\end{figure}

In Figure  \ref{fig:varDN} (main set), we show the order parameter $S_1$
as a function of $1/\Delta$ and in the vicinity of the critical point $\xi_c=1$ (number of counterions  $N=100$ is fixed). 
$S_1$, which represents the mean inverse localization length of counterions, 
gradually decreases with decreasing $1/\Delta$ as de-condensation of counterions becomes gradually 
more pronounced, but  
for Manning parameters as large as $\xi=1.05$ (open circles), the data quickly saturate to a finite 
value as $\Delta\rightarrow\infty$.  For sufficiently small  Manning parameter (e.g.,  $\xi<0.97$), on the other hand, 
$S_1$ converges to zero. In the vicinity of the critical point ($\xi=1$, diamonds), 
a non-saturating behavior is found suggesting a power-law decay as
$S_1 \sim \Delta^{-\gamma}$, where $\gamma>0$.  
As seen, the data at $\xi=1$ roughly coincide for both
small coupling ($\Xi=0.1$, open diamonds) and large coupling ($\Xi=10^2$, filled diamonds)
 indicating that electrostatic correlations do not influence the scaling behavior (see below).
 There still remain non-negligible deviations between the simulation data at the critical point (diamonds) 
and the PB power-law prediction (\ref{eq:Sn_PB_del}) with $\gamma_{\mathrm{PB}}=2$, which is 
shown in the figure by a straight dot-dashed line. These deviations arise from the
finiteness of the number of particles.

Interestingly, the data obtained for increasing number of counterions, $N$ (at fixed lateral extension parameter,  $\Delta$),  
also indicate a power-law decay near the critical point, i.e. as $S_1 \sim N^{-\nu}$,  where $\nu>0$.
This is shown in the inset of Figure \ref{fig:varDN}, where the scaling exponent 
$\nu$ appears to be about $2/3$ (represented by a dashed line). 
In fact, for sufficiently large $N$, the data deviate from this power-law behavior since finite-size effects
due to lateral extension of the system, $\Delta$, are simultaneously present.
Thus in order to determine the exponents $\gamma$ and $\nu$,  a more systematic approach is required, which 
should  incorporate both lateral-size and ion-number effects.

%%----Fig S1 vs 1/Delta and 1/N

%--------------------------------------------------------
\subsubsection{Generalized finite-size scaling relations} 
\label{subsubsec:GFSS} 

In brief,  our data suggest that at the critical point ($\zeta=1-\xi_c/\xi=0$) and 
for a bounded system (finite $\Delta$) in the thermodynamic limit $N\rightarrow \infty$, the order parameter
$S_n(\zeta, \Delta, N)=\langle 1/\tilde r^n\rangle$ decays  as
\begin{equation}
  S_n(0, \Delta, \infty)\sim \Delta^{-\gamma},
\end{equation} 
while  in an unbounded system ($\Delta\rightarrow \infty$) and for finite $N$, we expect a power-law 
decay as 
\begin{equation}
  S_n(0, \infty, N)\sim N^{-\nu}. 
\end{equation} 
In thermodynamic infinite-system-size limit  ($\Delta\rightarrow \infty$ and $N\rightarrow \infty$), the 
true critical transition sets in with $S_n(\zeta<0, \infty, \infty)=0$, and we
anticipate the scaling behavior with the reduced Manning $\zeta$ as 
\begin{equation}
  S_n(\zeta, \infty, \infty)\sim \zeta^{\chi}
\end{equation}
in a sufficiently small neighborhood {\em above} $\xi_c=1$.

These scaling relations may all be deduced from a general finite-size scaling hypothesis for $S_n$,
i.e. assuming that $S_n(\zeta, \Delta, N)$ takes a {\em homogeneous scale-invariant}
 form with respect to its arguments in the vicinity of the
transition point, $\xi_c$,  when 
both $N$ and $\Delta$ are sufficiently large. 
In other words,  for any positive number $\lambda>0$,
\begin{equation}
  S_n(\lambda\zeta, \lambda^{-b}\Delta, \lambda^{-c} N)=
        \lambda^{a}S_n(\zeta, \Delta, N), 
\label{eq:GFSS1}
\end{equation}
where $a, b$ and $c$ are a new set of exponents associated with $\zeta, \Delta$ and $N$
respectively. The above relation implies that when the reduced Manning parameter, $\zeta$, 
is rescaled with a factor $\lambda$, 
the size parameters, $N$ and $\Delta$, can be rescaled such that the order parameter remains 
invariant up to a scaling prefactor. Finite-size scale-invariance is a common feature in critical 
phase transitions \cite{critical,Fisher72} and provides an accurate tool to estimate critical exponents 
in numerical simulations \cite{Landau91,Binder88}.  
The exponents  in Eq. (\ref{eq:GFSS1}) 
can be calculated directly from MC simulations.  These exponents are in fact 
related to and give the values of  the desired critical exponents 
$\gamma$, $\nu$ and $\chi$, as will be shown below. 
Note that the exponents  may in general depend on $n$ (the index of $S_n$), 
the coupling parameter, $\Xi$, or the space dimensionality, which are not explicitly incorporated 
in the proposed scaling hypothesis, but their influence will be examined  later.

Given Eq. (\ref{eq:GFSS1}), the following relations are obtained
by suitably choosing $\lambda$. For $\lambda=N^{1/c}$, one finds
\begin{equation}
  S_n(\zeta, \Delta, N)=N^{-a/c}
  {\mathcal C}_n(\zeta N^{1/c}, \Delta N^{-b/c}),
\label{eq:GFSS_N_del}
\end{equation} 
where ${\mathcal C}_n(u, v)$ is the scaling function corresponding to a
system with both finite $N$ and $\Delta$. 
The above expression
is useful for a system with finite $N$ in the limit 
$\Delta\rightarrow \infty$. 
Thus assuming that ${\mathcal C}_n(u, v)$ exists for $v=\Delta N^{-b/c}\rightarrow \infty$, the 
relation (\ref{eq:GFSS_N_del}) reduces to
\begin{equation}
  S_n(\zeta,  \infty, N)=N^{-a/c}
  {\mathcal N}_n(\zeta N^{1/c}),
\label{eq:GFSS_N}
\end{equation} 
where the scaling function ${\mathcal N}_n(u)={\mathcal C}_n(u, \infty)$. 
The critical exponent $\nu$ follows by considering this relation right at the 
critical point, $\zeta=0$, i.e. 
\begin{equation}
  S_n(0, \infty, N)={\mathcal N}_n(0)\, N^{-\nu},
\end{equation} 
where $\nu$ is obtained as
\begin{equation}
  \nu = \frac{a}{c}.
\label{eq:nu}
\end{equation}
On the other hand, we assume that in the vicinity of (and above) the critical point  (i.e. for 
small but finite $\zeta$), $S_n(\zeta,  \infty, N)$ is only a finite function of the reduced
Manning parameter $\zeta$ when the limit
$N\rightarrow \infty$ is taken. Hence the scaling function ${\mathcal N}_n(u)$
 is required to behave as 
${\mathcal N}_n(u)\sim u^a$ for $u\rightarrow \infty$, which yields
\begin{equation}
  S_n(\zeta, \infty, \infty)\sim \zeta^{\chi},
\end{equation}  
where the critical  exponent associated with $\zeta$ reads
\begin{equation}
  \chi=a.
\label{eq:chi}
\end{equation}

To determine the critical  exponent associated with $\Delta$ in terms of the exponents 
$\{a, b, c\}$, we need to consider Eq. (\ref{eq:GFSS1}) for $\lambda=\Delta^{1/b}$.
We thus have  
\begin{equation}
  S_n(\zeta, \Delta, N)=\Delta^{-a/b}
  {\mathcal C'}_n(\zeta\Delta^{1/b}, N\Delta^{-c/b}),
\label{eq:GFSS_del_N}
\end{equation} 
where ${\mathcal C'}_n(u, v)$ is a new scaling function. 
This relation is useful for a system with finite $\Delta$ in the limit 
$N\rightarrow \infty$, where assuming that  ${\mathcal C'}_n(u, v)$ exists, 
we obtain 
\begin{equation}
  S_n(\zeta, \Delta, \infty)=\Delta^{-a/b}
  {\mathcal D}_n(\zeta\Delta^{1/b})
\label{eq:GFSS_del}
\end{equation} 
with a new scaling function ${\mathcal D}_n(u)={\mathcal C'}_n(u, \infty)$. 
The critical exponent $\gamma$ follows by considering this relation right at the 
critical point, $\zeta=0$, that yields 
\begin{equation}
  S_n(0, \Delta, \infty)={\mathcal D}_n(0)\,\Delta^{-\gamma},
\end{equation} 
where $\gamma$ reads 
\begin{equation}
  \gamma = \frac{a}{b}.
\label{eq:gamma}
\end{equation}
Therefore, we have a complete set of relations (\ref{eq:nu}), (\ref{eq:chi}) and (\ref{eq:gamma}) 
from which the critical exponents $\gamma$, $\nu$ and $\chi$ may be obtained using the 
exponents $a, b$ and $c$. 

Equation (\ref{eq:GFSS_del}) compares directly with the mean-field result, Eq. (\ref{eq:Sn_PB_del}), where
we showed that $\gamma_{\mathrm{PB}}=2$ and $\chi_{\mathrm{PB}}=2$.  
Note also that the exponent $\nu$ is not defined within the mean-field theory.

%%------Fig NFS 
\begin{figure}[t]
\begin{center}
\includegraphics[angle=0,width=8cm]{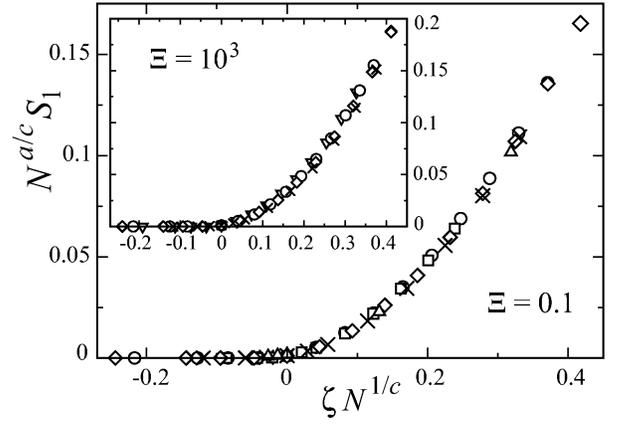}
\caption{Rescaled order parameter, $N^{a/c}S_1$, as a function of the rescaled reduced
Manning parameter, $\zeta N^{1/c}$, in the vicinity of the critical point, 
$\xi_c=1.0$, and for small and large coupling parameters 
$\Xi=0.1$ (main set) and $\Xi=10^3$ (inset). Symbols show data for various number of particles
$N=50$ (triangle-downs),  70 (circles),
75 (squares), 100 (diamonds), 200 (cross symbols), 300 (triangle-ups), and fixed $\Delta=300$.  In these plots, 
the exponents are chosen here as $a/c=2/3$ and $1/c=1/3$. 
Error-bars are smaller than the symbol size.} 
\label{fig:NFS}
\end{center}
\end{figure}

%%%%%%%%%%%%%%%%%%%%%%%%%%%%%%%%%%%%%%
%       SUBSECTION:    EXPONENTS
%%%%%%%%%%%%%%%%%%%%%%%%%%%%%%%%%%%%%%
\subsection{Critical  exponents: the CCT universality class}
\label{subsec:sim_exp}

%----------------------------------
%   SUBSUBSECTION:    Chi & Nu
%----------------------------------
\subsubsection{The exponents $\chi$ and $\nu$}
\label{subsubsec:chi_nu}

In order to verify  the generalized finite-size scaling hypothesis (\ref{eq:GFSS1})
and estimate the critical exponents numerically,  we shall adopt the standard 
data-collapse scheme used widely in literature \cite{Binder88}. 

We begin with the exponents  $\chi$ and $\nu$ that can be calculated 
using Eq. (\ref{eq:GFSS_N_del}),  which involves  a scaling
function, ${\mathcal C}_n(u, v)$, of two arguments $u=\zeta N^{1/c}$ and $v=\Delta N^{-b/c}$.
In our simulations, $N$ ranges from 25 up to 300 and $\Delta$
ranges from 50 up to 300; thus assuming that the exponent $b/c$ is small, which will be verified
later on, we deal with a typically large value for $v\sim 10-10^2$. Therefore the limiting 
relation (\ref{eq:GFSS_N}) is approximately valid and yields
\begin{equation}
  N^{a/c}S_n\simeq {\mathcal N}_n(\zeta N^{1/c}).
\label{eq:NS_n}
\end{equation} 

Now if the data for $S_n$ are plotted as function of $\zeta=1-\xi_c/\xi$ for {\em various} $N$ (but at fixed sufficiently  large 
$\Delta$), equation (\ref{eq:NS_n}) predicts that by rescaling the reduced Manning parameter $\zeta$ by the factor
of $N^{1/c}$ and the order parameter by the factor $N^{a/c}$, all data should collapse onto a single
curve. Numerically, this procedure allows to determine 
the exponents $a/c$ and $1/c$ in such a way that the best data collapse is achieved. 
We show the results in Figure \ref{fig:NFS} for various $N$ (symbols) and for the coupling parameter $\Xi=0.1$ (main set). 
The collapse of the data onto each other is indeed achieved within the numerical error-bars for the exponents in the range 
$1/c\simeq 1/3\pm 0.05$ and $a/c\simeq 2/3\pm 0.1$. 
This yields the critical exponents 
$\nu$ and $\chi$ from Eqs. (\ref{eq:nu}) and (\ref{eq:chi}) as
\begin{eqnarray}
  \nu\simeq 2/3\pm 0.1,\\
  \chi \simeq 2.0 \pm 0.4,   
\end{eqnarray}
where the errors are estimated using the standard error propagation methods.  The value obtained for 
$\chi$ agrees with the mean-field result, Eq. (\ref{eq:chi_PB}).

In order to check whether the exponents vary with the electrostatic coupling parameter, $\Xi$, we repeat 
this procedure for a wide of range of values for $\Xi$. We find the 
same values for the exponents for coupling parameters up to $\Xi=10^5$. For
comparison, the results for  $\Xi=10^3$ are shown in the inset  
of Figure \ref{fig:NFS}, where the data collapse is demonstrated for $1/c= 1/3$ and $a/c= 2/3$. 
 
%%------Fig S1 vs 1/Delta and var N
\begin{figure}[t]
\begin{center}
\includegraphics[angle=0,width=8cm]{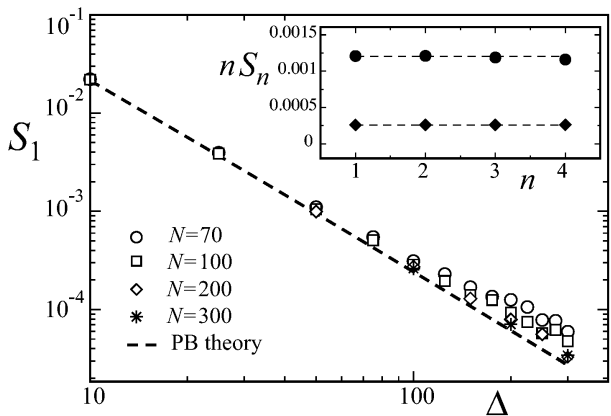}
\caption{Main set: Order parameter $S_1$ as a function of the lateral extension parameter, $\Delta$,
for increasing number of counterions from $N=70$  up to 300 as indicated on the graph (Manning 
parameter here is $\xi=1.0$ and the coupling parameter $\Xi=0.1$). Dashed line shows
the PB power-law, Eq. (\ref{eq:Sn_PB_del}). Inset: 
Rescaled order parameters $S_n=\langle \tilde r^{-n}\rangle$ as a function of
$n$ for Manning parameters  close to the critical point (from top): $\xi=1.03$ (filled
circles) and  $\xi=1.01$ (filled diamond). Here the coupling parameter is 
$\Xi=0.1$, the number of counterions $N=100$, and $\Delta=300$. 
%Error-bars are smaller than the symbol size.
}
\label{fig:s1vsDvarN}
\end{center}
\end{figure}

%----------------------------------
%   SUBSUBSECTION:    Gamma
%----------------------------------
\subsubsection{The exponent $\gamma$}
\label{subsubsec:gamma}

Given the exponents $a$ and $c$ calculated above and making use of the finite-size scaling relation
(\ref{eq:GFSS_del_N}), we can estimate the exponent $b$,
and thereby the scaling exponent $\gamma$, associated with the lateral extension parameter. 
In this case, however, the second argument $v=N\Delta^{-c/b}$ 
in the scaling function ${\mathcal C'}_n(u,v)$ defined in 
Eq. (\ref{eq:GFSS_del_N}) may not be considered as large
within our simulations (since as shown below the ratio $c/b$ is large).
 But it turns out that the dependence of 
${\mathcal C'}_n(u,v)$ on $v$ is quite weak such that the finite-size scaling relation 
(\ref{eq:GFSS_del}) is approximately valid and can thus give  
the desired exponent. To examine this latter property of ${\mathcal C'}_n(u,v)$, we consider
the relation (\ref{eq:GFSS_del_N}) right at the critical Manning parameter ($\zeta=0$), i.e. 
\begin{equation}
  S_n(0, \Delta, N)=\Delta^{-a/b}
  {\mathcal C}_n(0, N\Delta^{-c/b}).
\end{equation}
In Figure \ref{fig:s1vsDvarN}, $S_n(0, \Delta, N)$ is plotted as a function of $\Delta$ in a log-log plot
for increasing $N$ from 70 up to 300 and for $\Xi=0.1$. 
As clearly seen, the order parameter varies quite  weakly with the 
number of particles, and the variations are already within
the error-bars (equal to symbol size)  for $N>100$.

%%-----Fig DFS
\begin{figure}[t]
\begin{center}
\includegraphics[angle=0,width=8cm]{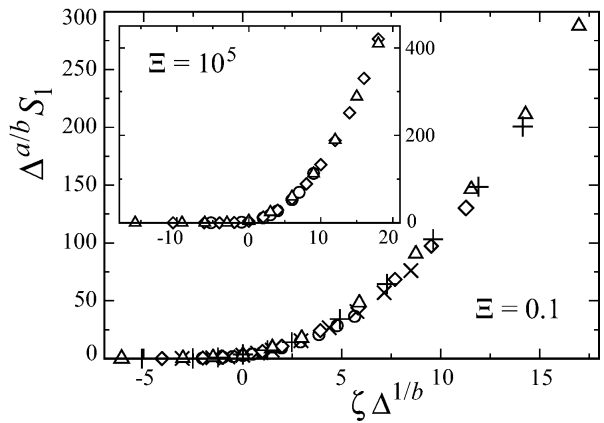}
\caption{Rescaled order parameter, $\Delta^{a/b}S_1$, as a function of 
the rescaled reduced Manning parameter, $\zeta \Delta^{1/b}$, in the vicinity of the critical  
point, $\xi_c=1.0$, and for small and large coupling parameters 
$\Xi=0.1$ (main set) and $\Xi=10^5$ (inset). Symbols show data for 
various lateral extension parameters $\Delta=100$ (circles),
150 (cross symbols), 200 (diamonds), 250 (plus symbols) and 300 (triangle-ups). 
Number of counterions is fixed ($N=200$ for the coupling parameter $\Xi=0.1$
and $N=100$ for  $\Xi=10^5$), and the exponents are here 
chosen as $a/b=2.0$ and $1/b=1.0$.
%Error-bars are smaller than the symbol size.
} 
\label{fig:DFS}
\end{center}
\end{figure}

Thus multiplying both sides of Eq. (\ref{eq:GFSS_del}) with $\Delta^{a/b}$, we have 
\begin{equation}
  \Delta^{a/b}S_n\simeq {\mathcal D}_n(\zeta\Delta^{1/b}),
\end{equation} 
in which the exponent $a$ previously determined  as $a=\chi=2.0\pm 0.4$. 
We thus plot the order parameter $S_n$ as a function of $\zeta$ for {\em various} $\Delta$ (but at fixed sufficiently 
large $N$), and rescale both $S_n$ and $\zeta$ values with the scaling factors $\Delta^{a/b}$ and $\Delta^{1/b}$
respectively; the exponent $b$ is chosen in such a way that the best data collapse is obtained within the error-bars. 
The result is shown in Figure \ref{fig:DFS} for $\Delta^{a/b}S_1$
as a function of  $\zeta \Delta^{1/b}$, where the coupling parameter is chosen as $\Xi=0.1$. 
The collapse of the data onto each other is obtained  only for the exponent $1/b$ in the range 
$1/b\simeq 1.0\pm 0.2$, yielding  the critical exponent $\gamma$
from Eq. (\ref{eq:gamma}) as
\begin{equation}
  \gamma \simeq 2.0 \pm 0.6, 
\end{equation} 
which agrees with the mean-field exponent, Eq. (\ref{eq:gamma_PB}). 
We find the same value for  $\gamma$ by repeating the above procedure 
for larger coupling parameters. For instance, 
the results for  $\Xi=10^5$ are shown in the inset  
of Figure \ref{fig:DFS}, where we have chosen $1/b= 1.0$.

Note that the estimated values of $b$ and $c$ show that the ratio
$b/c$ is as small as 0.3, which is consistent with the assumption made in using the asymptotic forms
(\ref{eq:GFSS_N}) and (\ref{eq:GFSS_del})  in the foregoing 
data-collapse procedure. 

As a main result, our numerical data 
confirm the existence of characteristic scaling relations 
associated with the counterion-condensation transition in 3D and show that  the values of 
the critical exponents  are {\em universal}, i.e. independent of the coupling parameter, $\Xi$, and
agree with the mean-field universality class. 

Also, in agreement with mean-field results, the exponents are found to be  independent of $n$,  the index of the 
order parameters $S_n=\langle 1/\tilde r^n\rangle$.  In fact, we find that the higher-order
moments are related to the first-order moment, $S_1$, via
\begin{equation}
  S_n\simeq \frac{S_1}{n}
\end{equation}
in the {\em vicinity} of the critical point, which indicates that $nS_n$ is independent of $n$, 
as demonstrated in the inset of  Figure \ref{fig:s1vsDvarN} (compare with the mean-field relation (\ref{eq:S_n_PB_approx})).

%%%%%%%%%%%%%%%%%%%%%%%%%%%%%%%%%%%%%%%%%%%%%%%%%%%%%%%%%%%%%%%%%%%%%

%                                2D SYSTEM                      

%%%%%%%%%%%%%%%%%%%%%%%%%%%%%%%%%%%%%%%%%%%%%%%%%%%%%%%%%%%%%%%%%%%%%%
\section{Counterion-condensation transition (CCT) in two dimensions}
\label{sec:2D}

In this section, we shall investigate the role of space dimensionality in the asymptotic
behavior of counterions at a charged cylindrical boundary, by considering  a 2D 
counterion-cylinder model. As a typical trend in bulk critical phenomena, the effects of fluctuations near the critical point  
are known to grow with diminishing dimension \cite{critical},
and cause large deviations from mean-field theory.  It is therefore interesting to study
the CCT in a lower spatial dimension.

%%%%%%%%%%%%%%%%%%%%%%%%%%%%%%%%
\subsection{The two-dimensional model}
\label{subsec:2D_model}

In 2D, we use a primitive cell model similar to the 3D model described in Section \ref{subsec:model}. 
It consists of a 2D central charged cylinder (central  ``disk'') of radius $R$ confined co-axially
and together  with its neutralizing point-like counterions in an outer cylinder (outer  ``ring'') of radius $D$. 
In order to construct the two-dimensional interaction Hamiltonian, 
we use the fact that the Coulomb interaction
between two elementary charges  in 2D (the 2D Green's function) is of the form 
\begin{equation}
  v_{\mathrm{2D}}({\mathbf x})=-\ln |{\mathbf x}|.
  \label{eq:2D_Green}
\end{equation} 
This follows directly from the solution of the 2D Poisson equation for a point charge, that is
\begin{equation}
  \nabla^2 v_{\mathrm{2D}}({\mathbf x})= -2\pi \delta^2({\mathbf x}).
\end{equation}

The configurational Hamiltonian of the 2D system may thus be written as
\begin{equation}
   \frac{{\mathcal H}_N}{k_{\mathrm{B}}T} = 
         \lambda_{c}\lambda_{r} \sum_{i=1}^{N} \ln \bigg(\frac{r_i}{R}\bigg) - 
         \lambda_c^2 \sum_{\langle ij \rangle} \ln\bigg|\frac{{\mathbf x}_i-{\mathbf x}_j}{R}\bigg|    
\label{eq:2DHamilt}
\end{equation}
with ${\mathbf x}_i=(r_i, \phi_i)$ being the position vector of the $i$-th counterion 
(in polar coordinates), and $\lambda_{c}$ and $\lambda_{r}$ being dimensionless charges
of the counterions and the cylinder respectively. 
The first term gives the counterion-cylinder
attraction and the second term gives mutual repulsions between counterions. 
Clearly, the present 2D model is equivalent to
a 3D system comprising an infinitely-long central cylinder (of radius $R$) in the presence of 
mobile {\em parallel} lines of opposite charge as ``counterions", which may be applicable 
to a system of oriented
cationic and anionic polymers  \cite{Jason}. Using this 3D analogy, the prefactors 
$\lambda_{r}$ and $\lambda_{c}$ may be related to the linear charge density of the cylinder
and counterion lines respectively.

Taking the  logarithmic interaction (\ref{eq:2D_Green}) will also ensure 
that the general form of the field-theoretic representation for the system 
remains the same as in the 3D case \cite{Kardar_rev}, and in particular, the mean-field Poisson-Boltzmann theory,
which follows from a saddle-point analysis, 
is represented exactly by the same equations and results as discussed in 
Section \ref{sec:CCT_PB}.

%%%%%%%%%%%%%%%%%%%%%%%%%%%%%%%%%%
\subsection{Rescaled representation}
\label{subsec:2D_d_less}

In analogy with the 3D system, we shall refer to the dimensionless prefactor
of the counterion-cylinder interaction in Eq. (\ref{eq:2DHamilt}) as the {\em Manning parameter}, that is
\begin{equation}
   \xi =\lambda_{c}\lambda_{r}/2. 
\end{equation}
Also the prefactor of the counterion-counterion interaction is defined as the {\em coupling parameter}
\begin{equation}
   \Xi =\lambda_{c}^2/2.
\end{equation}
These definitions can be justified systematically when the Hamiltonian of the system is mapped to 
an effective field theory, where $\Xi$ and $\xi$ formally appear
in the same role as in 3D \cite{Burak_unpub}.  We shall conventionally rescale the spatial coordinates as 
${\tilde {\mathbf x}}={\mathbf x}/\mu_{\mathrm{2D}}$ using the length scale
$\mu_{\mathrm{2D}}\equiv R/\xi$, which is the 2D analogue of Eq. (\ref{eq:xi_Rtilde}).

The Hamiltonian in rescaled units reads
\begin{equation}
    \frac{{\mathcal H}_N}{k_{\mathrm{B}}T} = 
          2\xi \sum_{i=1}^{N} \ln \bigg(\frac{{\tilde r}_i}{\tilde R}\bigg) - 
          2\Xi \sum_{\langle ij \rangle} \ln\bigg|\frac{{\tilde {\mathbf x}}_i-{\tilde {\mathbf x}}_j}{\tilde R}\bigg|.  
\label{eq:2DHamilt_res}
\end{equation}
The electroneutrality condition  implies  
$\lambda_r= N\lambda_c$,
where $N$ is the number of counterions in the system. This relation may also be written as 
\begin{equation}
  \xi= N\Xi.
\label{eq:2DENC}
\end{equation}
Thus an important consequence of electroneutrality in  2D 
is that the coupling parameter and the Manning parameter
are related only via the number of counterions. In particular, in the thermodynamic limit
$N\rightarrow \infty$, the coupling parameter tends to zero,
$\Xi\rightarrow 0$, suggesting that the mean-field prediction should become exact!

We use a similar simulation method as devised for the 3D system using
 the transformed coordinates $(y, \phi)$ with $y=\ln ({\tilde r}/{\tilde R})$ being the logarithmic radial 
 distance of particles from the central cylinder. As explained in Section \ref{sec:sim}, this transformation 
leads to the centrifugal sampling method appropriate for equilibration of 
systems with large lateral extension parameter 
$\Delta=\ln (D/R)\gg 1$, where the critical behavior associated with the CCT emerges. 
The 2D partition function thus reads
\begin{equation}
   {\mathcal Z}_N=\frac{R^{2N}}{N!}
         \int_{\tilde V} \left[\prod_{i=1}^{N} {\mathrm{d}}\phi_i \, {\mathrm{d}} y_i\right]
                  \exp\bigg\{-\frac{{\mathcal H}_N^\ast}{k_{\mathrm{B}}T}\bigg\},
\label{eq:2DZ_y}
\end{equation}
where $0\leq y\leq \Delta$ and the transformed Hamiltonian,
\begin{equation}
   \frac{{\mathcal H}_N^\ast}{k_{\mathrm{B}}T}=(2\xi-2) \sum_{i=1}^{N} y_i
                     -2\Xi\sum _{\langle ij \rangle}
                      \ln\bigg|\frac{{\tilde {\mathbf x}}_i-{\tilde {\mathbf x}}_j}{\tilde R}\bigg|.  
\label{eq:2DH_virtual}
\end{equation}
The minimal set of dimensionless parameters in 2D  is 
given by the Manning parameter, $\xi$,
total number of counterions, $N$,  and  the lateral extension parameter $\Delta$. 
The range of simulation parameters and other details
are consistent with those given in Section \ref{subsec:sim_para}.

%%%%%%%%%%%%%%%%%%%%%%%%%%%%%%%%%%%%%%%%%%%%%%%%%%%%%%%%%%%%%%%%%%%%%

%                                2D SIMULATIONS                     

%%%%%%%%%%%%%%%%%%%%%%%%%%%%%%%%%%%%%%%%%%%%%%%%%%%%%%%%%%%%%%%%%%%%%%
\section{Simulation results in 2D}
\label{sec:2D_sim}

%%%%%%%%%%%%%%%%%%%%%%%%%%%%%%%%%%
\subsection{The order parameters}
\label{subsec:2D_op}

We consider the same set of order 
parameters $S_n=\langle 1/\tilde r^n\rangle$ 
as defined in Eq. (\ref{eq:S_n_def}) to characterize the CCT in 2D. 
They  can be measured in the simulations as
\begin{equation}
    S_n=\frac{1}{N}\sum_{i=1}^{N}{\overline{{\tilde r}_i^{-n}}} 
  \label{eq:S_n_2Ddef}
\end{equation}
for $n>0$, where the bar sign denotes the MC time average after 
proper equilibration of the system. 
Of particular interest is the behavior of $S_n$ as a function of Manning parameter, $\xi$,
which identifies the two regimes of complete de-condensation (with vanishing $S_n$) and
partial condensation (with $S_n>0$) 
as $\Delta\rightarrow \infty$. Unlike in 3D, where $\Xi$ can be varied as an independent parameter, 
various coupling regimes in the 2D system are spanned by changing the number of particles,
$N$,  for a given $\xi$ (see Eq. (\ref{eq:2DENC})).

\begin{figure}[t]
\begin{center}
\includegraphics[angle=0,width=8cm]{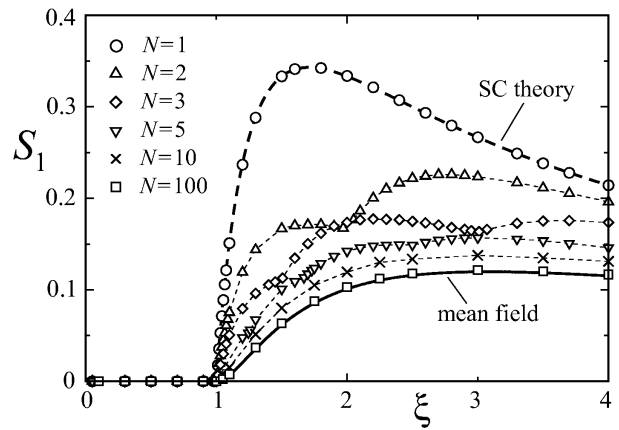}
\caption{Order parameter $S_1=\langle 1/{\tilde r}\rangle$ as a function of Manning parameter, $\xi$,
for the 2D counterion-cylinder system. Symbols show simulation data for different number of particles
as indicated on the graph. 
The mean-field PB (solid curve) and the strong-coupling (dashed curve)
predictions are obtained from Eqs. (\ref{eq:S_n_PB}) and (\ref{eq:Sn_SC_full}) respectively, which are valid
in 2D as well. The lateral extension parameter here is $\Delta=300$. 
Thin dashed curves are guides to the eyes. 
 }
\label{fig:2Ds1}
\end{center}
\end{figure}

The 2D simulation results for the order parameter $S_1$ are shown in Figure \ref{fig:2Ds1}
for increasing number of particles $N=1, 2, 3, 5, 10$ and 100 (symbols)
and for a large lateral extension parameter $\Delta=300$.
As seen for the smallest number of counterions, $N=1$, the data trivially follow the strong-coupling
prediction, Eq. (\ref{eq:Sn_SC_full}), shown by the 
dashed curve (Section \ref{sec:CCT_SC}). 
For increasing $N$, $S_1$ decreases and for sufficiently large values,  the 
data converge to the mean-field
PB prediction, Eq. (\ref{eq:S_n_PB}), shown by the solid curve.  
This in fact occurs for the whole range of Manning parameters
and thus confirms the trend predicted from the 2D
electroneutrality condition (\ref{eq:2DENC}). 
Accordingly, scaling analysis of the order parameters for large $N$ gives identical results for the critical  exponents
as in 3D (Sections \ref{subsec:scaling} and \ref{subsec:sim_exp}) and thus 
in agreement with the mean-field theory, which we shall not discuss here any further. 
The result that the mean-field theory for the counterion-cylinder system 
is {\em exact} in 2D for $N\rightarrow \infty$ is in striking contrast
with the typical trend in bulk  phase transitions \cite{critical}, and also  with the situation in 3D, where
the strong-coupling effects become important in the condensation phase ($\xi>1$) for growing $\Xi$ 
(Section \ref{subsubsec:corr}).

%%--------Fig S1 in 2D

The order-parameter data  in  Figure \ref{fig:2Ds1}, on the other hand,  reveal
a peculiar set of cusp-like singularities, that are quite  pronounced 
for small number of particles. These points become strictly singular in the limit
$\Delta\rightarrow \infty$ and represent the Manning parameters at which individual
counterions successively condense (or de-condense). 
We will demonstrate this point using 
an analytical approach in Section \ref{subsec:2D_singular}. 
(A similar singular behavior is also found in 3D for small $N$,
but the behavior in 3D appears to be more complex 
and will not be considered in this paper).

%%--------Fig E in 2D
\begin{figure*}[t]
\begin{center}
\includegraphics[angle=0,width=10cm]{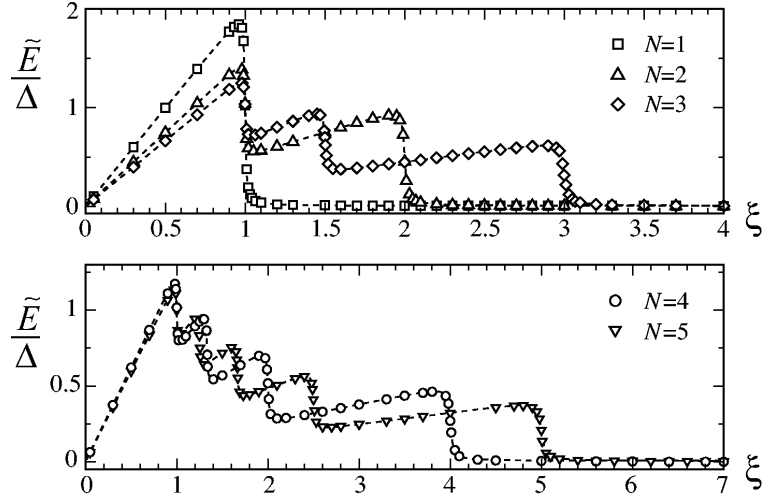}
\caption{The rescaled (internal) energy of the 2D counterion-cylinder system, $\tilde E=E_N/(Nk_{\mathrm{B}} T)$, as 
a function of Manning parameter, $\xi$, for different number of particles as indicated on the graph. 
Singular regions represent successive condensation (de-condensation) of counterions--see Section 
\ref{subsubsec:2D_Zsing}. The lateral extension parameter for these data is  $\Delta=300$. 
The dashed curves are the analytical results given by Eq. (\ref{eq:2D_energy_asymp}).
}
\label{fig:2Denergy}
\end{center}
\end{figure*}

%%--------Fig C in 2D
\begin{figure*}[t]
\begin{center}
\includegraphics[angle=0,width=10.7cm]{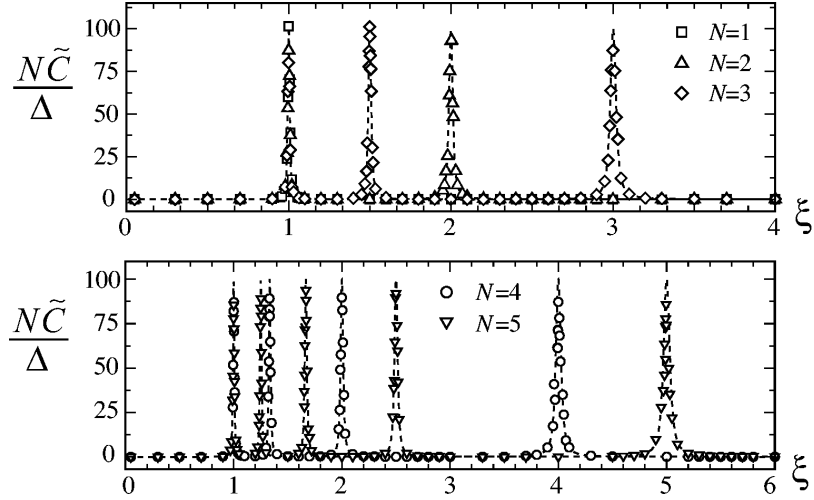}
\caption{The rescaled excess heat capacity of the 2D system, $\tilde C=C_N/(Nk_{\mathrm{B}} )$,  as 
a function of Manning parameter, $\xi$, for different number of particles, $N$, as indicated on the graph (for
clarity,  the data are here multiplied by $N$). 
The peaks represent successive condensation (de-condensation) of counterions--see Section 
\ref{subsubsec:2D_Zsing}. For these data $\Delta=300$. The dashed curves are
the analytical results given by Eq. (\ref{eq:2D_heat_asymp}).
}
\label{fig:2Dheat}
\end{center}
\end{figure*}

%%%%%%%%%%%%%%%%%%%%%%%%%%%%%
\subsection{Energy and heat capacity}

The singularities at small particle number,  $N$,  appear also  in the internal energy and the heat capacity. 
In Figures  \ref{fig:2Denergy} and \ref{fig:2Dheat}, we plot the rescaled 
energy, $\tilde E=E_N/(Nk_{\mathrm{B}} T)$, and excess heat capacity, 
$\tilde C=C_N/(Nk_{\mathrm{B}})$, obtained from the simulations using Eqs. (\ref{eq:E_def}) and (\ref{eq:C_def}) and the 2D Hamiltonian
(\ref{eq:2DHamilt_res}),  as a function of $\xi$ and  for $N=1, 2, 3, 4$ and 5. 
As seen, the energy shows a sawtooth-like structure for increasing $\xi$ consisting of wide regular 
regions, in which the energy almost linearly increases, and narrow singular regions, 
where the energy rapidly drops. Recalling the thermodynamic relation,
\begin{equation}
   \xi\frac{\partial \tilde E}{\partial \xi}={\tilde E}-{\tilde C},
\end{equation}  
it follows that the excess heat capacity vanishes in the regular regions, but develops
highly localized peaks in the singular regions,  as also 
seen from the simulation data in Figure \ref{fig:2Dheat}.

%%%%%%%%%%%%%%%%%%%%%%%%%%%%%%%%%%%%
\subsection{Condensation singularities in 2D: an analytical approach}
\label{subsec:2D_singular}

In what follows, we present an approximate (asymptotic) analysis of 
the 2D partition, which elucidates the physical mechanism behind the singular behavior
in 2D. The rigorous analysis of the 2D problem  is still missing
and more systematic approximations have been developed recently \cite{Burak_unpub}.

%------------------------------------------------------
\subsubsection{The partition function}
\label{subsubsec:2D_Zsing}

Suppose  that  the Manning parameter is such that $N-m$ counterions  are firmly bound 
to the central cylinder (disk), while $m$ counterions have 
de-condensed to infinity, where $m=1,\ldots, N$. Using the 2D Hamiltonian 
(\ref{eq:2DHamilt_res}), the partition function  can exactly be written as
\begin{equation}
		{\mathcal Z}_N=\int \left[\prod_{i=m+1}^{N}  {\mathrm{d}} { {\mathbf x}}_i\right] 
			\exp\bigg\{-\frac{{\mathcal H}_{N-m}}{k_{\mathrm{B}}T}\bigg\}
				\times \prod _{l=1}^m{\mathcal Z}_N^{(l)}
\label{eq:2D_Z_split}
\end{equation}
in actual units, where ${\mathcal H}_{N-m}$ represents interactions among condensed counterions (labeled by 
$i=m+1,\ldots, N$),  and
\begin{equation}
	{\mathcal Z}_N^{(l)}=\int  {\mathrm{d}} { {\mathbf x}}_l\,
		\exp\bigg\{-2\xi \ln \bigg(\frac{r_l}{R}\bigg)+\frac{2\xi}{N}\sum_{i=l+1}^{N}\ln 
		\bigg|\frac{{ {\mathbf x}}_i-{ {\mathbf x}}_l }{R}\bigg|\bigg\}
\label{eq:Z_Nl_def}
\end{equation}
is the contribution from individual de-condensed counterions (labeled by 
$l=1,\ldots, m$). Assuming that the de-condensed
counterions are de-correlated form each other and also from the condensed counterions 
as they diffuse to infinity for $\Delta\rightarrow\infty$ (i.e. using 
$|{ {\mathbf x}}_i-{ {\mathbf x}}_l |\simeq  r_l$),  
${\mathcal Z}_N^{(l)} $ approximately factorizes as
\begin{eqnarray}
   {\mathcal Z}_N^{(l)} & \simeq & 2\pi \! \int \!r_l \, {\mathrm{d}} r_l 
			\exp\bigg\{\! -2\xi \ln  \bigg(\frac{r_l}{R}\bigg)+\frac{2\xi}{N}\sum_{i=l+1}^{N}\ln 
				 \bigg(\frac{r_l}{R}\bigg) \bigg\}
				\label{eq:Z_Nm_sing}
				 \nonumber\\
	& = & 2\pi  R^2\, \frac{\exp\big[2\big(1-\xi l /N\big)\Delta\big]-1}{2\big(1-\xi l/N\big)}. 
	\label{eq:Z_Nl_approx}
\end{eqnarray}
In the limit $\Delta\rightarrow \infty$,
${\mathcal Z}_N^{(l)}$ 
diverges for Manning parameters 
\begin{equation}
    \xi\leq \xi_l^s\equiv \frac{N}{l}, 
\end{equation}
which indicates de-condensation of the $l$-th counterion from the charged cylinder (see Section \ref{subsec:onsager}). 
Repeating the above argument for various number of de-condensed counterions, 
one finds a set of singular Manning parameters, 
\begin{equation}
\xi_l^s=\frac{N}{l}\,\,\,\,\,\,\,\,\,\,\,\,\,\,\,\,\,\,\,\,\,{\mathrm{for}}\,\,\,l=1,\ldots,N, 
\label{eq:sing_points}
\end{equation}
at which individual 
counterions de-condense from the charged cylinder.
These singular points coincide with the values obtained from our simulations
based on the full partition function (\ref{eq:2DZ_y}) for very large $\Delta$
(see Figures \ref{fig:2Ds1}-\ref{fig:2Dheat} and Table \ref{table:sing} 
for the numerical values).

%------------------------------------------------------
\subsubsection{Energy and heat capacity}
\label{subsubsec:2D_EC}

In general the partition function (\ref{eq:2D_Z_split}) can exactly  be written as 
\begin{equation}
   {\mathcal Z}_N= \prod _{l=1}^N{\mathcal Z}_N^{(l)}, 
  \label{eq:ZN_prod}
\end{equation} 
where ${\mathcal Z}_N^{(l)}$ is defined in Eq. (\ref{eq:Z_Nl_def}). 
For $\Delta\gg 1$, the dominant contribution to the internal energy and the heat capacity  comes from 
de-condensed counterions. Thus, in order to derive analytical expressions for 
energy and heat capacity on the leading-order for large $\Delta$, we shall use Eq. (\ref{eq:ZN_prod}) together
with the approximate expression  (\ref{eq:Z_Nl_approx}) for ${\mathcal Z}_N^{(l)}$.  
Hence, we obtain the leading-order contribution to the free energy ${\mathcal F}_N/(k_{\mathrm{B}}T)=-\ln {\mathcal Z}_N$ as 
\begin{equation}
   {\tilde {\mathcal F}}\equiv \frac{{\mathcal F}_N}{Nk_{\mathrm{B}}T}\simeq 
   				-\frac{1}{N}\sum_{l=1}^N \ln \frac{e^{2(1-\xi/\xi_l^s)\Delta}-1}{2(1-\xi/\xi_l^s)}
\label{eq:2D_free_asymp}
\end{equation}
for  $\Delta\gg 1$, and thereby the (rescaled) internal energy $\tilde E=\xi\partial \tilde {\mathcal F}/\partial \xi$ 
and the rescaled heat capacity 
$\tilde C=-\xi^2\partial^2 \tilde {\mathcal F}/\partial \xi^2$  are obtained as 
\begin{equation}
{\tilde  E}\simeq  \sum_{l=1}^N \frac{\xi}{N\xi_l^s}\bigg( \frac{2\Delta\,\exp [2(1-\xi/\xi_l^s)\Delta]}{\exp [2(1-\xi/\xi_l^s)\Delta]-1}  
  							-\frac{1}{1-\xi/\xi_l^s}\bigg), 
\label{eq:2D_energy_asymp} 
\end{equation}
\begin{equation}
 {\tilde  C}\simeq  \sum_{l=1}^N \bigg(\frac{\xi}{\xi_l^s}\bigg)^2\bigg(\frac{1}{(1-\xi/\xi_l^s)^2}
  						-\frac{\Delta^2} {\sinh^2 [2(1-\xi/\xi_l^s)\Delta]}   \bigg). 
\label{eq:2D_heat_asymp}
\end{equation}
The above expressions are shown in Figures \ref{fig:2Denergy} and \ref{fig:2Dheat}  for $\Delta=300$   
and for various  number of particles (dashed curves), which as seen
closely reproduce the behavior obtained in the simulations (symbols). 

Note that as an individual counterion de-condenses at $\xi_l^s$,  the internal energy suddenly jumps, since
 the de-condensing counterion gains a large amount of energy due to its logarithmic interaction
with the central cylinder.  The {\em regular} regions (between two successive jumps) in the energy curve  
are dominated by de-condensed counterions and thus exhibit linear scaling 
with $\Delta=\ln (D/R)$.  The asymptotic form of the energy in
these regions for $\Delta\rightarrow \infty$ is obtained from Eq. (\ref{eq:2D_energy_asymp}) as
\begin{equation}
 \lim_{\Delta\rightarrow \infty} \frac{\tilde E}{\Delta}=\frac{l(l+1)}{N^2}\xi\,\,\,\,\,\,\,\,\,\,\,\,\,\,{\mathrm{for}}\,\,\,  \xi_{l+1}^s<\xi<\xi_l^s.
\label{eq:2D_E_reg}
\end{equation}

The {\em singular} part of the energy corresponds to a narrow region around
each $\xi_l^s$, which (except for the uppermost singularity) is bounded between
a local minimum (slightly above $\xi_l^s$) and a local maximum (slightly below $\xi_l^s$).   
The approximate locations of these extrema are obtained  as
\begin{equation}
  \frac{ \xi_l^{\mathrm{min}}}{ \xi_l^s}\simeq 1+ \frac{1}{\sqrt{\Delta(l-1)}}\,\,\,\,{\mathrm{and}}\,\,\,\,
   \frac{ \xi_l^{\mathrm{max}}}{ \xi_l^s} \simeq 1 -\frac{1}{\sqrt{\Delta(l+1)}} 
\label{eq:2D_Emaxmin}
\end{equation}
using Eq. (\ref{eq:2D_energy_asymp}), and for large $\Delta$. 
The {\em energy jump}, $\delta \tilde E_l$, upon de-condensation of a counterions at 
$\xi=\xi_l^s$ is then given by 
\begin{equation}
  \delta \tilde E_l \equiv \tilde E(\xi_l^{\mathrm{max}})-\tilde E(\xi_l^{\mathrm{min}}) \simeq \frac{2\Delta}{N}. 
\label{eq:2D_Ejump}
\end{equation}
Note that this value can also  be obtained directly from Eq. (\ref{eq:2D_E_reg}).
For $\Delta\rightarrow \infty$ but  at {\em fixed} and {\em finite} $N$, the energy curve tends to a sharp 
sawtooth-like form as both $\xi_l^{\mathrm{min}}$ and $\xi_l^{\mathrm{max}}$ tend to $\xi_l^s$ producing $N$ 
strictly step-like singular points, at which the limiting  energy jump is 
\begin{equation}
  \lim_{\Delta\rightarrow \infty} \frac{\delta \tilde E_l}{\Delta}=\frac{2}{N}. 
\label{eq:2D_Ejump_limit}
\end{equation}

The heat capacity expression  (\ref{eq:2D_heat_asymp}), on the other hand, exhibits $N$ 
isolated peaks for $\Delta\gg 1$. The heat capacity at $\xi_l$ diverges as $\tilde C(\xi=\xi_l^s) \simeq \Delta^2/3$
with increasing $\Delta$, giving rise to $N$ limiting algebraic divergencies as 
\begin{equation}
  \lim_{\Delta\rightarrow \infty }\tilde C = \sum_{l=1}^N \left(1-\frac{\xi_l}{\xi}\right)^{-2}. 
\label{eq:2D_heat_limit}
\end{equation}

%%%%%%%%%%%%%%%%%%%%%%%%%%%%%%%%%%%%%%%%%%%%%%%%%%%%%%%%%%%%%%%%%%%%%%%
\subsection{Critical point and the continuum limit}
\label{subsec:2D_continuum}

The lower-most singularity located at $\xi=\xi^s_N$  
is associated with the de-condensation of  the ``last" counterion from the
charged cylinder. 
As shown above,  this singularity occurs at unity ($\xi^s_N=1$) when $\Delta\rightarrow \infty$ and is thus
independent of the number of counterions.  It therefore 
gives the exact location of the 2D critical point as $\xi_c=1$ when the continuum (thermodynamic) 
limit $N\rightarrow\infty$ is approached, which coincides with the mean-field prediction. Note that 
in analogy with the method used in Section \ref{subsec:xi_c}, 
$\xi_c$ can also be derived from the asymptotic value of the energy maximum location,  
Eq. (\ref{eq:2D_Emaxmin}),  for $l=N$, when $\Delta$ and $N$ both tend to infinity.

Equations (\ref{eq:2D_Ejump})-(\ref{eq:2D_heat_limit}) represent the asymptotic results when the system
size increases to infinity but the number of particles, $N$, is finite. In the converse limiting case, i.e. when
$\Delta$ is {\em large} and {\em fixed} but $N$ increases to infinity (continuum limit), all singularities
smoothen except for the one, which represents the critical point.  The limiting
 energy curve for $N\rightarrow \infty$ may be obtained as follows. First note that 
the width of the energy jump around each singularity tends to zero as indicated by Eq. (\ref{eq:2D_Ejump}). 
Secondly, the spacing between singular points $\xi_l^s$ (and thus the  width of regular regions for $\xi>1$)
tends to zero  (as $\sim 1/N$)  with increasing $N$.
Therefore, the energy at a given Manning parameter $\xi$ between two successive singularities,
$\xi_{l+1}^s<\xi<\xi_l^s$, is  approximately given by $\tilde E\simeq \tilde E(\xi=\xi_l^s)$, where the right hand side
is obtained from Eq. (\ref{eq:2D_energy_asymp}) as $\tilde E(\xi=\xi_l^s) = \Delta/\xi_l^s$. This implies that   
\begin{equation}
  \lim_{N\rightarrow \infty} {\tilde E} = \frac{\Delta}{\xi} 
\end{equation}
for $\xi\geq 1$ and sufficiently large $\Delta$. 
For small Manning parameter  $\xi< 1$, there are no singularities 
and from Eq. (\ref{eq:2D_energy_asymp}), we obtain 
\begin{equation}
  \lim_{N\rightarrow \infty} {\tilde E} =  2\xi\Delta \times \lim_{N\rightarrow \infty} \sum_{l=1}^N\frac{1}{N\xi_l^s}=\xi\Delta
\end{equation}
for large $\Delta$. These limiting results can also be obtained 
using Eq. (\ref{eq:2D_E_reg}). 

The energy curve in the continuum limit therefore coincides with the universal form 
obtained within the mean-field theory in Section \ref{subsubsec:EC_PB} (see Eq. (\ref{eq:energyPB_asym})). 
The heat capacity in this limit
follows from  Eq. (\ref{eq:E_C_relation}), and exhibits a universal jump at $\xi_c=1$ in agreement
with Eq. (\ref{eq:heatPB_asym}).

%-------------------------------------------------------------------------
%  Table 1
%-------------------------------------------------------------------------
\begin{table}[t]
\begin{center}
\begin{tabular*}{5.5cm}{l|lllll} 
\hline\hline
$N$ &  & $\{\xi^s_N,\ldots,\xi^s_1\}$& & & 
\\ \hline
1 & 1 &&&&\\
2 & 1 & 2 &&&\\
3 & 1 & 3/2 & 2 &&\\
4 & 1 & $4/3\simeq 1.33$ & 2 & 4 &\\
5 & 1 & $5/4=1.25$ & $5/3\simeq 1.67$ & 5/2 & 5\\
\hline
\end{tabular*}
\end{center}
\caption{\label{table:sing} 
Numerical values of the location of the singularities, Eq. (\ref{eq:sing_points}),  for the 2D counterion-cylinder
system for several number of particles (compare 
Figures  \ref{fig:2Ds1}-\ref{fig:2Dheat}). }
\vspace*{5mm}
\end{table}
%-------------------------------------------------------------------------

%%%%%%%%%%%%%%%%%%%%%%%%%%%%%%%%%%%%%%%%%%%%%%%%%%%%%%%%%%%%%%%%%%%%%%%
\subsection{The condensed fraction}
\label{subsec:2D_alpha_continuum}

The preceding results enable us to calculate  the limiting condensed
fraction of counterions, $\alpha(\xi)$, as well when $\Delta\rightarrow \infty$
and $N\rightarrow\infty$. 
For a given Manning parameter, $\xi$, and number of particles, $N$, the condensed fraction
$\alpha_N(\xi)$ is given by the number of singularities
located below  $\xi$, i.e.
\begin{equation}
    \alpha_N(\xi)\equiv 1-\frac{l}{N}\,\,\,\,\,\,\,\,\,\,\,\,\,\,\,\,\,{\mathrm{for}}\,\,\, \xi_{l+1}^s<\xi<\xi_l^s.
\label{eq:2D_alpha_def}
\end{equation}   
This fraction is trivially zero for $\xi<\xi_c=1$ as $\Delta\rightarrow \infty$. 
Using  Eqs. (\ref{eq:sing_points}) and (\ref{eq:2D_alpha_def}), we obtain the condition
\begin{equation}
     \alpha_N(\xi)-\frac{1}{N} <1-\frac{1}{\xi}< \alpha_N(\xi), 
 \label{eq:2D_alpha_noneq}
\end{equation}
which in the limit of infinite number of counterions  yields
\begin{equation}
    \alpha(\xi)\equiv \lim_{N\rightarrow \infty}\alpha_N(\xi)=1-\frac{1}{\xi}. 
\end{equation}
This is nothing but the mean-field or Manning condensed fraction, $\alpha_{\mathrm{M}}=1-1/\xi$ 
(Section \ref{subsubsec:CondRatio}).  
The finite-ion-number correction to this limiting value follows from Eq. (\ref{eq:2D_alpha_noneq}) as 
\begin{equation}
 \alpha_N(\xi)-\alpha(\xi)  \sim N^{-1}. 
\end{equation}

%%%%%%%%%%%%%%%%%%%%%%%%%%%%%%%%%%%%%%%%%%%%%%%%%%%%%%%%%%%%%%%%%%%%%

%                                DISCUSSION , CONCLUSION                  

%%%%%%%%%%%%%%%%%%%%%%%%%%%%%%%%%%%%%%%%%%%%%%%%%%%%%%%%%%%%%%%%%%%%%%
\section{Conclusion and Discussion}

In this paper, we present an extensive numerical analysis of 
the critical behavior of counterions  at a charged cylinder (counterion-condensation transition) 
in both two and three spatial dimensions.  Analytical results for the critical behavior are also
derived using the asymptotic theories of mean field (Poisson-Boltzmann) 
and strong coupling. 
The counterion-condensation transition (CCT) 
is regulated by  the dimensionless Manning parameter (rescaled inverse temperature), 
$\xi=q\ell_{\mathrm{B}} \tau$, and occurs at a critical threshold $\xi_c$, below which  
counterions completely unbind (de-condense) to infinity, but above
$\xi_c$, a finite fraction of counterions binds (or condenses) in the vicinity of the charged cylinder. 
Since the CCT criticality emerges asymptotically in the limit of infinite system size and infinitely many particles, 
we have employed  Monte-Carlo  (MC) simulation of a periodic 
cylindrical cell model in {\em logarithmic radial coordinates}, which gives rise to a 
powerful (centrifugal) sampling method for extremely large lateral system sizes within reasonable equilibration times. 
This constitutes the key part of the present numerical investigation, since the critical and universal aspects of the CCT 
within the cell model can only be captured for large {\em logarithmic} system size
$\Delta=\ln(D/R)\gg1 $ (with $D$ and $R$ being the outer boundary and 
the charged cylinder radii respectively). 

As main results, we have determined the precise location of the critical Manning parameter, $\xi_c$, 
the scaling universality class of the CCT and the singular behavior of  energy and heat capacity on a systematic
level and without suppressing inter-particle correlations. 
As shown both the mean internal energy and the excess heat capacity become singular
at the critical point. The excess heat capacity, which vanishes in the de-condensation 
phase,  shows a universal discontinuity (jump)
at the critical point indicating that the CCT is  a second-order 
transition, as also suggested in a recent mean-field study \cite{Rubinstein01}. 
In a finite system, these singularities 
appear in the form of a pronounced peak, the asymptotic behavior of which is  
used to determine the critical Manning parameter, $\xi_c$, in the simulations.  
On the other hand, the critical  exponents associated with the CCT are 
obtained using a combined finite-ion-number, $N$, and finite-size, $\Delta$, analysis 
of the order parameters $S_n= \langle {\tilde r}^{-n} \rangle$
(with $\tilde r=r/\mu$ being the radial distance from the cylinder axis in units of the Gouy-Chapman length, $\mu$).
These order parameters  represent  the mean inverse localization length of counterions.  
For $\xi<\xi_c$ and in an infinitely large system, $S_n$ vanishes, but takes a finite value above $\xi_c$, which
exhibits the scaling relation $S_n\sim \zeta^\chi$, where $\zeta=1-\xi_c/\xi$ 
is the reduced Manning parameter (the reduced temperature) and the exponent $\chi$
is determined as $\chi=2.0\pm 0.4$ within our MC simulations. 
In finite systems, $S_n$ does not vanish at $\xi_c$ and displays a power-law decay 
with increasing size parameters, $\Delta$ and $N$, 
as $S_n(\xi=\xi_c)=\Delta^{-\gamma}$ (when number of particles, $N$, is fixed) 
and $S_n(\xi=\xi_c)=N^{-\nu}$ (when lateral extension parameter, $\Delta$, is fixed), where
the critical exponents are determined  as $\gamma=2.0\pm 0.6$ and $\nu=2/3\pm 0.1$. 

The critical exponents are demonstrated to be universal, i.e. independent of the 
coupling strength, $\Xi$ (varied over several decades $0.1<\Xi<10^5$), 
and agree with the values we obtained from  the mean-field PB theory as  
 $\chi_{\mathrm{PB}}=2.0$ and $\gamma_{\mathrm{PB}}=2.0$ (note that the exponent $\nu$ is not
defined within mean-field theory).  Interestingly, 
the critical Manning parameter both in 3D and 2D is also found to be universal 
and given by the mean-field value $\xi_c=1$. 
Therefore,  in striking contrast with the typical situation in bulk critical phenomena, where 
deviations from mean-field theory due to fluctuations grow with 
decreasing space dimensionality, the CCT criticality is found to be described 
by the mean-field universality class both in 3D and 2D. 
Surprisingly, the mean-field theory is found to be {\em exact} in 2D in the whole range of Manning parameters (or temperatures) 
when $N\rightarrow \infty$. In 3D,  however, correlation effects in the condensation phase ($\xi>\xi_c$)
lead to strong deviations from mean-field theory and support the strong-coupling predictions 
characterized by an excessive accumulation of counterions near the cylinder surface in agreement
with previous numerical studies \cite{Deserno00,Henle04,Bratko82,Zimm84,Rossky85} 
and  experiments \cite{Zakha99}.   
An important result is that the large-distance form of the density profile remains unaffected by
these correlations and  thereby a universal condensed fraction is obtained when the 
inflection-point criterion is applied.

To our knowledge, rigorous analytical derivation of the critical Manning parameter or the
scaling exponents of the CCT in 3D is not yet available and our simulations provide the first numerical 
results for the universal and critical features of this transition in the large-system-size limit \cite{Naji_CC}.  
In 2D, we have shown that the simulation results can be 
understood using an approximate analytical theory.  
The present predictions for  order parameters and thermodynamic quantities can be examined in experiments. In particular, the order parameters may be obtained 
from the distribution of counterions around charged polymers,
which has been directly measured using anomalous scattering techniques \cite{Das}.

In this study, we have made use of an idealized (primitive) cell model  \cite{Alfrey51,Fuoss51,Lifson54}
 in order to bring out the main {\em universal} aspects of the CCT.  
It  is interesting to  examine possible  effects due to more specific factors that exist in realistic situations, namely, the 
discrete charge pattern of polymers \cite{Liu_Muthu02,Liao03,de-la-Cruz95,Manning01,Henle04,Zimm84,Jayaram90}, 
chain flexibility and finite contour length \cite{Ramanathan_Wood82a,Ramanathan_Wood82b,Winkler98,Liu_Muthu02,Liao03,de-la-Cruz95,Schiess98,Nyquist99,Muthu04,Jayaram90,Stevens95} as well as the influence of non-uniform dielectric boundaries \cite{Muthu04}
on the critical behavior.  
However, the present results already indicate that short-range effects such as electrostatic correlations
do not affect the properties of the system near the critical region ($\xi\sim \xi_c$), since most  of counterions
are de-condensed and  the critical behavior is predominately determined by long-range features. 
Future studies will be useful to address the possible influence of short-range specific
effects on the critical behavior  \cite{Naji_unpub}. 

In this work, we have not investigated the role of additional salt and co-ions, which lead to screening
of electrostatic interactions (see,  e.g.,  Refs. \cite{Deserno00,Liu_Muthu02,Zimm84,Rossky85}). 
It is known that the Debye screening length, $r_s$, plays the role of an upper bound cut-off (similar to the
outer boundary $D$ in the cell model): 
 the CCT occurs for vanishing salt concentration, i.e. when $\ln (r_s/R)\rightarrow \infty$ \cite{Manning69,Macgillivray,Ramanathan,Shaugh}. 
Thus we can expect similar asymptotic behaviors arise near $\xi_c=1$  and within a similar model as used here 
when  the vanishing-salt limit is approached \cite{Naji_unpub}.

%%%%%%%%%%%%%%%%%%%%%%%%%%%%%%%%%%%%%%%%%%%%%%%%%%%%%%%%%%%%%%%%

%                                        ACKNOWLEDGEMENT

%%%%%%%%%%%%%%%%%%%%%%%%%%%%%%%%%%%%%%%%%%%%%%%%%%%%%%%%%%%%%%%%
\begin{acknowledgments}
We are grateful to H. Boroudjerdi, Y. Burak, G.S. Manning, A.G. Moreira,  H. Orland, R. Podgornik, and Y.Y. Suzuki
for useful discussions and valuable comments. We acknowledge financial support from 
the DFG German-French Network. 
\end{acknowledgments}

%%%%%%%%%%%%%%%%%%%%%%%%%%%%%%%%%%%%%%%%%%%%%%%%%%%%%%%%%%%%%%%%

%                                       APPENDICES

%%%%%%%%%%%%%%%%%%%%%%%%%%%%%%%%%%%%%%%%%%%%%%%%%%%%%%%%%%%%%%%%
\appendix

%%%%%%%%%%%%%%%%%%%%%%%%%%%%%%%%%%%%%%%%%%%%%%%%%%%%%%%%%%%%%%%%
\section{Singularities associated with the Onsager instability}
\label{app:Onsager}

The rescaled energy, $\tilde E=E_N/(Nk_{\mathrm{B}}T)$, and  
excess heat capacity, $\tilde C=C_N/(Nk_{\mathrm{B}})$, associated with the Onsager instability
can be calculated from Eq.  (\ref{eq:Z_single}). The results coincide 
with those obtained in Section \ref{subsec:2D_singular}  by choosing $N=1$ (see
also Figures \ref{fig:2Dheat} and \ref{fig:2Denergy}). 
In brief, the approximate location of the energy peak $\xi_\ast^E$ is obtained as
\begin{equation}
  \xi_\ast^E \simeq 1-\frac{1}{\sqrt{2\Delta}},
\end{equation}
for finite large $\Delta$, which shows a different asymptotic convergency (from below) 
to the critical value $\xi=1$ as compared with the mean-field result, Eq. (\ref{eq:Emax_asymp}).   
The heat capacity peak is located above the critical point at
\begin{equation}
  \xi_\ast^C \simeq 1+\frac{5}{\Delta^2}. 
\end{equation}
In the limit $\Delta\rightarrow \infty$, the heat capacity 
diverges at the critical point from above and below as 
\begin{equation}
  {\tilde C}\sim \zeta^{-2}
\end{equation}
where $\zeta=1-1/\xi$. The left tail of energy (for $\xi<\xi_\ast^E$) goes to infinity linearly with $\Delta$
as $\tilde E\simeq 2\xi \Delta +\xi/(\xi-1)$, but its right tail shows a power-law divergency as
\begin{equation}
  {\tilde E}\sim \zeta^{-1}. 
\end{equation}
Note that these behavior are distinctly different from
those obtained in the simulations with many particles (Section \ref{subsec:xi_c}) and within the mean-field 
theory (Section \ref{subsubsec:EC_PB}). 
   
%%%%%%%%%%%%%%%%%%%%%%%%%%%%%%%%%%%%%%%%%%%%%%%%%%%%%%%%%%%%%
\section{Rescaled PB equation}
\label{app:PB_rescal}

The Poisson-Boltzmann equation for mean electrostatic potential $\psi_{\mathrm{elec}}$
in  actual units reads 
\begin{equation}
  \nabla^2_{\mathbf x} \psi_{\mathrm{elec}}=
        \frac{\sigma({\mathbf x}) e}{\varepsilon \varepsilon_0}
         -\frac{q e \rho({\mathbf x})}{\varepsilon \varepsilon_0},
\end{equation}
where the density profile of counterions is given by 
\begin{equation}
  \rho({\mathbf x})=\rho_0\, \Omega({\mathbf x}) \,
      e^{-q e \beta \psi_{\mathrm{elec}}}, 
\label{eq:app_actual_den}
\end{equation}
with $\rho_0$ being a normalization prefactor and $\Omega({\mathbf x})$ specifying the volume
accessible to counterions (Eq. (\ref{eq:Omega})). The rescaled PB equation (\ref{eq:PB}) 
is recovered using above equations and the dimensionless parameters as
${\tilde {\mathbf x}}={\mathbf x}/\mu$,
${\tilde \sigma}({\tilde {\mathbf x}})=\mu \sigma({\mathbf x})/\sigma_{\mathrm{s}}$, 
${\tilde \rho}({\tilde {\mathbf x}})=\rho({\mathbf x})/(2\pi\ell_{\mathrm{B}}\sigma_{\mathrm{s}}^2)$, 
and also $\psi=qe \psi_{\mathrm{elec}}/(k_{\mathrm{B}} T)$, 
where $\mu=R/\xi=1/(2\pi q \ell_{\mathrm{B}}\sigma_{\mathrm{s}})$ is the Gouy-Chapman length (Eq. (\ref{eq:mu})).
The prefactor ${\rho_0}$ is related to ${\tilde \kappa}$ in Eq. (\ref{eq:PB}) 
through ${\tilde \kappa}=\kappa\mu$ where $\kappa^2=4\pi q^2 \ell_{\mathrm{B}}  \rho_0$.
One can show that the density profile  
in Eq. (\ref{eq:app_actual_den}) is also mapped to 
the rescaled density (\ref{eq:density}) in the text. 

Normalization of the density (to the total number of counterions) 
in actual units reads $ \int {\mathrm{d}}{\mathbf x} \, \rho({\mathbf x})=N$,
and in rescaled units, we have  $\int {\mathrm{d}}{\tilde {\mathbf x}}\, {\tilde \rho}({\tilde {\mathbf x}}) =2\pi\Xi N$,
which is equivalent to Eq. (\ref{eq:Lambda_kappa})
in the text, when Eqs.  (\ref{eq:ENC}) and (\ref{eq:density}) are used. 

%%%%%%%%%%%%%%%%%%%%%%%%%%%%%%%%%%%%%%%%%%%%%%%%%%%%%%%%%%%%%%
\section{Asymptotic results within the PB theory}
\label{app:PB}

%------------------------------------------------------------
\subsection{Limiting behavior of $\beta$ for large $\Delta$}
\label{app:PB_beta}

A general discussion has been given by Fuoss et al.
\cite{Fuoss51} for the overall behavior of $\beta$ using Eq. (\ref{eq:beta}). Here, 
we first review some of their results as quoted in the text, and then 
derive the finite-size scaling relations for $\beta$ near the PB critical point ($\xi_c^{\mathrm{PB}}=1$) 
as used in Section \ref{subsec:PB_exp}. 

%---------------
\subsubsection{Small  Manning parameter $\xi < \Lambda_{\mathrm{AF}}$:}

The integration constant $\beta$ 
vanishes at $\xi=\Lambda_{\mathrm{AF}}$ and   tends to unity, $\beta\rightarrow 1^-$, 
for small $\xi\rightarrow 0^+$, as it can be checked easily from Eq.  (\ref{eq:beta})
(we arbitrarily choose $\beta\geq 0$) \cite{Fuoss51}. 
Further inspection shows that 
in this regime, $\beta\rightarrow (\xi-1)^-$ when 
$\Delta\rightarrow \infty$ \cite{Fuoss51}.
Hence for $\Delta\gg 1$, one can propose the following form
\begin{equation}
    \beta^2  \simeq 
                    (\xi-1)^2(1-x), 
\label{eq:app_beta_x}
\end{equation} 
where $x$ is a small function of $\xi$ and $\Delta$. To determine $x$, 
we may rearrange the first equation in (\ref{eq:beta}) as  
\begin{equation}
    \beta \Delta=  \frac{1}{2}\ln 
                                \frac{1-\beta}{1+\beta}
                      - \frac{1}{2}\ln 
                                \frac{(\xi-1)+\beta}{(\xi-1)-\beta}.
\label{eq:app_beta_new}
\end{equation}
and use this together with Eq. (\ref{eq:app_beta_x}) to obtain  
\begin{equation}
         x \simeq 
                \frac{4\xi}{2-\xi} e^{2(\xi-1)\Delta},
\label{eq:app_x}
\end{equation}
which reproduces  Eq. (\ref{eq:beta_exp_low}) in the text. 

%----------------
\subsubsection{Large  Manning parameter $\xi > \Lambda_{\mathrm{AF}}$:}

For large Manning parameter $\xi > \Lambda_{\mathrm{AF}}$,
 $\beta$, Eq. (\ref{eq:beta}), tends to a finite upper bound 
$\beta_\infty=\pi/\Delta$, when $\xi\rightarrow \infty$ \cite{Fuoss51}. 
Thus for $\Delta\rightarrow \infty$,
$\beta$ vanishes for the whole range of Manning parameters 
$\xi > \Lambda_{\mathrm{AF}}\simeq 1$ as used frequently in the text 
(see e.g. in Eq. (\ref{eq:Sn_PB_zeta})).

%----------------
\subsubsection{Finite-size scaling for $\beta$ close to $\xi_c^{\mathrm{PB}}$:}

Of particular importance in our analysis is the behavior of $\beta$ close to $\xi_c^{\mathrm{PB}}=1$.
(Since always $\Lambda_{\mathrm{AF}}\leq 1$, we restrict the discussion only to 
the regime $\xi\geq\Lambda_{\mathrm{AF}}$). Analysis of Eq. (\ref{eq:beta})
shows that for sufficiently large $\Delta$,
we have $\beta\simeq \pi/(2\Delta)$ right at the critical point $\xi_c^{\mathrm{PB}}=1$. We may
then perform a Taylor expansion around $\xi_c^{\mathrm{PB}}$ to obtain the approximate
form of $\beta$ for small $\zeta=1-\xi_c^{\mathrm{PB}}/\xi$. We find
\begin{equation}
  \beta(\zeta, \Delta)=\frac{\pi}{2\Delta}+\frac{2}{\pi}\zeta
               -\frac{8\Delta}{\pi^3}\zeta^2+{\mathcal O}(\zeta^3),
\label{eq:app_beta_exp_reg}
\end{equation}
which remains valid for $\zeta\Delta<\pi^2/4$. This relation clearly indicates a
scale-invariant form for $\beta$ when $\Delta\gg 1$. Comparing this with Eq. (\ref{eq:beta_FSS}), 
we find the approximate form of the scaling function $  {\mathcal B}(u)$ as
\begin{equation}
  {\mathcal B}(u)\simeq \frac{\pi}{2}+\frac{2}{\pi}u
               -\frac{8}{\pi^3}u^2,
\label{eq:app_exp_B}
\end{equation}
where $u=\zeta\Delta<\pi^2/4$. In particular, we have 
${\mathcal B}(u)\rightarrow \pi/2$ as $u\rightarrow 0$. 

The asymptotic behavior of ${\mathcal B}(u)$ for  $u\rightarrow \infty$ (or equivalently 
$\Delta\rightarrow \infty$ for finite $\zeta$) can 
be obtained using a different series expansion, since in this limit 
$\beta$ becomes singular at $\xi_c^{\mathrm{PB}}=1$ and the above expansion breaks down. 
This is because $\beta$ is always singular (with an infinite first
derivative) at $\Lambda_{\mathrm{AF}}$ which tends to the critical Manning parameter  \cite{Fuoss51}.
We thus perform an expansion around $\xi=\Lambda_{\mathrm{AF}}$, which yields
$\beta\simeq \sqrt{3\zeta/\Delta}$. This gives the asymptotic
form of the scaling function as
\begin{equation}
  {\mathcal B}(u)\simeq \sqrt{3u}\,\,\,\,\,\,\,\,\,\,\,\,\, u\rightarrow \infty. 
\end{equation}

%-----------------------------------------------------
\subsection{The PB cumulative density profile}
\label{app:PB_cumu}

The PB cumulative density of 
counterions, $n_{\mathrm{PB}}(y)$, is defined through Eq. (\ref{eq:PB_cumu}). It
 can be easily shown that $n_{\mathrm{PB}}(y)$ 
is a monotonically increasing function of $y=\ln (r/R)$, i.e. ${\mathrm{d}}n_{\mathrm{PB}}/{\mathrm{d}}y\geq 0$ \cite{Deserno00}.
It is therefore bounded by its boundary values $n_{\mathrm{PB}}(0)=0$ and $n_{\mathrm{PB}}(\Delta)=N$
(see Figure \ref{fig:cumu}). 

For $\xi>\Lambda_{\mathrm{AF}}$, $n_{\mathrm{PB}}(y)$ has an inflection point the location of 
which, $y_\ast=\ln(r_\ast/R)$, follows from equation ${\mathrm{d}}^2 n_{\mathrm{PB}}/{\mathrm{d}}y^2=0$ as 
$y_\ast= \tan^{-1}[(\xi-1)/\beta]/\beta$. 
It is easy to check (using the results in Appendix \ref{app:PB_beta}) 
that $y_\ast\simeq \Delta/2$ for large $\xi$  and that $y_\ast\rightarrow 0$ 
for  $\xi\rightarrow 1$ ($y_\ast$ becomes negative for smaller values). 
For $\xi<\Lambda_{\mathrm{AF}}$, on the other hand, 
the cumulative density, $n_{\mathrm{PB}}(y)$, vanishes  
for $y<\Delta$ as $\Delta\rightarrow \infty$, as can be checked by inserting 
the approximate expression (\ref{eq:app_beta_x}) for $\beta$ into Eq. (\ref{eq:PB_cumu}). 

We note that the main quantities of interest within the PB theory can be expressed solely
in terms of the cumulative density profile. This includes the PB potential field 
$\psi_{\mathrm{PB}}$ and the order parameters
$S_n^{\mathrm{PB}}$. 
Using the definitions of these quantities (Section \ref{sec:CCT_PB}), we obtain the following relations
(which are valid for all $\xi$)
\begin{eqnarray}
 &&  \psi_{\mathrm{PB}}(y)= 2\xi \bigg(y-\frac{1}{N}\int_0^y\, n_{\mathrm{PB}}(y')\, {\mathrm{d}}y' \bigg) \\
  && S_n^{\mathrm{PB}}= \frac{1}{\xi^n}\int_0^\Delta\, {\mathrm{d}}y\, e^{-ny}\bigg(\frac{1}{N}
  	\frac{{\mathrm{d}} n_{\mathrm{PB}}}{{\mathrm{d}}y}\bigg). 
 \label{eq:app_Sn_cumu_rel}
\end{eqnarray}

%-----------------------------------------------------
\subsection{Asymptotic behavior of  $S_n$ within PB theory}
\label{app:PB_S_n}

%-----------------
\subsubsection{Large Manning parameter $\xi>\Lambda_{\mathrm{AF}}$:}

Consider the exact mean-field expression  for $S_n^{\mathrm{PB}}$, Eq. (\ref{eq:S_n_PB}). 
The integrand in Eq. (\ref{eq:S_n_PB}) is the product of an exponentially decaying factor 
with an inverse-squared sine-function, which has a series of divergencies at
$y_m=(m\pi-\epsilon)/\beta$ for integer $m$ and $\epsilon\equiv \cot^{-1}[(\xi-1)/\beta]$. 
For  $\Delta\rightarrow \infty$, we know from Appendix \ref{app:PB_beta} that 
$\beta\rightarrow 0$  when $\xi>\Lambda_{\mathrm{AF}}\simeq 1$ implying
$\epsilon\rightarrow 0$. In this limit, $\epsilon$ may be expanded as 
\begin{equation}
   \epsilon\simeq \frac{\beta}{\xi-1}-\frac{\beta^3}{3(\xi-1)^3}+{\mathcal O}(\beta^5).
\label{eq:eps_beta}
\end{equation}
The location of singularities, $y_m$, tend to infinity with increasing $\Delta$
except for $m=0$ for which we have $y_0=-\epsilon/\beta\rightarrow -1/(\xi-1)$ using Eq. (\ref{eq:eps_beta}). 
The quantity $S_n^{\mathrm{PB}}$ in Eq. (\ref{eq:S_n_PB}) 
is therefore dominated by the lower bound of the integral (around $y=0$)
due to the exponentially-decaying integrand.  To derive the asymptotic form  of 
$S_n^{\mathrm{PB}}$ for large $\Delta$, one can expand the integrand either around the lower limit of the integral
$y=0$ or around the singular point $y_0$. Both procedures lead to the same
scaling relation (\ref{eq:Sn_PB_zeta}) for $S_n$ in the strict limit of $\Delta\rightarrow \infty$ when $\xi$ is {\em close} 
to the critical value $\xi_c^{\mathrm{PB}}=1$. 
But only the second procedure leads to a  correct result when $\xi$ increases beyond  
the critical value. This is because for large $\xi$,  the singularity at $y_0\sim-1/(\xi-1)$ approaches  
zero rendering the expansion around $y=0$ a poor approximation.

By expanding the integrand around $y=0$ (up to the first order
in $y$), we obtain from  Eq. (\ref{eq:S_n_PB}) that
 \begin{eqnarray}
   S_n^{\mathrm{PB}}&\simeq&\frac{\beta^2}{\xi^{n+1}\sin^{2} \epsilon}
             \int_0^\Delta {\mathrm{d}}y \, e^{-(n+2\xi-2)y}\nonumber\\
             &\simeq&\frac{\beta^2+(\xi-1)^2}{\xi^{n+1}(n+2\xi-2)}.
\end{eqnarray} 
This relation reproduces Eq. (\ref{eq:S_n_PB_approx}) 
in the text, which, as explained above, is valid for $\xi$ close to $\xi_c^{\mathrm{PB}}=1$. 

For larger values of $\xi$, we expand the integrand in 
(\ref{eq:S_n_PB}) around $y_0=-\epsilon/\beta$, which yields
\begin{equation}
 S_n^{\mathrm{PB}}\simeq \frac{\beta/\epsilon}{\xi^{n+1}}\left[1-\frac{n \epsilon}{\beta}e^{n\epsilon/\beta}
               \Gamma(0,\frac{n\epsilon}{\beta})\right], 
\end{equation}  
where $\Gamma(a, b)$ is the incomplete Gamma function. 
This relation provides a quite good approximation for $S_n$ for large $\Delta$ and 
in the whole range of $\xi>\xi_c^{\mathrm{PB}}=1$. In particular, 
when the limit $\Delta\rightarrow \infty$ is taken, it yields the correct 
result for $S_n(\xi, \Delta\rightarrow \infty)$ (see Eq. (\ref{eq:app_Sinf_Sinf}) below).

%-------------------
\subsubsection{Small Manning parameter $\xi<\Lambda_{\mathrm{AF}}$:}

For $\xi<\Lambda_{\mathrm{AF}}\simeq 1$, the order parameters, $S_n$, vanish as $\Delta\rightarrow \infty$ 
indicating complete de-condensation of counterions. 
To demonstrate this, we use Eq. (\ref{eq:app_Sn_cumu_rel}), which,  for $\Delta\gg 1$, can be written as 
\begin{equation}
	\xi^n S_n^{\mathrm{PB}} = \frac{ n}{N} \int_0^\Delta\, {\mathrm{d}}y\, e^{-ny} n_{\mathrm{PB}}(y)+{\mathcal O}(e^{-n \Delta}).
\end{equation}
Since the cumulative density is bounded by the number  of counterions, $N$, 
and tends to zero at any finite $y$ for $\xi<\Lambda_{\mathrm{AF}}\simeq 1$ (Appendix \ref{app:PB_cumu}), 
we  get $S_n^{\mathrm{PB}} \rightarrow 0$ in this regime as $\Delta\rightarrow \infty$.

%--------------------------------------------------
\subsection{PB solution in an unbounded system ($\Delta=\infty$)}
\label{app:PB_Delta_infty}

In the present study, we have assumed that the counterion-cylinder system 
is bounded laterally ensuring the normalization of density profile, ${\tilde \rho}_{\mathrm{PB}}({\tilde r})$, 
to the total number of counterions, $N$, even in the limit $\Delta\rightarrow \infty$. 
In a strictly unbounded system (with $\Delta=\infty$),
the normalization property of density is not preserved, since a finite fraction of counterions escape to infinity. 
In this case, the PB equation (\ref{eq:PB}) can be solved by relaxing the normalization condition 
(\ref{eq:Lambda_kappa}). Assuming the boundary conditions at the cylinder surface as 
$\psi_{\mathrm{PB}}^{\infty}({\tilde R})=0$ and 
${\tilde R}[{\mathrm{d}}\psi_{\mathrm{PB}}^{\infty}({\tilde r}={\tilde R})/{\mathrm{d}} {\tilde r}]=2\xi$, one finds \cite{Netz_Joanny}
\begin{eqnarray}
  \psi_{\mathrm{PB}}^{\infty}({\tilde r}) = 
            \left\{
             \begin{array}{ll}
              2\xi\ln \frac{\tilde r}{\tilde R}
              & {\,\,\,\xi\leq\xi_c^{\mathrm{PB}}=1,}\\
            \\
                     2\ln \frac{\tilde r}{\tilde R}
                     + 2\ln \left[1+(\xi-1)\ln \frac{\tilde r}{\tilde R}\right]
              & {\,\,\,\xi\geq\xi_c^{\mathrm{PB}}=1.}
      \end{array}
        \right. 
\label{eq:app_kappa}
\end{eqnarray}
Also ${\tilde \kappa}^2={\tilde \rho}_{\mathrm{PB}}^{\infty}({\tilde R})=0$ for $\xi\leq\xi_c^{\mathrm{PB}}=1$ and 
${\tilde \kappa}^2/2=(\xi-1)^2/\xi^2$ otherwise.
Hence using Eq. (\ref{eq:density}), we have
 the density profile in a strictly unbounded system (for $\tilde R\leq \tilde r\leq \tilde D$)
\begin{eqnarray}
    {\tilde \rho}_{\mathrm{PB}}^{\infty}({\tilde r})= 
            \left\{
             \begin{array}{ll}
              0
              & {\xi\leq 1,}\\
            \\
               \frac{(\xi-1)^2}{\xi^2}
                        \left[\frac{\tilde r}{\tilde R}\right]^{-2}
                        \left[1+(\xi-1)\ln \frac{\tilde r}{\tilde R}\right]^{-2}
              & {\xi\geq 1,}
      \end{array}
        \right. 
\end{eqnarray}
which has the same form as given in Eqs. (\ref{eq:density_asyp_low}) 
and (\ref{eq:density_asyp_high}). But now ${\tilde \rho}_{\mathrm{PB}}^{\infty}({\tilde r})$
is normalized to the condensed fraction of counterions,  
$\alpha_{\mathrm{M}}$ (Eq. (\ref{eq:alpha_PB})), i.e.
\begin{eqnarray}
   \int_{\tilde R}^{\infty}{\mathrm{d}}{\tilde r}\,\,{\tilde r}\,
                   {\tilde \rho}_{\mathrm{PB}}^{\infty}({\tilde r})
           =\alpha_{\mathrm{M}}\xi
            = \left\{
             \begin{array}{ll}
              0
              & {\,\,\xi\leq\xi_c^{\mathrm{PB}}=1,}\\
            \\
               \xi-1
              & {\,\,\xi\geq\xi_c^{\mathrm{PB}}=1}
      \end{array}
        \right. 
\label{eq:app_norm}
\end{eqnarray}
(compare with Eq. (\ref{eq:density_norm})). 
The order parameters in the unbounded system, 
$S_n^{{\mathrm{PB}}, \infty}$, may be calculated using ${\tilde \rho}_{\mathrm{PB}}^{\infty}$. 
For $\xi>\xi_c^{\mathrm{PB}}=1$, we obtain
\begin{equation}
  S_n^{{\mathrm{PB}}, \infty}=\frac{1}{\xi^n}\left[1-\frac{n}{\xi-1}e^{n/(\xi-1)}
               \Gamma(0,\frac{n}{\xi-1})\right].
\label{eq:app_S_n_infty}
\end{equation}  
In the vicinity of the critical point ($\xi\rightarrow 1^+$),  we get
\begin{equation}
  S_n^{{\mathrm{PB}}, \infty}(\zeta)\sim \frac{\zeta}{n},
\label{eq:app_S_n_chi}
\end{equation}
which exhibits a different exponent as compared with the quantity
$S_n^{\mathrm{PB}}(\zeta, \Delta\rightarrow \infty)$ in Eq. (\ref{eq:Sn_PB_zeta}). 
This is again due to the difference in normalization factor, which enters in $S_n$ 
through Eq. (\ref{eq:S_n_def})  (note the order in which the integration and 
the infinite-system limit are taken). 
In general, the order parameter $S_n^{\mathrm{PB}}(\zeta, \Delta\rightarrow \infty)$
is obtained by multiplying $S_n^{{\mathrm{PB}}, \infty}(\zeta)$ with the condensed fraction 
$\alpha_{\mathrm{M}}$, as
\begin{equation}
 S_n^{\mathrm{PB}}(\zeta, \Delta\rightarrow \infty)=\alpha_{\mathrm{M}}S_n^{{\mathrm{PB}}, \infty}(\zeta).
\label{eq:app_Sinf_Sinf}
\end{equation}

%%%%%%%%%%%%%%%%%%%%%%%%%%%%%%%%%%%%%%%%%%%%%%%%%%%%%%%%%%%%
\section{Hamiltonian of a periodic cylindrical cell model (3D)}
\label{app:H_periodic}

As stated in Section \ref{subsec:sim_para}, the periodic boundary conditions used in the simulations in 3D lead to
summation of Coulombic interactions ($v_{\mathrm{3D}}({\mathbf x})=1/|{\mathbf x}|$)
over all periodic images. The resultant summation series 
are convergent for an {\em electroneutral} system  and can be mapped
to fast-converging series, which can be handled easily in the simulations \cite{Lekner,Sperb}. In what follows, we
derive the convergent expressions for the configurational Hamiltonian  (\ref{eq:H_full}). 

The main simulation box (of height $L$ and containing $N$ counterions) is replicated infinitely many times in  $z$ 
direction generating a series of $M\rightarrow \infty$ image boxes labeled with $m=-M/2,\ldots,-1,0,1,\ldots,+M/2$
(with $m=0$ being the main box). 
The Hamiltonian  (\ref{eq:H_full}) consists of three parts 
${\mathcal H}_N={\mathcal H}_{\mathrm{ci}}+{\mathcal H}_{\mathrm{int}}+{\mathcal H}_{\mathrm{self}}$, namely,   
counterion-counterion interaction, ${\mathcal H}_{\mathrm{ci}}$, counterion-cylinder interaction,  ${\mathcal H}_{\mathrm{int}}$, 
and the cylinder self-energy, ${\mathcal H}_{\mathrm{self}}$, that will be analyzed separately.  
(Here we use actual units and in the end, transform the results to the rescaled form.)

%---------------------------------------------------------
\subsection{${\mathcal H}_{\mathrm{int}}$ and ${\mathcal H}_{\mathrm{self}}$ terms}

The counterion-cylinder interaction part per simulation box reads
${\mathcal H}_{\mathrm{int}}/(Mk_{\mathrm{B}}T)=\sum_{\alpha=1}^N u(r_\alpha)$, where using
$\sigma({\mathbf x})=\sigma_{\mathrm{s}}\delta(r-R)$, we have
\begin{equation}
	u(r_\alpha)=-q \ell_{\mathrm{B}} \int v_{\mathrm{3D}}({\mathbf x}-{\mathbf x}_\alpha) \sigma({\mathbf x})
	  {\mathrm{d}}{\mathbf x} = 2\xi\ln \bigg(\frac{r_\alpha}{R}\bigg)+c_0
\end{equation}
with $\alpha$ running only over the counterions within the main box.
The constant term is given by 
\begin{equation}
  	c_0=-q \ell_{\mathrm{B}} \int v_{\mathrm{3D}}({\mathbf x}-{\mathbf x}_0) \sigma({\mathbf x})\, {\mathrm{d}}{\mathbf x},
\end{equation}
where ${\mathbf x}_0$ belongs to the cylinder surface. 
$c_0$ may be written in terms of the cylinder self-energy,
\begin{equation}
         \frac{{\mathcal H}_{\mathrm{self}}}{Mk_{\mathrm{B}}T} =  \frac{\ell_{\mathrm{B}}}{2} \int \, \sigma({\mathbf x})
         v_{\mathrm{3D}}({\mathbf x}-{\mathbf x}') \sigma({\mathbf x}') \, {\mathrm{d}}{\mathbf x}\, {\mathrm{d}}{\mathbf x}'. 
 \label{eq:app_self}
\end{equation}
Using the electroneutrality condition $\tau L=q N$ (per box), one can show that
$\beta {\mathcal H}_{\mathrm{self}}/(Mk_{\mathrm{B}}T) =-Nc_0/2$. Thus, we have  
\begin{equation}
         \frac{1}{Mk_{\mathrm{B}}T}\bigg( {\mathcal H}_{\mathrm{int}}+ {\mathcal H}_{\mathrm{self}}\bigg) =
         2\xi\sum_{\alpha=1}^N  \ln \bigg(\frac{r_\alpha}{R}\bigg)+C_0,
 \label{eq:app_Hint_Hself}
\end{equation}          
where $C_0=-\beta {\mathcal H}_{\mathrm{self}}/M$ is  a constant (see Eq. (\ref{eq:Hamilt})), which 
diverges logarithmically with $M$. This can be seen from the asymptotic behavior of the
self-energy for large $M$ (or large $ML/R$), i.e.
\begin{eqnarray}
	 \frac{ {\mathcal H}_{\mathrm{self}} /(k_{\mathrm{B}}T )}{\tau^2 \ell_{\mathrm{B}} M L }\!&=&\!\!
  	   \int_0^{2\pi} \! \frac{{\mathrm{d}}\theta{\mathrm{d}}\theta'}{4\pi^2} \!\!
		\int_{-\frac{ML}{2}}^{\frac{ML}{2}} \!  \frac{{\mathrm{d}}z\, {\mathrm{d}}z'/(2ML)}
				{\sqrt{(z-z')^2+4R^2\sin^2\frac{\theta-\theta'}{2}}}
		\nonumber\\
	&\simeq& a_0+\ln \bigg(\frac{ML}{R}\bigg)+{\mathcal O}\bigg (\frac{R}{ML}\bigg)^2,
\label{eq:app_Hself_approx}
\end{eqnarray}
where $a_0\simeq \ln 2 -1$. This logarithmic divergency
is canceled by a similar divergent term coming from the interaction between counterions (see below). 

%---------------------------------------------------------
\subsection{ ${\mathcal H}_{\mathrm{ci}}$ term and the Lekner-Sperb formulae}
\label{subsec:LS}

The contribution from counterionic interactions (per box and for large $M$) can be written as
\begin{eqnarray}
  \frac{ {\mathcal H}_{\mathrm{ci}}}{Mk_{\mathrm{B}}T}&=&\frac{q^2  \ell_{\mathrm{B}} }{2M}\sum_{i\neq j}
  		v_{\mathrm{3D}}({\mathbf x}_i-{\mathbf x}_j)=\\
				&=&\frac{q^2  \ell_{\mathrm{B}} }{2L}\bigg[
			\sum_{\alpha\neq \beta=1}^{N} S_M\bigg(\frac{{\mathbf x}_\alpha-{\mathbf x}_\beta}{L}\bigg)
			+NS_M^0\bigg], \nonumber
	\label{eq:app_Hcc}
\end{eqnarray}	
where $i$ and $j$ run over {\em all} counterions (including periodic images), 
while $\alpha$ and $\beta$ run over counterions
in the main simulation box. We also have $S_M^0=2\sum_{m=1}^{M/2} m^{-1}$ and 
\begin{equation}
	S_M\bigg(\frac{{\mathbf x}_\alpha-{\mathbf x}_\beta}{L}\bigg)=
		\sum_{m=-M/2}^{M/2}\bigg[\rho_{\alpha\beta}^2+(\zeta_{\alpha\beta}+m)^2\bigg]^{-1/2},
\end{equation}
where $\rho_{\alpha\beta}=[(x_\alpha-x_\beta)^2+(y_\alpha-y_\beta)^2]^{1/2}/L$ and 
$\zeta_{\alpha\beta}=(z_\alpha-z_\beta)/L$.
Note that $S_M^0$, in particular, represents the interaction between a counterion and its
periodic images, which are lined up in $z$ direction. This series diverges and may be written as
$S_M^0=2\ln (M/2)+2{\mathbf C}_e$ for $M\rightarrow \infty$,
where ${\mathbf C}_e=0.577215\ldots$ is the Euler's constant. 
In this limit,  $S_M$ is also divergent, but it may be split into a
convergent and a divergent part as 
\begin{equation}
  S_M=S_M^0+S_{\mathrm{LS}}(\rho_{\alpha\beta},\zeta_{\alpha\beta})+2(\ln 2- {\mathbf C}_e),
 \label{eq:ser_id}
\end{equation} 
in which the convergent series $S_{\mathrm{LS}}(\rho_{\alpha\beta},\zeta_{\alpha\beta})$
can be expressed in terms of special functions.

Now inserting the above results for $S_M$ and $S_M^0$ into Eq. (\ref{eq:app_Hcc}), we have
for $M\rightarrow \infty$,
\begin{equation}
   \frac{ {\mathcal H}_{\mathrm{ci}}}{Mk_{\mathrm{B}}T}=
   	\frac{q^2  \ell_{\mathrm{B}} }{2L}\sum_{\alpha\neq \beta=1}^N 
		S_{\mathrm{LS}}(\rho_{\alpha\beta},\zeta_{\alpha\beta})
	+\xi({\mathbf C}_e-\ln 2)+\xi N\ln M,
\label{eq:app_Hcc_LS}
\end{equation}
with a logarithmic divergent term (last term) from the one-dimensional
periodicity of the system. 
Using Eqs. (\ref{eq:app_Hself_approx}) and (\ref{eq:app_Hcc_LS}), it immediately follows that
the divergencies in Eqs. (\ref{eq:app_Hint_Hself}) and (\ref{eq:app_Hcc_LS}) 
cancel each other when the electroneutrality condition is assumed. 
Thus we have the well-defined expression 
\begin{equation}
   \frac{ {\mathcal H}_N}{Mk_{\mathrm{B}}T}= 2\xi\sum_{\alpha=1}^{N}\ln \bigg(\frac{r_\alpha}{R}\bigg)+
   	\frac{q^2  \ell_{\mathrm{B}} }{2L}\sum_{\alpha\neq \beta=1}^N 
		S_{\mathrm{LS}}(\rho_{\alpha\beta},\zeta_{\alpha\beta})+\xi N h_0,
\end{equation}
where $h_0=1+\ln (R/2L)+({\mathbf C}_e-\ln 2)/N$; equivalently,
\begin{equation}
   \frac{ {\mathcal H}_N}{Mk_{\mathrm{B}}T}= 2\xi\sum_{\alpha=1}^{N}\ln \bigg(\frac{\tilde r_\alpha}{\tilde R}\bigg)+
   	\frac{\Xi }{2\tilde L}\sum_{\alpha\neq \beta=1}^N 
		S_{\mathrm{LS}}(\tilde \rho_{\alpha\beta},\tilde \zeta_{\alpha\beta})+\xi N h_0,
\label{eq:app_H_LS}
\end{equation}
in rescaled units, where $h_0=1+\ln (\xi^2/2\Xi N)+({\mathbf C}_e-\ln 2)/N$. The above expression 
is used to obtain the internal energy
and the heat capacity of the system in our simulations (Section \ref{sec:sim}). 
The term $S_{\mathrm{LS}}(\tilde \rho_{\alpha\beta},\tilde \zeta_{\alpha\beta})$ 
may be obtained from Eq. (\ref{eq:ser_id})  using mathematical identities proposed 
by Lekner  \cite{Lekner} and Sperb \cite{Sperb}. It may be written in the form of 
two formally {\em identical} series expansion 
\begin{equation}
   S_{\mathrm{LS}}
            = \left\{
             \begin{array}{ll}
             {\mathrm{I:}}\, -2\ln \tilde \rho_{\alpha\beta}+4\sum_{m=1}^{\infty} K_0(2\pi m  \tilde \rho_{\alpha\beta})
              \cos(2\pi m  \tilde \zeta_{\alpha\beta}).\\
            \\
              {\mathrm{II:}}\, \sum_{m=1}^{\infty} {{-1/2} \choose m} \, (\tilde \rho_{\alpha\beta})^{2m}\, \bigg[ Z(2m+1, 1+\tilde \zeta_{\alpha\beta})\\
            	 \,\,\,\,\,\,\,\,\,\,\,\,\,\,\,-Z(2m+1, 1-\tilde \zeta_{\alpha\beta})\bigg]+( \tilde \rho_{\alpha\beta}^2+  \tilde \zeta_{\alpha\beta}^2)^{-1/2}\\
	   	\,\,\,\,\,\,\,\,\,\,\,\,\,\,-\Psi(1+\tilde \zeta_{\alpha\beta})-\Psi(1-\tilde \zeta_{\alpha\beta})-c_\ast.
                     \end{array}
        \right. 
 \label{eq:app_S_LS}
\end{equation}
In the above relations, $K_0(x)$ is the modified Bessel function of the second kind, 
$Z(n, x)=\sum_{k=0}^\infty 1/(k+x)^n$ is the Hurwitz Zeta function, $\Psi(x)$ is the
Digamma function, and $c_\ast=2\ln 2\simeq 1.386294\ldots$. 

The series in Eq. (\ref{eq:app_S_LS}) can be evaluated numerically up to the desired accuracy.
Note that the first series (Lekner scheme) involves the Bessel function $K_0(x)$, which
decays exponentially for large $x$ (as $\sim \exp(-x)/\sqrt{x}$) but diverges logarithmically
for small $x$. It is therefore rapidly converging when the radial distance between two given
particles, $\tilde \rho_{\alpha\beta}$, is sufficiently large. We use the following recipe to truncate series I:
For $\tilde \rho_{\alpha\beta}>3$, we truncate after the third term, for  $1/3\leq\tilde \rho_{\alpha\beta}\leq 3$,
we include $2+[3/\tilde \rho_{\alpha\beta}]$ terms in the sum (where $[x]$ refers to the integer part of $x$),
and for $1/4\leq \tilde \rho_{\alpha\beta}<1/3$, we sum at least 12 terms. 
This recipe ensures a {\em relative} truncation error of about $|e_r|\sim 10^{-10}$. 
For small radial separation $\tilde \rho_{\alpha\beta}<1/4$ between two particles,  series I
becomes inefficient and slow. We thus employ  the second series expression (Sperb scheme). 
This series is rapidly converging for small $\tilde \rho_{\alpha\beta}$ provided
that $\tilde \zeta_{\alpha\beta}$, which enters in the argument of the Hurwitz Zeta function, is sufficiently
small, namely for $|\tilde \zeta_{\alpha\beta}|\leq 1/2$ \cite{Sperb} (note that in general we have
$|\tilde \zeta_{\alpha\beta}|\leq 1$). In fact due to the periodicity of the system, 
the energy expression (\ref{eq:app_Hcc}) remains invariant under the transformations  
$\tilde \zeta_{\alpha\beta}\rightarrow 1-\tilde \zeta_{\alpha\beta}$  and 
$\tilde \zeta_{\alpha\beta}\rightarrow -\tilde \zeta_{\alpha\beta}$, and thus $\tilde \zeta_{\alpha\beta}$
can be always restricted  to the range  $|\tilde \zeta_{\alpha\beta}|\leq 1/2$. In this case, we use
up to 8 terms in series II. The relative truncation error, $|e_r|$, varies for
different $\tilde \zeta_{\alpha\beta}$, e.g., for $\tilde \zeta_{\alpha\beta}\simeq0.4$ and 
$\tilde \rho_{\alpha\beta}\simeq 0.25$, one has $|e_r|\sim 10^{-7}$. The error substantially 
decreases for smaller  $\tilde \zeta_{\alpha\beta}$. 
The above  truncation recipes yield accurate estimates for the interaction energies
within the statistical error-bars of the simulations.

%%%%%%%%%%%%%%%%%%%%%%%%%%%%%%%%%%%%%%%%%%%%%%%%%%%%%%%%%%%%%

%						References

%%%%%%%%%%%%%%%%%%%%%%%%%%%%%%%%%%%%%%%%%%%%%%%%%%%%%%%%%%%%%


\begin{thebibliography}{99}
\bibitem{Note_Ons} 
The Onsager divergency may equivalently occur when the lower bound of the integral 
(cylinder radius) tends to zero $R\rightarrow 0$ \cite{Manning69}, since only the ratio $D/R$ is 
relevant (through $\Delta=\ln (D/R)$).
\bibitem{Note_Sn} Note that the density moments for $n<0$ are divergent since for finite $\xi$, 
a finite fraction of the counterions is always de-condensed.
\end{thebibliography}
\end{document}